\documentclass{aa}

\usepackage{graphicx}
\usepackage{txfonts}
\usepackage{subcaption}
\usepackage{lscape}
\usepackage{placeins}
\usepackage[colorlinks=true, allcolors=blue]{hyperref}
\usepackage{orcidlink}
\usepackage{amsmath}
\usepackage{natbib}
\usepackage{multirow}
\usepackage{xcolor}

\usepackage{enumitem}

\begin{document}

   \title{Deconstructing the cosmic baryon distribution \\ with fast radio bursts}

\authorrunning{Konar et al.}
\titlerunning{Deconstructing the cosmic baryon distribution with FRBs}
  \author{Koustav~Konar\orcidlink{0000-0002-8236-1605}\inst{1, 2}\corrauth{konar@astro.rug.nl}
        \and Yin-Zhe Ma\orcidlink{0000-0001-8108-0986}\inst{2}\corrauth{mayinzhe@sun.ac.za}
        \and L\'{e}on~V.~E. Koopmans\orcidlink{0000-0003-1840-0312}\inst{1}\email{koopmans@astro.rug.nl}
        \and P.~Daniel Meerburg\orcidlink{0000-0002-6080-6845}\inst{3}\email{p.d.meerburg@rug.nl}}

   \institute{Kapteyn Astronomical Institute, University of Groningen, PO Box 800, 9700 AV Groningen, The Netherlands
   \and  Department of Physics, Stellenbosch University, Stellenbosch, Western Cape, South Africa
   \and Van Swinderen Institute for Particle Physics and Gravity, Nijenborgh 3, 9747 AG Groningen, The Netherlands
   }

   \date{\today}

  \abstract
   {The precise distribution of baryons in the Universe remains an open question in cosmology, as their highly diffuse nature hinders direct observation. However, mapping this distribution is integral for precision cosmology in studies of gravitational lensing, galaxy evolution and the cosmic microwave background radiation with current and upcoming surveys.}
   {We use fast radio bursts (FRBs) and their dispersion measures (DMs) to probe and localise this baryon distribution in the large-scale structure (LSS).}
   {Using the IllustrisTNG simulations, we classify the LSS into eight density bins of increasing matter density, analogous to voids, filaments, and haloes. We model the DM contribution from each substructure and compare it with observed DMs in the simulation to infer the density estimator, the ratio of electron number to total matter density. The resulting overdetermined system is solved within a Bayesian framework using a simulation-physics-informed model. To demonstrate observational applicability, we create a \textit{Euclid}-like mock galaxy catalogue and build a density estimator based on luminous matter along the line of sight of an FRB. We apply a bias correction for the underlying dark matter distribution and compare the two density estimators to recover the electron number density across the eight density bins.}
   {Electrons are predominantly localised in low-density regions, while the ensemble mean across all bins aligns with the global theoretical estimate. Combining the FRB and galaxy density estimators, we recover the density of ionised baryons as $\Omega_\mathrm{b} f_\mathrm{ion} = 0.0455^{+0.0028}_{-0.0026}$ (against $0.0462$ in the simulation), corresponding to $\Omega_\mathrm{b} = 0.0479^{+0.0029}_{-0.0027}$ for $f_\mathrm{ion}=0.95$. Our bin-wise electron density estimate reproduces the simulation ground truth to within 10 per cent and showcases the distribution of baryons in three regions of the LSS as $f_\mathrm{gas}^\mathrm{void} = 0.186 \pm 0.012$, $f_\mathrm{gas}^\mathrm{filament} = 0.179 \pm 0.015$ and $f_\mathrm{gas}^\mathrm{halo} = 0.046 \pm 0.014$ against ground truths of 0.167, 0.173 and 0.067, respectively.}
    {}

   \maketitle
   \nolinenumbers

\section{Introduction}
Fast radio bursts (FRBs) are bright and typically millisecond transient signals detected in the radio band of 100 MHz to 10 GHz with observed flux in the range of 50 mJy to 100 Jy \citep{thornton_population_2013, petroff_dawn_2022}. After being discovered by \citet{lorimer_2007_first_frb_detection}, the number of FRB detections has increased by orders of magnitude due to surveys such as the Australian Square Kilometre Array Pathfinder (ASKAP) \citep{mcConnel_ASKAP_2016}, the Canadian Hydrogen Intensity Mapping Experiment Fast Radio Burst project (CHIME/FRB) \citep{ abbott_ChimeCat2_2026}, Apertif \citep{marazuela_Apertif_FRB_2025}, Five-Hundred-Meter Aperture Spherical Radio Telescope (FAST) \citep{niu_repeating_2022}, MeerKAT \citep{caleb_Meerkat_2023}, FLIMFLAM \citep{khrykin_FLIMFLAM_2024} and DSA-110 \citep{law_DSA110_2024}. An updated compilation of FRB sources is available in \citet{xu_blinkverse_2023}. These discoveries have revealed that FRBs are predominantly of extragalactic origin, with distances spanning from 3.6 Mpc \citep{bharadway_nearby_frb_2021} to a redshift of $z=1.34$ \citep{connor_cosmic_2025}, or more recently discovered at $z = 2.148$ by \citet{caleb_z_2_FRB_2025}.

One characteristic observational property of an FRB is its dispersion measure (DM), which arises from the frequency-dependent phase velocity of electromagnetic waves propagating through the plasma in the intergalactic medium (IGM) \citep{rybicki_1979_radiative}. In the local Universe ($z<3$), where the baryonic component is largely ionised, it is proportional to the electron column density along the line of sight (LoS) and hence traces the underlying baryon distribution and scales with redshift.

This DM-$z$ or Macquart relation has been used to estimate the total baryon content of the Universe with $<10$ localised FRBs \citep{macquart_census_2020}, establishing it as a pivotal probe of the missing baryon problem, where the total baryon estimate from Big Bang nucleosynthesis (BBN) and the cosmic microwave background (CMB) \citep{cyburt_BBN_2016, cooke_determination_2018, aghanim_planck_VI_2020} do not coincide with the late-time observations \citep{fukugita_cosmic_2004, shull_baryon_2012}. A recent review of the baryon problem and possible solutions can be found in \citet{priyadharshini_solving_2025}.

Further studies suggest that the majority of baryons are expected to be in a diffuse form, occupying the low-density regions outside dark matter haloes or the warm-hot intergalactic medium (WHIM). Traditionally, techniques like the kinetic (kSZ) and thermal Sunyaev-Zel’dovich (tSZ) effect \citep{bonjean_gas_2018, tanimura_density_2020}, X-ray emission and absorption \citep{eckert_warm_2015}, OVII absorption line \citep{nicastro_missing_2018} and Lyman-$\alpha$ forest \citep{penton_local_2004, hu_impact_2024} are employed to probe the diffuse gas in the WHIM. The critical parameter dictating the effectiveness of different probes is the temperature. The SZ effect is sensitive to higher average temperatures of $10^{5-8}$ K, whereas Lyman-$\alpha$ detects regions at $10^{4-5}$ K. Moreover, the WHIM region is very faint in electromagnetic radiation due to its diffuse nature. The absorption spectra also rely on bright background sources, such as quasars, or on stacking multiple observations to increase the signal.

The DM of FRBs, particularly the scatter in the DM-$z$ relation, serves as a complementary probe for localising the inhomogeneous baryon distribution \citep{inoue_2004_dm_derivation, walters_probing_2019, connor_cosmic_2025}.  As the DM arises from interactions with electrons along the LoS, it can effectively trace diffuse baryons regardless of temperature or density. While \citet{macquart_census_2020} used the homogeneous DM-$z$ relation to yield the total baryon estimate consistent with estimates from BBN and the CMB, its distribution in the Universe remains largely unknown. This mapping is critical for studies of galaxy evolution \citep{behroozi_correlation_2019, khrykin_cosmic_2024}. Particularly, FRBs can constrain feedback mechanisms like active galactic nuclei (AGN) and supernovae \citep{reischke_cosmological_2023, sharma_hydrodynamical_2025}, which in turn constrain the star formation rate \citep{booth_interaction_2013}. Similarly, weak lensing (WL) is dependent on comprehensive baryon distribution maps for unbiased inference, particularly at non-linear, smaller spatial scales, where matter distribution differs significantly from the collisionless dark matter distribution due to baryonic feedback effects \citep{ chisari_impact_2018, schneider_baryonic_2020}. The kSZ effect, similar to the DM, is the integrated effect of the electron number density and the velocity field, where a robust knowledge of the baryon distribution can isolate the velocity field to measure the growth parameter \citep{madhava_cosmology_ksz_2019, takahashi_measurement_2025, edmond_measurement_2026}.

Since the baryon distribution in the Universe is inhomogeneous, models focus on decomposing the total integrated DM into contributions from different components of the cosmic web. Prior studies with cosmological observations and simulations partition the total DM into structures like voids, filaments and haloes, where results indicate that the dominant contribution arises from filaments \citep[see e.g.][]{jaroszynski_fast_2019, zhang_intergalactic_2021, walker_dispersion_2024}. While they provide a qualitative or categorical estimate of the DM contribution, a quantitative description remains a significant challenge. Furthermore, as the DM is an integrated signal, for real data, it alone cannot localise the baryons in the cosmic web. We require additional tracers of the spatial density distribution between the FRB source and the observer to recover the density-dependent baryon information. Therefore, a framework utilising the observational data alongside simulation-based insights is essential.

In this paper, we propose such a method, using galaxy surveys as tracers of the density distribution, and explicitly linking the FRB DM distribution and the density distribution in the large-scale structure (LSS) via a simple linear density estimator. Here, density estimation is the process of inferring a continuous underlying field from discrete points. The classification of that continuous field into eight equidistant logarithmic bins of the underlying density, analogous to voids, filaments, and haloes, but without any ambiguous boundaries, disentangles the integrated DM value into individual contributions. This framework not only provides a global estimate but also describes the baryon distribution across different sub-regions of the LSS.

To validate this framework, we simulate data using snapshot information from the IllustrisTNG project, a large-scale cosmological hydrodynamical simulation. It consists of various tracers and sub-grid physical models, calibrated against observational data, to capture small-scale effects. For multiple FRB sightlines, the density estimator is defined as the ratio of electron number density to the total mass density of the Universe.

Our analysis integrates observational estimates with the simulated density field by comparing the FRB density estimator with matter density probed via (upcoming) spectroscopic galaxy surveys mimicking observations with e.g. the Dark Energy Spectroscopic Instrument \citep[DESI;][]{abareshi_overview_2022}, Vera Rubin Observatory Legacy Survey of
Space and Time \citep[LSST;][]{ivezic_LSST_2019}, \textit{Euclid} \citep{scaramella_Euclid_2022}, Nancy Grace Roman Space Telescope \citep[Roman;][]{spergel_ROMAN_2013, foley_ROMAN_2018} and SPHEREx \citep{crill_SphereX_2020}. Specifically, in this work, we create a \textit{Euclid}-like mock catalogue as a subset of all the subhaloes in the IllustrisTNG simulations.

For real observational data, we require localised or host-identified FRBs that fall within the footprint of galaxy surveys. In particular, with upcoming surveys such as the Canadian Hydrogen Observatory and Radio-Transient Detector \citep[CHORD;][]{vanderlinde_CHORD_2019} and the Deep Synoptic Array (DSA)-2000 \citep{hallinan_DSA_2019} detecting $10^4$ FRBs each year, the localised number is expected to increase by several orders of magnitude. Similarly, galaxy surveys with large sky coverage, such as \textit{Euclid} and LSST, will enable a precise comparison of the density estimators presented in our work.

The paper is structured as follows. In Section \ref{sec:frbs}, we describe the observable properties of FRBs. Section \ref{sec:sims} presents the simulation setup used to calculate the physical properties, followed by the density estimation in Section \ref{sec:inversion}. In Section \ref{sec:mocks}, we create a density estimator using a mock galaxy catalogue. Section \ref{sec:result} contains the principal results from this work. Finally, we discuss our findings in Section \ref{sec:discussion} and conclude in Section \ref{sec:conclusion}.

The fiducial cosmology in IllustrisTNG is set from the Planck-2015 cosmological results \citep{ade_planck_cosmology_2016} and in this work we use the same to maintain consistency, i.e., $\Omega_\mathrm{m} = 0.3089$, $\Omega_\mathrm{b} = 0.0486$, $\Omega_\Lambda = 0.6911$, $\sigma_8 = 0.8159$, $n_\mathrm{s} = 0.9667$ and $H_0 = 67.74 \;\mathrm{km \,s}^{-1}\,\mathrm{Mpc}^{-1}$.

\section{Fast radio bursts}
\label{sec:frbs}
The radio signal emanating from an FRB disperses due to interactions with ionised plasma, which is characterised by a frequency-dependent time delay in the arrival of the signal, i.e. $\Delta t \sim \mathrm{DM}/\nu^2$, where $\nu$ is the frequency, and DM is called the dispersion measure. This DM is a proxy for the integrated electron density between the FRB source and the observer, generally expressed as
\begin{align}
\label{eq:dm_1}
    \mathrm{DM} = \int_0^z n_\mathrm{e}(z') \,\frac{c}{(1+z')^2 \, H(z')}\,\mathrm{d}z' \;,
\end{align}
where $n_\mathrm{e}$ is the physical free electron number density at a redshift of $z$, $c$ is the speed of light and $H(z)$ is the Hubble function. The total dispersion is divided into three primary components based on the sky position $\hat{\boldsymbol{x}}$ and redshift $z$ via
\begin{align}
\label{eq:dm_with_mw_lss_host_comps}
   \mathrm{DM}_\mathrm{tot}(\hat{\boldsymbol{x}},z) = \mathrm{DM}_\mathrm{MW}(\hat{\boldsymbol{x}}) + \mathrm{DM}_\mathrm{LSS}(\hat{\boldsymbol{x}},z) + \frac{\mathrm{DM}_\mathrm{host}}{1+z}\;,
\end{align}
where $\mathrm{DM}_\mathrm{MW}$ is the contribution from the Milky Way, $\mathrm{DM}_\mathrm{LSS}$ is from the LSS of the Universe in the LoS towards the FRB, and finally $\mathrm{DM}_\mathrm{host}$ is the contribution from the host galaxy of the FRB progenitor.

Note that the Milky Way contribution is local and depends solely on the direction. The host contributions can vary based on the position of the FRB progenitor inside the host galaxy \citep{reischke_analytical_host_2024}, and it is redshifted with a $(1+z)^{-1}$ term for the cosmological redshift. As a result, the positional dependence of DM inside the host galaxy is suppressed in the observed DM value. There exist models like YMW16 \citep{yao_YMW16_2017}, NE2001 \citep{cordes_NE2001_2001}, or the updated NE2025 \citep{ocker_ne2025_2026} based on pulsar observations to calculate the Milky Way contribution. However, with a lack of a global FRB population, the host contribution is the least constrained, and studies focus on a statistical estimate with a log-normal distribution based on analytical and hydrodynamical insights \citep[see e.g.][]{jaroszynski_fast_2019, reischke_consistent_2023, medlock_probing_2024, reischke_analytical_host_2024, konar_constraining_2025}.
Despite these contributions, for an FRB at cosmological distances, most of the signal is encoded in the LSS part.

To study the electron distribution in the LSS, therefore, we focus on the $\mathrm{DM}_\mathrm{LSS}$ part and substitute the electron number density in Eq.~\eqref{eq:dm_1} as an integral over redshift \citep{ioka_2003_dm_derivation, inoue_2004_dm_derivation}
\begin{eqnarray}
\label{eq:dm_lss_full}
    \mathrm{DM}_\mathrm{LSS}(\hat{\boldsymbol{x}},z) &=& \frac{3 c H_0 \Omega_\mathrm{b} }{8 \pi G m_\mathrm{p}} f_\mathrm{ion} f_\mathrm{e} \nonumber \\
     & \times & \int_0^z \frac{1+z'}{E(z')} \, \left[1+\delta_\mathrm{e}(\hat{\boldsymbol{x}},z')\right] \, \mathrm{d}z' \;,
\end{eqnarray}
$H_0$ is the Hubble constant, $\Omega_\mathrm{b}$ is the baryon density in the present Universe, $G$ is the gravitational constant, $m_\mathrm{p}$ is the proton mass, $\delta_\mathrm{e}$ is the unitless electron density contrast and $E(z) = H(z)/H_0$ is the expansion function. Additionally, $f_\mathrm{ion}$ is the fraction of baryons in the diffuse ionised phase, i.e. excluding baryons in stars, compact objects, and the dense neutral interstellar medium, and $f_\mathrm{e}$ is the electron fraction calculated via \citep{deng_wei_2014_dm_IGM_derivation, walters_probing_2019, macquart_census_2020}
\begin{align}
    f_\mathrm{e} = Y_\mathrm{H} + \frac{1}{2} Y_\mathrm{He} \;,
\end{align}
where $Y_\mathrm{H} \simeq 0.76$ and $Y_\mathrm{He} \simeq (1-Y_\mathrm{H}) \simeq 0.24$ are the hydrogen and helium abundances \citep{aghanim_planck_VI_2020, erik_new_Yhe_2026}. Note that we are considering a redshift of $z<3$, where the IGM is effectively fully ionised \citep{ becker_detection_2011}.

The mean of Eq.~\eqref{eq:dm_lss_full} is called the DM-$z$ or Macquart relation, which probes the total baryon content of the Universe \citep{macquart_census_2020}. The inhomogeneous part, on the other hand, leads to sample variance between FRBs observed at the same redshift and encodes the distribution of baryons in the Universe along with baryon feedback effects via AGN and supernovae \citep{reischke_cosmological_2023, sharma_hydrodynamical_2025}

\section{Simulation}
\label{sec:sims}
Classifying the LSS and studying its contribution to the DM observations requires density contrast, which cannot be probed directly via observations. However, if we have a proxy for the density along the LoS together with a relation between total density and electron density, we can use an ensemble of FRBs to infer the relation statistically. To validate such a relation, we require the ground truth. Hence, we use cosmological hydrodynamical simulations, where the underlying relation is directly accessible. Generally, such simulations evolve a large number ($\gtrsim 10^8$) of particles in time in a three-dimensional box of cosmological dimensions. The evolution is tracked via snapshots at multiple redshifts. Various simulations differ in their scale, box size, resolution, particle mass, and implementation of physical laws.

Specifically, simulations such as Magneticum \citep{dolag_constraints_2015}, MICE \citep{pol_estimates_2019}, EAGLE \citep{batten_cosmic_2021}, RAMSES \citep{zhu_contributions_2021}, and CROCODILE \citep{zhang_probing_2025} exist and have been used in cosmological and astrophysical studies of FRBs.

\subsection{IllustrisTNG}
\label{sub:illustris}
Here, we use TNG300-1 simulations from the IllustrisTNG project \citep{nelson_illustris_galaxy_2018, pillepich_illustris_simulating_2018, springel_illustris_matter_2018, naiman_illustris_chemical_2018, marinacci_illustris_haloes_2018}, which is a large-scale, cosmological, magnetohydrodynamical (MHD) simulation that includes gas, dark matter, tracers, stars, and black holes. In addition, it contains observationally calibrated sub-grid models for star formation, gas cooling, stellar and AGN feedback \citep{weinberger_simulating_2017, pillepich_illustris_simulating_2018}. The simulation evolves the particles from an initial redshift of $z=127$ to a final redshift of $z=0$ while capturing the evolution with 100 snapshots. There are two types of snapshots, full and mini, where mini contains a subset of the full information\footnote{\url{https://www.tng-project.org/data/docs/specifications/}}.

The choice of IllustrisTNG is motivated by the advantage of a higher resolution over Magneticum, better gas dynamics than MICE, larger box sizes compared to EAGLE, more precomputed fields than RAMSES and different combinations of the kinetic and thermal feedback models based on AGN accretion, which is lacking in CROCODILE \citep{zhang_probing_2025}. Moreover, since we are concerned with the LSS contribution to the DM in Eq.~\eqref{eq:dm_with_mw_lss_host_comps}, we adopt the largest-volume simulation, i.e., TNG300-1. The redshift range is set to $z \in [0, 1.5]$ to match the current host-identified FRB sample at a redshift of $z=1.34$ \citep{connor_cosmic_2025}. Note that we choose to exclude the highest entry at $z = 2.148$ discovered by \citet{caleb_z_2_FRB_2025} because of the lack of localised FRB samples in between. Choosing to do so would anchor the extrapolation to a single data point, which is not ideal because we would lose information about scatter. With that in mind, we use full snapshots at redshifts 0.0, 0.1, 0.2, 0.3, 0.4, 0.5, 0.7, 1.0 and 1.5. With a box size of 205 $h^{-1}$ Mpc, there are empty regions or gaps in comoving distance, where we reuse sections of the snapshots to interpolate (more in Section \ref{sub:dm}). The following sections describe the calculation of all the required physical quantities in these snapshots.

\subsection{Electron density}
Simulating the DM values requires the electron density in Eq.~\eqref{eq:dm_1}, or equivalently, the electron density contrast in Eq.~\eqref{eq:dm_lss_full}. However, IllustrisTNG does not natively provide electron density as a field. Hence, we follow the formalism laid out in \citet{jaroszynski_fast_2019, zhang_intergalactic_2021, walker_dispersion_2024}. The ``ElectronAbundance'' field, defined as the fractional electron number density with respect to the total hydrogen number density, can be utilised. Moreover, IllustrisTNG includes both star-forming and non-star-forming Voronoi cells, which are the spatial discretisation units on which the MHD fluid dynamics are run. The star-forming cells, probed by ``StarFormationRate'', consist of gases in both warm and cold phases, for which the simulation output needs to be inverted to extract the free electrons from the subgrid model \citep{springel_sfr_illustris_2003}. However, \citet{konietzka_ray_tracing_2025} found that neglecting the contribution from star-forming cells yields only a sub-per cent deviation from the total DM estimate and is therefore omitted here. Then, the free electron density of each cell is
\begin{align}
\label{eq:ele_density}
    n_\mathrm{e} = \eta_\mathrm{e} \,Y_\mathrm{H} \,\frac{\rho_\mathrm{gas}}{m_\mathrm{p}} \, (1+z)^3 \;,
\end{align}
where $\eta_\mathrm{e}$ is the electron abundance given by the ``ElectronAbundance'' field, $\rho_\mathrm{gas}$ is the gas cell comoving mass density and $m_\mathrm{p}$ is proton mass. Here, electron abundance represents the number of free electrons per hydrogen nucleus, which implicitly accounts for the contribution from ionised helium, i.e., in a fully ionised region, $\eta_\mathrm{e} \approx 1.16$. As we compute the electron density using the ``ElectronAbundance'' field, no additional ionised baryon fraction is applied to Eq.~\eqref{eq:ele_density}.

\subsection{Dispersion measure}
\label{sub:dm}
With the electron density for each Voronoi cell, we can approximate the integral in Eq.~\eqref{eq:dm_1} as a sum of incremental distance elements. Specifically, the discretised version of the line-of-sight integral is
\begin{align}
\label{eq:dm_as_sum}
    \mathrm{DM} = \int_0^z \frac{n_\mathrm{e}}{1+z} \mathrm{d}l \approx \sum \, \frac{n_\mathrm{e}}{1+z} \,\Delta l \;,
\end{align}
where $n_{\mathrm{e}}$ and $\Delta l$ are the physical electron number density and differential physical path length on the LoS. The physical path length can be calculated by multiplying the comoving distance in simulation units by the snapshot's scale factor. Note that in each IllustrisTNG snapshot, the redshift or cosmic time is constant, i.e., there is no evolution along the line of sight inside one box. Hence, the redshift dependence of the DM is not written explicitly. As for the spatial vector (i.e., direction), we define it implicitly and drop the $\hat{\boldsymbol{x}}$ notation for brevity.

The full TNG300-1 snapshots are only available at a specific redshift with a length of 205 $h^{-1}$ Mpc. Hence, the discrete sum in Eq.~\eqref{eq:dm_as_sum} is not continuous in redshift, i.e., the full snapshots used in Section \ref{sub:illustris} do not overlap in comoving distance. One of the solutions is the trapezoidal rule with interpolation as presented in \citet{zhang_intergalactic_2021, walker_dispersion_2024}. This interpolation or stretching, however, distorts the DM distribution, specifically affecting the higher-order moments, leading to a biased estimate of the cosmic variance of the DM, i.e., the interpolation produces a lower scatter in the DM–$z$ relation than otherwise expected from different LoS through the LSS \citep{konietzka_ray_tracing_2025}. Improvements include placing the periodic boxes parallel to each other, where a LoS traces the same structure multiple times, or randomly rotating the boxes before placing them. The latter avoids some level of repetition but introduces incorrect transitions at the simulation boundary due to random rotations. However, we choose this method because it is computationally feasible and yields a better estimate of cosmic variance than the trapezoidal method.

Note that because the primary goal of this work is to demonstrate that the electron-to-total mass density profile can be successfully reconstructed from a set of DM observations, the spatial correlation introduced by reusing simulation boxes is not critical. This approach, however, results in an underestimation of the error bars, since some LoS in the simulations might not be fully independent. Nevertheless, it does not affect the core results. For any missing region between two snapshots, we use a slice of the snapshot parallel to the $z$ axis that has a lower redshift than the next available snapshot. As the underlying cosmology is known and the LoS vector is parallel to the $z$ axis, this slicing is exact and implicitly avoids overlap.

Having established a continuous LoS in the simulation, we focus on the mock FRB population. In our work, we study the precise cosmological contribution of the LSS to the DM. Therefore, the redshifts of the FRBs are required. While the true redshift distribution of FRBs is unknown at this point, we create the mock redshift distribution following the works by \citet{chen_formation_2024, zhang_revisiting_2025}. The unnormalised probability distribution function for the redshifts is parametrised by
\begin{align}
    p(z) \propto (1+z)^{-5.9} \, \frac{D_\mathrm{c}(z)^2}{E(z)} \;,
\end{align}
where $D_\mathrm{c}(z)$ is the comoving distance, and we combine the number density evolution scale of $(1+z)^{-4.9}$ with a cosmological factor of $(1+z)^{-1}$, along with the differential comoving volume element. We restrict the final sampled redshifts to $z \leq 1.5$. This is a catalogue selection motivated by the currently available localised FRB sample rather than a physical claim that the intrinsic FRB rate vanishes above this redshift.

The sky positions are randomly sampled from $\mathrm{RA} \in \mathcal{U}[0, 2\pi)$ and $\mathrm{DEC} \in \arcsin(\,\mathcal{U}[-1,1]\,)$, leading to a random sky distribution, which is then converted to Cartesian coordinates using the underlying cosmology. We apply periodic boundary conditions to ensure the FRBs remain within the simulation box.

Starting with a snapshot at a given redshift, we use the coordinates of the full box to select a LoS mask, with the vector parallel to the $z$-axis and spanning the whole box. The centre of each LoS is the $(x,y)$ coordinates of the FRB sample. Afterwards, we use all necessary fields inside this mask to calculate the electron density in Eq.~\eqref{eq:ele_density}. The distance element, $\Delta l$, is simply the derivative of the increasing $z$-coordinates due to the parallel LoS vector. Finally, the DM is the sum over segments of the physical electron density times the physical path length, divided by (1+redshift) of the corresponding snapshot. We derive the expression in the following section.

\subsection{Decomposition}
\label{sub:decomposition}
In each LoS, the measured DM contains contributions of ionised electrons from many over- and under-dense regions of the LSS. Here, we decompose this integrated signal into contributions from specific parts of the cosmic structure. Generally, in the works by \citet{haider_illustris_lls_2016, martizzi_baryons_2019, artale_illustris_classification_2022}, the LSS is classified based on the ratio of dark matter density to critical density, $\rho_\mathrm{c}/\rho_\mathrm{crit}$. In this scheme, typically three distinct regions are defined, namely voids ($\rho_\mathrm{c}/\rho_\mathrm{crit}$< 0.1), filaments ($\rho_\mathrm{c}/\rho_\mathrm{crit} = 0.1-57$) and haloes ($\rho_\mathrm{c}/\rho_\mathrm{crit} > 57$).

However, these boundaries are a little arbitrary and ambiguous, with significant overlap, and the definition itself is method-dependent \citep{libeskind_tracking_2018}. Hence, in this work, we define 8 logarithmic bins on the underlying density without labelling them as any morphological structures. These 8 bins are equidistant in log space, where the minima and maxima are the two extrema of the snapshot density. This ensures we account for the full range of densities within the snapshot.

Next, we derive the general expression of the discretised DM for an arbitrary number of LoS and density bins. Let there be $N$ sightlines that are indexed by $j \, (j = 1, \dots, N)$. The differential path segment on a single LoS is indexed via $k \, (k=1, \dots, K_j)$ and finally, we have $M$ logarithmic density bins indexed by $i \, (i = 1, \dots, M)$. For the $j$-th sightline, the observer frame DM is written as
\begin{align}
\label{eq:dm_discret}
    \mathrm{DM}_j = \sum_{k=1}^{K_j} \; \frac{n_{\mathrm{e},\,jk}\,\Delta l_{jk}}{1 + z_k} \;,
\end{align}
where $z_k$ is the redshift of the $k$-th path segment, which is constant within the individual IllustrisTNG snapshots. Both electron density and the distance increment are in physical units here. Note that for real observations, we cannot directly probe the electron density; instead, we measure the mass overdensity traced by galaxies. Hence, we introduce the ratio
\begin{align}
\label{eq:r_defintion}
    r_{jk} = \frac{n_{\mathrm{e},\,jk}}{\rho_{jk}} \;,
\end{align}
where $n_\mathrm{e}$ is the electron density and $\rho$ is the total mass density. In a later section, we discuss how we derive $\rho$ from the galaxy population.

An analytical expression for the mean of this ratio across all density bins can be derived. Starting with the definitions of electron number density and matter density, one finds
\begin{align}
\label{eq:r_theo}
    r_\mathrm{theo} = \frac{\Omega_\mathrm{b}}{\Omega_\mathrm{m} \,\mu_\mathrm{e}\,m_\mathrm{p}} \; f_\mathrm{ion} \;,
\end{align}
where $\Omega_\mathrm{b}$ $(\approx0.0486 \pm 0.0005)$ is the baryon density, $\Omega_\mathrm{m}$ $(\approx 0.3089 \pm 0.0062)$ is the matter density and $\mu_\mathrm{e}\simeq2/(1+Y_\mathrm{H}) \approx 1.14$ is the mean molecular weight per free electron. Note that the DM arises due to contributions from diffuse ionised baryons expressed via $f_\mathrm{ion}$ in Eq.~\eqref{eq:dm_lss_full}.

Here, we adopt $f_\mathrm{ion} = 0.95$, the value as given for TNG300-1 at $z=0$ \citep{takahashi_statistical_2021, konietzka_ray_tracing_2025}. Further, we keep this value constant and assume no redshift evolution. Substituting the values yields $r_\mathrm{theo} \approx (7.84 \pm 0.18) \times 10^{22} \;\mathrm{gram}^{-1}$, where the uncertainty propagates the errors on $\Omega_\mathrm{b}$ and $\Omega_\mathrm{m}$ (neglecting their positive correlation). Within the simulation, these parameters are exact inputs, so this uncertainty is only relevant for observational data. Since $r_\mathrm{theo}$ is a global ratio, it corresponds to the mass-weighted mean of the bin values, i.e., $\langle r\rangle_M = \sum_i r_i M_i/ \sum_i M_i$, where $M_i$ is the total matter mass in bin $i$. We also note that this value is used as a reference, and not imposed in the inference. It enters neither the mock DMs (Eq.~\eqref{eq:ele_density}) nor the configuration matrix that we derive in the following.

As stated earlier, with this definition of the ratio and 8 log bins, no classification of the LSS into voids, filaments, and haloes is necessary. First, we convert the total matter densities, accessed via the ``SubfindDensity'' field, from simulation units to physical units. Then, the minimum and maximum bin edges are the extrema of the matter density on a logarithmic scale. Finally, there are $M(=8)$ equidistant bins.

Substituting Eq.~\eqref{eq:r_defintion} into Eq.~\eqref{eq:dm_discret}, the DM for a single LoS is
\begin{align}
    \mathrm{DM}_j =\sum_{k=1}^{K_j} \frac{r_{jk}\,\rho_{jk}\,\Delta l_{jk}}{1+z_k} \;,
\end{align}
where $r_{jk}$ is the realisation of the electron-to-matter ratio in cell $k$ of the simulation snapshot. All the quantities here are in physical units, related to the comoving quantities via $\rho_{j k} = \rho_{\mathrm{com}, \,j k} \cdot (1+z_k)^3$ and $l_{j k} = l_{\mathrm{com}, \, jk} \cdot (1+z_k)^{-1}$, with $z_k$ being the redshift. To pivot the calculation on the density bins, we partition the set of cells $1,\dots, K_j$ into bins $B_i$,
\begin{align}
    B_i \;=\;\{\,k : \rho_{j k}\in\mathrm{bin} \; i\,\} \;.
\end{align}
As the ratio $r$ is modelled as a function of density, we approximate the cell-specific value $r_{jk}$ by a common bin-averaged value $r_i$ for all cells belonging to density bin $B_i$. Therefore, we can write
\begin{align}
    \mathrm{DM}_j &\approx \sum_{i=1}^{M} \sum_{k\in B_i} \frac{r_i\,\rho_{jk}\,\Delta l_{jk}} {1+z_k} \nonumber\\
    &=\sum_{i=1}^{M} r_i \sum_{k\in B_i} \frac{\rho_{jk}\,\Delta l_{jk}}{1+z_k} = \sum_{i=1}^{M} r_i \, A_{ji} \;,
\end{align}
where we have defined the configuration matrix as
\begin{align}
\label{eq:configuration_matrix}
    A_{ji} = \sum_{k\in B_i} \frac{\rho_{jk}\,\Delta l_{jk}} {1+z_k} \;.
\end{align}
Finally, for all $N$ sightlines and $M$ logarithmic density bins, we can stack them in matrices with the compact notation
\begin{align}
\label{eq:frb_decopose_1}
    \mathbf{d} = \mathbf{A}\,\mathbf{r} + \mathbf{n} \;,
\end{align}
where $\mathbf{d} = \{\mathrm{DM}_1, \cdots, \mathrm{DM}_N\}^\mathrm{T}$ is the observer-frame DM, $\mathbf{A}$ is the configuration matrix, and $\mathbf{r}$ is the density estimator as the ratio between electron number density and total matter density. The last term, $\mathbf{n}$, is the measurement error. As the total DM has three distinct contributions in Eq.~\eqref{eq:dm_with_mw_lss_host_comps}, the Milky Way and host parts are simply subtracted from the total DM. Here, we simulate the LSS contribution and omit the remaining components because the density estimator is designed to detect contributions from various regions of the LSS. Adding and then subtracting the Milky Way and host DM values results in a net-zero effect.

Assuming the noise is independent and uncorrelated, we model the measurement error as a zero-mean Gaussian with standard deviation 20. Then the entries are random draws from such a distribution, i.e., $\mathbf{n} \in \mathcal{N}(0, \sigma_n^2)$ pc cm$^{-3}$, where $\sigma_n = 20$ pc cm$^{-3}$. The choice is a combined effect of the average instrumental noise of $\mathcal{O}(10^{-1})$ pc cm$^{-3}$ in the second CHIME catalogue \citep{abbott_ChimeCat2_2026}, a larger $\mathcal{O}(10)$ pc cm$^{-3}$ coming from the galactic DM models like NE2001, NE2025 and YMW16, and finally, $\mathcal{O}(10)$ pc cm$^{-3}$ from the host model. In comparison, the sightline-to-sightline noise dominates at $\mathcal{O}(10^2)$ and contains the cosmological information of the inhomogeneous LSS \citep{reischke_cosmological_2023}.

Note that Eq.~\eqref{eq:frb_decopose_1} is a linear system in $\mathbf{r}$, and one can estimate the least-squares or the maximum likelihood solution when a trivial inversion is ill-posed due to over-determinism. That is the principal approach of our work, where given the observed values of $\mathbf{d}$ and $\mathbf{A}$ for an FRB dataset, the solution yields a density estimator $\mathbf{r}$.

\section{Density estimation with FRB}
\label{sec:inversion}
To ensure that the sightlines are largely independent, we quantify how similar the density structure traversed by each pair of LoS is. For the $j$-th LoS, we compute the mean gas density $\rho_j(z_k)$ in $N_z \,(=500)$ slices along the LoS vector, averaging over the Voronoi cells within a transverse column of width $L_\mathrm{box}/100$; $L_\mathrm{box}$ is the length of the snapshot box. Then, the correlation is calculated as
\begin{align}
\label{eq:los_corr}
    R_{ij} = \frac{1}{N_z}\sum_{k=1}^{N_z}
    \frac{\left[\rho_i(z_k)-\bar{\rho}_i\right]\left[\rho_j(z_k)-\bar{\rho}_j\right]}{s_i\,s_j} \;,
\end{align}
where $\bar{\rho}_j$ and $s_j$ are the mean and standard deviation of $\rho_j(z_k)$ along the LoS. Two sightlines that cross through the same structures have $|R_{ij}|$ close to unity, whereas statistically independent LoS yield $R_{ij} \approx 0$. We compute $R_{ij}$ in the {$z=0$} snapshot and retain a LoS only if $|R_{ij}| < 0.3$. This choice preserves $\sim 72$ per cent of the simulated data while minimising redundant, closely clustered LoS. Based on this threshold, 470 LoS are selected for the final analysis. We discuss the effects of this choice on our results in Section \ref{sub:eff_corr}. Physically, a higher correlation implies two sightlines probe the same cosmic structure and therefore bias the expected variance. Note that in real observations, the angular separation between FRBs is typically much larger than in our simulation snapshot, which is 205 $h^{-1}$ Mpc.

In the general case, if $N > M$, i.e., we have more FRB DM values than logarithmic bins, Eq.~\eqref{eq:frb_decopose_1} is over-determined. Here, we have $N=470$ and $M=8$. Assuming the signal and noise are Gaussian, the maximum likelihood estimate (MLE) is
\begin{align}
\label{eq:max_likelihood}
    \mathbf{r}_\mathrm{ML} = \left(\mathbf{A}^\mathrm{T} \mathbf{C}^{-1} \mathbf{A}\right)^{-1} \, \mathbf{A}^\mathrm{T} \mathbf{C}^{-1} \mathbf{d} \;,
\end{align}
where $\mathbf{C}$ is the covariance matrix and the term $\mathbf{A}^\mathrm{T} \mathbf{C}^{-1} \mathbf{A}$ is the Fisher information matrix. The full covariance matrix has four distinct contributions: three corresponding to the quantities in Eq.~\eqref{eq:dm_with_mw_lss_host_comps} and a noise covariance. The covariance for the Milky Way can be obtained from models like NE2001, NE2025, or YMW16, while the host component comes from the variance of a log-normal distribution \citep{konar_constraining_2025}. Note that the Milky Way and the measurement uncertainties are subdominant compared to the variance in the LSS, and the host contribution gets suppressed with the $(1+z)^{-1}$ factor. Consequently, to a good approximation, the final covariance is dominated by the variance coming from the LSS part, i.e.,
\begin{align}
    \mathbf{C} \approx \mathbf{C}_\mathrm{LSS} = \mathrm{diag} \left[\sigma^2_\mathrm{LSS}(z_j)\right] \;.
\end{align}
We therefore retain only the diagonal terms, approximating the LoS as statistically independent after imposing the $|R_{ij}| < 0.3$ selection.

For a real FRB dataset, the observed DM already contains all the components in Eq.~\eqref{eq:dm_with_mw_lss_host_comps}. The full covariance, in that case, is the summation of the individual covariances, which can be estimated following \citet{reischke_cosmological_2023} for a Gaussian likelihood, or one can implement a likelihood-free inference using forward simulations \citep{konar_constraining_2025}. Note that the covariance depends on the angular separation $\theta$ between FRBs on the sky, with sightlines converging on the observer. However, in our case, the LoS vectors are parallel to the $z$ axis, so we adopt a diagonal covariance after imposing the $|R
_{ij}|<0.3$ selection. Even though it is not critical with the current observational dataset, as the number of localised FRBs increases, precise cosmological inference would rely on accurate subtraction of the Milky Way and host contributions from the total DM.

The form of Eq.~\eqref{eq:frb_decopose_1} and its inversion for an over-determined system is a well-documented problem in statistics and regression modelling \citep[see e.g.][]{cox_analysis_1993, ruymgaart_unified_1993, donoho_statistical_1994, bissantz_convergence_2007, stuart_inverse_2010}. As there may be multiple solutions for $\mathbf{r}$ that reproduce the data $\mathbf{d}$ for a given $\mathbf{A}$ matrix, the inversion can be ill-posed, especially for low-density bins that contribute less information to the measured DMs. The MLE solver in Eq.~\eqref{eq:max_likelihood} for such a case is noisy. Therefore, a maximum a posteriori (MAP) solution is preferred that includes a regularisation scheme like Tikhonov regularisation to the Fisher matrix in Eq.~\eqref{eq:max_likelihood} \citep{golub_tikhonov_1999}. In such a  regularisation scheme, the solution reads
\begin{align}
\label{eq:tikhonov_regul_method}
    \mathbf{r}_\mathrm{tikh} = \left(\mathbf{A}^\mathrm{T} \mathbf{C}^{-1} \mathbf{A} + \lambda \,\mathbf{R}^\mathrm{T}\,\mathbf{R}\right)^{-1} \, \mathbf{A}^\mathrm{T} \mathbf{C}^{-1} \mathbf{d} \;,
\end{align}
where $\mathbf{R}$ is the regularisation operator, which in its simplest form can be an identity matrix $\mathbf{I}$ (i.e., Tikhonov regularisation). Alternatively, a finite-difference matrix can be used to obtain a smoother solution than Tikhonov. The term $\lambda$, a scalar, is called the regularisation scale or `ridge' and essentially controls the amplitude of Tikhonov regularisation. This is equivalent to providing a prior probability distribution next to the likelihood function. The choice of $\lambda$ can be optimised from the condition number of the Fisher matrix \citep{cline_condition_number_1979, belsley_1980_regression}. Similar to MLE, we find that the solution is also prone to noise, particularly at low-density bins and produces unphysical negative values.

Therefore, we implement a Bayesian prior with a parametric form on $\mathbf{r}$ \citep[see e.g.][addressing a similar problem with a hierarchical approach]{dries_hierarchical_2016}. Recalling that we define 8 logarithmic bins on the density, let us define the bin centres as $x_i$, i.e., $x_i \in \{\log\rho_{\mathrm{m,1}}, \dots, \log\rho_\mathrm{m,8}\}$. Then, the elements in $\mathbf{r}$ are modelled using a scaled skewed Gaussian function as
\begin{align}
\label{eq:r_prior}
    r_\mathrm{FRB}(x) = M \, \,\mathcal{S} \, r_\mathrm{theo} \, f(x) \;,
\end{align}
with $\mathcal{S}$ as the scale relative to the theoretical mean value $r_\mathrm{theo}$, and $M\,(=8)$ as the number of density bins, a fixed constant included only so that $\mathcal{S}$ is of order unity. It is fully degenerate with $\mathcal{S}$ and does not affect the reconstructed $\mathbf{r}$. In other words, $\mathbf{r}$ is a function of density with a shape function that is defined as\footnote{We note that this functional form is empirically motivated from simulations and flexible enough to capture the range of trends in the shape of $\mathbf{r}$, but it has no physical origin.}
\begin{eqnarray}
    f(x) & \equiv & \frac{1}{w_\rho \sqrt{2 \pi}} \exp \left[ - \, \frac{1}{2}  \left(\frac{x - \log \rho_\mathrm{tran}}{w_\rho}\right)^2 \right]  \nonumber \\ & \times & \left[ 1 + \mathrm{erf}\left( \frac{\alpha_\rho}{\sqrt{2}} \, \frac{x - \log \rho_\mathrm{tran}}{w_\rho} \right) \right],
\end{eqnarray}
where $\rho_\mathrm{tran}$ is the location parameter of the profile, which approximately marks the transition from $r_\mathrm{FRB} \ga r_\mathrm{theo}$ at low densities to $r_\mathrm{FRB} \la r_\mathrm{theo}$ at high densities, due to feedback. Similarly, $w_\rho$ is the width of the profile in $\log \rho$, which dictates how rapidly the value of $r_\mathrm{FRB}$ changes. Finally, the skewness parameter, $\alpha_\rho$, being positive or negative, determines whether the resulting distribution is right- or left-skewed, respectively, while $\alpha_\rho=0$ corresponds to a Gaussian shape. Here, $f(x)$ is normalised to unit area in $x = \log_{10}(\rho_\mathrm{m}/\mathrm{g\,cm^{-3}})$. No normalisation condition is imposed on $\mathbf{r}$, so $\mathcal{S}$ is a free parameter of the fit. Moreover, since the profile extends beyond the sampled density range, $\mathcal{S}$ is neither the arithmetic nor the mass-weighted mean of $\mathbf{r}/r_\mathrm{theo}$ over the bins. Therefore, these are the four free parameters in the inference.

This functional form is an empirical choice and represents a higher electron-to-matter density ratio in the LSS regions below $\log \rho_\mathrm{tran}$, scaled by $\mathcal{S}$. At very low densities, we expect the baryons to follow the underlying dark matter distribution more closely. Alternatively, at densities above $\log \rho_\mathrm{tran}$, it drops to lower values following the tail of a Gaussian distribution. Such a choice is motivated by the feedback effect via AGN and supernovae, which expels hot gas from the central part of the halo to the outskirts and is expected to be dominant in high-density regions, as shown in several hydrodynamical simulations \citep[see e.g.][]{chisari_modelling_2019, ayromlou_feedback_2023, khrykin_cosmic_2024}.

However, note that treating the four parameters independently implies that only the data dictate the amplitude and overall shape. As such, we impose wide uniform priors and list the prior ranges in Table~\ref{tab:prior_table} with a brief description of how each parameter alters the shape of the distribution in Eq.~\eqref{eq:r_prior}. As $\mathcal{S}$ scales the theoretical value, an upper limit of four is extreme enough to describe all plausible DMs. In principle, one could impose narrower prior boundaries on $\mathcal{S}$ from previous observational values of $\Omega_\mathrm{b}$ from CMB studies, as the mass-weighted mean of $\mathbf{r}$ must reproduce the cosmic ionised baryon fraction. We choose to keep a wider boundary as an implicit validation test, avoiding a circular argument. A narrow prior can be implemented when observational data is used with unknown ground truth. The prior for the logarithm of the transition density (or the distribution's location) is also set to be uniform over the two extrema. Similarly, the width of the distribution of densities controls the curvature of the profile, with the prior range allowing for a sharp drop or a flat profile. The skew parameter has a prior distribution over both positive and negative values, allowing flexibility for left- or right-skewed distributions. Combined, the prior choices provide a general profile to characterise the baryon distribution across the density bins. Alternatively, in Section \ref{sub:eff_non_para}, we assess how a non-parametric model on $\mathbf{r}$ performs with our dataset.

\begin{table}[ht!]
\caption{Parameters and the prior ranges used in the inversion step.}
\label{tab:prior_table}
\centering
\renewcommand{\arraystretch}{1.5}
\begin{tabular}{c c p{3.0cm}}
\hline\hline
Parameter & Prior & Description \\
\hline
   $\mathcal{S}$ & $\mathcal{U}[0.5, 4]$ & Amplitude relative to $r_\mathrm{theo}$ \\
   $\log \rho_\mathrm{tran}$ & $\mathcal{U}[(\min, \max) \log\rho_i]$ & Location of the bin where feedback starts affecting \\
   $w_\rho$ & $\mathcal{U}[0.01,8]$ & Determines the steepness of the drop from the flat profile \\
   $\alpha_\rho$ & $\mathcal{U}[-5,5]$ & Determines the skew \\
\hline
\end{tabular}
\end{table}

As the functional form of the density profile is chosen in Eq.~\eqref{eq:r_prior} based on empirical knowledge, we constrain the four free parameters that best describe the DM vector $\mathbf{d}$ for a fixed $\mathbf{A}$ matrix. Given an instance of the prior $\theta$, the forward prediction is then
\begin{align}
\label{eq:d_tilde}
    \tilde{\mathbf{d}}(\theta) = \mathbf{A}\,\mathbf{r}(\theta) \;,
\end{align}
where $\theta = \{\mathcal{S}, \log \rho_\mathrm{tran}, w_\rho, \alpha_\rho\}$ is the parameter in Eq.~\eqref{eq:r_prior}. Then, the Gaussian log-likelihood can be written as
\begin{align}
    \ln\mathcal{L}(\theta) = -\frac{1}{2}\Big[(\mathbf{d}-\Tilde{\mathbf{d}})^\mathrm{T} \mathbf{C}^{-1}(\mathbf{d}-\Tilde{\mathbf{d}}) + \ln\det \mathbf{C} + N\ln(2\pi)\Big] \;.
\end{align}
Together with the uniform priors, the final log posterior is
\begin{align}
\label{eq:posterior}
    \ln p(\theta\mid\mathbf{d}) = \ln p(\theta) + \ln\mathcal{L}(\theta) \;.
\end{align}
We utilise an affine-invariant ensemble sampler in $\texttt{emcee}$ \citep{mackey_emcee_2013} to sample the posterior in Eq.~\eqref{eq:posterior}. Specifically, we use 64 walkers, with a burn-in of 2000 steps, a reset, and a production phase of 4000 steps. We use the least-squares solution as the starting position for the walkers to achieve faster convergence.

\section{Mock catalogue}
\label{sec:mocks}
The inversion presented in Section \ref{sec:inversion} is implemented using simulated data, in which we used the total matter density directly from the simulations. This is inaccessible in a real scenario and requires a proxy. Hence, in this section, we create a mock galaxy catalogue using the subhaloes from IllustrisTNG.

In IllustrisTNG, dark matter haloes are first identified by the Friends-of-Friends (FOF) algorithm using a linking length parameter. Subsequently, the collapsed subhaloes are located using the Subfind algorithm \citep{springel_illustris_matter_2018}. The subhalo catalogue in each snapshot contains all such subhaloes and their baryonic properties. Next, we treat these subhaloes or galaxies as luminous tracers of the LSS density field and compute a more realistic observations-driven density estimator.

\subsection{Selection}
\label{sub:mock_photometry}
IllustrisTNG natively provides the absolute magnitudes in photometric bands, namely U (300-400 nm), B (350-540 nm), V (380-455 nm), K(1800 - 2580 nm), r (570-590 nm), i (620-740 nm) and z (750 - 1400 nm) via the ``SubhaloStellarPhotometrics'' field in the subhalo catalogues. We remove the IllustrisTNG samples flagged with ``SubhaloFlag'', which implies a non-cosmological origin\footnote{\url{https://www.tng-project.org/data/docs/background/\#subhaloflag}}.

Next, we create a mock \textit{Euclid}-like survey catalogue from these samples. \textit{Euclid} is a stage-IV galaxy survey with a wide $\sim 14\,000$ deg$^2$ field \citep{scaramella_Euclid_2022, mellier_Euclid_2025}. It consists of four photometric bands, including the Visible Camera (VIS), which covers the 550-900 nm range, and the Near IR Spectrometer and Photometer (NISP). It has three bands: Y (950 - 1192 nm), J (1192 - 1544 nm), and H (1544 - 2000 nm).

As the \textit{Euclid} filters do not exactly match the IllustrisTNG bands, we need to convert the photometric values into specific \textit{Euclid} bands. In that regard, we calculate the flux and the corresponding K correction of each galaxy using \texttt{Kcorrect}\footnote{\url{https://kcorrect.readthedocs.io/en/stable/}} \citep{blanton_K_correction_2007}. Then, for a given flux and the \textit{Euclid} filter response curves, as presented in \citet{schirmer_euclid_NISP_2022}, the AB magnitudes in \textit{Euclid} bands are obtained using \texttt{Speclite}\footnote{\url{https://speclite.readthedocs.io/en/latest/index.html}}. We select the galaxies based on \textit{Euclid} selection criteria in the VIS band, i.e., $I_\mathrm{E} \leq 24.5$ and signal-to-noise (SNR) $\geq 10$ \citep{scaramella_Euclid_2022}. Finally, we randomly select 60 per cent of the galaxies to imitate the sample incompleteness \citep{hamaus_forecasts_2022, kraljic_Euclid_3D_2026}.

We provide a more detailed description of each step in converting magnitudes from IllustrisTNG simulation output to the final selection based on the \textit{Euclid} specifications in Appendix \ref{ap:photometry}. Essentially, this selection creates a subset of galaxies from the full subhalo catalogue.

The last required ingredient for density estimation is the halo mass, which is provided by the ``SubhaloMass'' field. However, in observations, the halo mass is a derived quantity. There exist multiple techniques to infer mass via scaling relations and indirect methods, e.g. velocity dispersion \citep{bernardi_galaxy_2010, zahid_scaling_2016}, gravitational lensing \citep{hoekstra_lensing_2008}, mass-to-light ratio \citep{taylor_galaxy_2011}, and photometric information \citep{lower_how_2020, leja_new_2020}.

For \textit{Euclid} data, stellar mass can be estimated, for example, from spectral energy distribution (SED) fitting or machine learning \citep{bisigello_deep_2023}. Here, we use the colour information and luminosities calculated using the apparent magnitudes in the four bands. For a given luminosity, the mass can be estimated using the mass-to-light ratio $M/L$. In Appendix \ref{ap:stellar_and_halo_mass}, we present more specific details. Next, with a catalogue of mock galaxy positions and masses, we build a density estimator that coincides with the FRB LoS.

\subsection{Density estimation using galaxies}
\label{sub:galactic_density_estimator}
Density estimation attempts to reconstruct a continuous underlying field using discrete data points or tracers of that field.  Specifically, density estimation using galaxies assumes that luminous matter is a biased tracer of the underlying dark matter distribution \citep{verde_2dF_bias_2002, desjacques_large_2018}. Mathematically, we can write
\begin{align}
\label{eq:gal_bias}
    \delta_\mathrm{gal} = b \; \delta_\mathrm{m} \;,
\end{align}
where $\delta_\mathrm{gal}$ and $\delta_\mathrm{m}$ are the galaxy and matter overdensities, respectively, related by the bias parameter $b$.

A large body of work exists on various methods of density estimation and is broadly divided into three categories. Firstly, in the grid-based method, galaxies are distributed on a 3D Cartesian grid and weighted by their mass before applying a smoothing kernel \citep{darvish_comparative_2015}. Next, in tessellation techniques, the kernel is made self-adaptive and has two major algorithms: Voronoi Tessellation Field Estimator \citep[VTFE;][]{rien_fragmenting_1994, hung_establishing_2020} and Delaunay Tessellation Field Estimator \citep[DTFE;][]{schapp_continous_2000, romano_DTFE_2007, feldbrugge_phase_2025}. Finally, the Bayesian Origin Reconstruction from Galaxies \citep[BORG;][]{jasche_bayesian_2013} method applies Hamiltonian Monte Carlo (HMC) sampling to create an initial field and then evolves it to reproduce the observed galaxy distribution. An extensive list of available tools implementing these methods, along with a broader overview and application, is presented in \citet{contarini_era_2026}.

In our work, we use Python libraries for the analysis of numerical simulations \citep[\texttt{\texttt{Pylians}}\footnote{\url{https://pylians3.readthedocs.io/en/master/index.html}},][]{pylians}. It creates a continuous field on a 3D grid and progressively smooths the field from large to small spatial scales. This preserves the natural hierarchical structure of the LSS. We use Cartesian coordinates and the galaxies' masses as inputs to the code. The \{x,y\} values are the sky positions of a galaxy, while the $z$ coordinate is the comoving distance calculated from the redshift of the galaxy for the Planck-2015 cosmological model. In the fiducial mock catalogue, we use the galaxy positions in real space and do not include the peculiar-velocity contributions, i.e., redshift-space distortion (RSD). The effect of including RSD is discussed in Section \ref{sub:rsd_and_grid}.

One of the largest sources of error here stems from uncertainties in the galaxies' redshifts. We assume that we have spectroscopic redshifts for all the galaxy samples and include the upper limit of \textit{Euclid}'s spectroscopic redshift error of $\sigma_{\mathrm{spec}} = 0.001 (1+z)$ \citep{mellier_Euclid_2025}. As the LoS vectors are parallel to the $z$ axis, this results in uncertainties in the galaxies' $z$ positions, which are calculated using the underlying cosmology. The assumption of a full spectroscopic sample is not realistic, as most wide-field surveys capture photometric information of the galaxies, where the uncertainties are larger. Therefore, we systematically evaluate the effect of spectroscopic and photometric redshift uncertainties on the galactic density estimator in Section \ref{sub:eff_redshift}. Similarly, uncertainties in the mass estimates derived from luminosities are included in the final error of the density estimator.

We choose $512^3$ as the grid resolution, deposit the galaxies on that grid, and add the mass via a mass assignment scheme like cloud-in-cell (CIC), triangular-shaped cloud (TSC), nearest grid point (NGP), and piecewise cubic spline (PCS) \citep[see e.g.][for comparisons]{cui_ideal_2008}. We test the effect of the algorithm via cross-spectra and coherence for all possible pair combinations and find no noticeable influence. Finally, we use CIC. To test the sensitivity to grid resolution, we used two additional grids at $256^3$ and $1024^3$, while keeping the galaxy catalogue, CIC assignment and density-bin definition fixed. We discuss the findings in Section \ref{sub:rsd_and_grid}.

Furthermore, we compare the quality of the \texttt{\texttt{Pylians}} cube with estimates from the Void Analysis Software Toolkit \citep[\texttt{VAST}\footnote{\url{https://vast.readthedocs.io/en/latest/}},][]{douglass_VAST_2022} package, implementing both VoidFinder and V2 algorithms. For the same galaxy sample, the voids identified by \texttt{VAST} occupy $\sim 86$ per cent of the total simulation box volume, where the choice of algorithm has no noticeable effect. For the \texttt{Pylians}' estimate, we defined a void as $\rho_\mathrm{c}/\rho_\mathrm{crit} < 0.1$ following \citet{haider_illustris_lls_2016, martizzi_baryons_2019, artale_illustris_classification_2022} and the fraction is 83 per cent. Additionally, the centres of the voids in the \texttt{VAST} catalogue coincide with the centre of the voxels in the \texttt{Pylians} cube. This comparison holds as we use the same grid resolution in both cases.

For a given LoS vector for an FRB, the contributing voxels are divided into the same density bin definition used for FRBs. This creates a density estimator of the visible matter along the LoS. As galaxies are biased tracers of the underlying dark matter distribution, the density obtained from the galaxy survey is also a biased proxy for it. We can translate the bias in Eq.~\eqref{eq:gal_bias} as
\begin{align}
\label{eq:r_gal_definition}
    \mathbf{r}_\mathrm{gal} \equiv \Bar{\rho}_\mathrm{m} \left[1 + \frac{\delta_{\mathrm{gal},i}}{b_i} \right] \;,
\end{align}
where $\Bar{\rho}_\mathrm{m}$ is the mean matter density, $\delta_{\mathrm{gal},i}$ is the galaxy overdensity, and $b_i$ is the bias in the $i$-th bin. We define $\mathbf{r}_\mathrm{gal}$ as the density estimator for galaxies.

The bias correction term is the last required quantity and remains an active area of research, with various pre-existing methods available. Traditionally, the galaxy power spectrum is used to estimate the linear bias. The non-linear bias estimation requires independent probes. For example, the weak lensing signal, particularly galaxy-galaxy lensing, can be used, as gravitational lensing is sensitive to both visible and dark matter \citep{bartelmann_weak_2001, prat_galaxy_2018, giannini_DES_gg_2026}. Similarly, the velocity field is affected by the total matter, and the effect, redshift-space distortion (RSD), quantifies the bias \citep{kaiser_clustering_1987, chuang_clustering_2016}. Beyond the two-point statistics, galaxy bispectra and higher-order statistics can infer the bias, as shown in works by \citet{gaztanaga_bias_1994, pollack_new_2014, ivanov_precision_2022}. Cross-correlation of galaxies with other tracers, such as cosmic microwave background (CMB) lensing maps \citep{schmittfull_parameter_2018} and the tSZ effect \citep{yan_probing_2021}, also constrains the bias parameter.

Here, we calibrate the bias using values taken from the simulation itself. We calculate both the total matter and galaxy number overdensities, as in Eq.~\eqref{eq:gal_bias}, on the same 512-grid and apply logarithmic binning. The result is the bias parameter in each bin. This calibration ensures the estimator is unbiased. However, for real observational data, when the galaxy distribution is converted to a density distribution, the noise is larger compared to our simulation-derived values.

\section{Results}
\label{sec:result}
This section contains the primary findings of this work. We start with a density estimator based on FRBs and investigate how its shape encodes both gravitational and non-gravitational information that affects the baryons. Further, we estimate and compare electron densities across the eight binned regions by incorporating density estimators from FRBs and galaxies.

\subsection{Reconstructed density estimator}
First, we present the results from the density estimation using FRBs, described in Section \ref{sec:inversion}. Within the Bayesian framework, we used a parametric model in Eq.~\eqref{eq:r_prior} with four free parameters, i.e., the scale $\mathcal{S}$, transition density $\log \rho_\mathrm{tran}$, the width $w_\rho$, and the skew parameter $\alpha_\rho$.

\begin{figure}[!ht]
    \centering
    \includegraphics[width=0.9\linewidth]{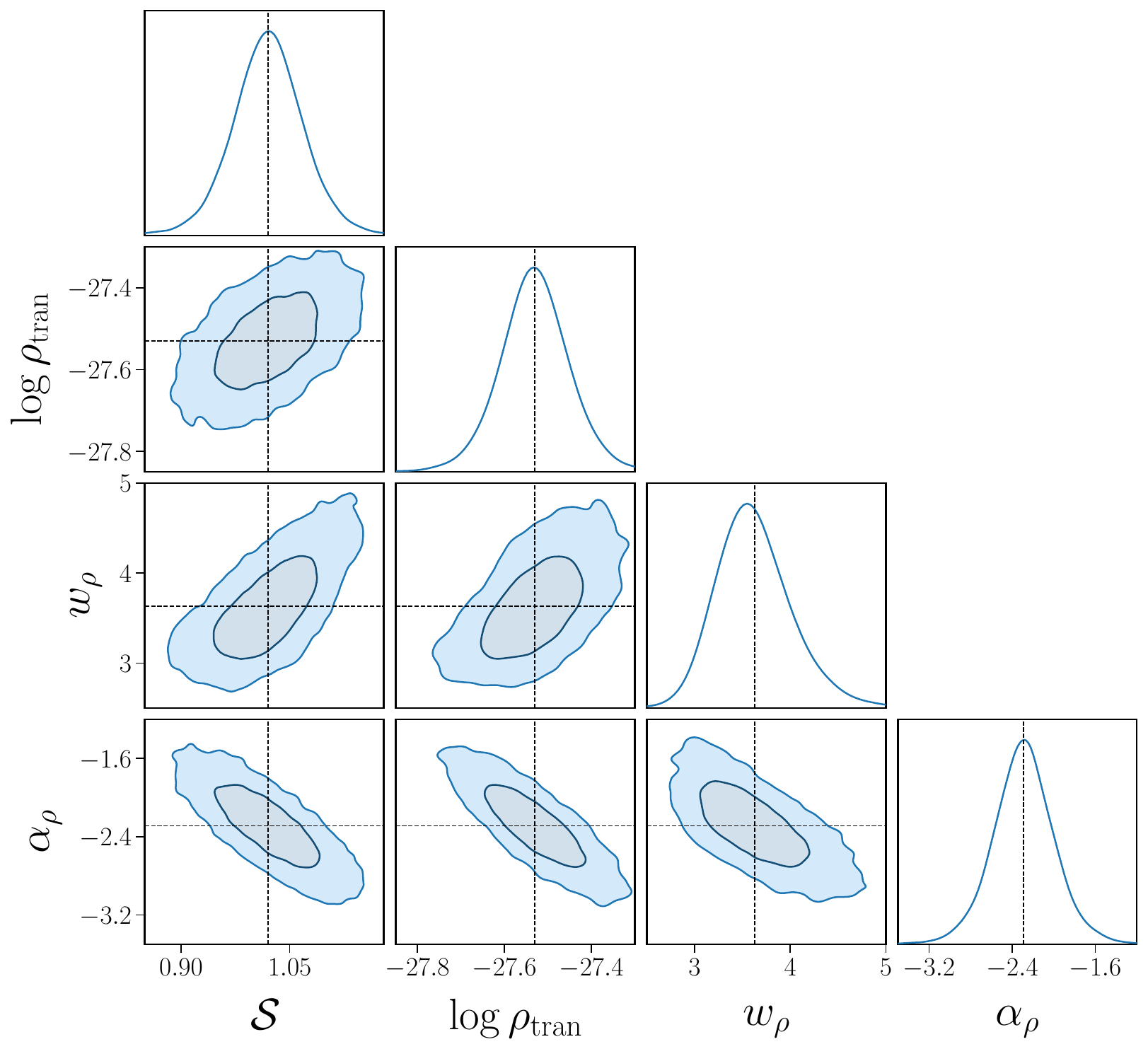}
    \caption{\small Marginalised contour plots for all the parameters in our inference process. The posterior mean values are shown as grey dotted lines at $\{1.02, -27.53, 3.59, -2.29\}$.}
    \label{fig:contour_mcmc}
    \vspace{.1cm}
\end{figure}

In Fig.~\ref{fig:contour_mcmc}, we show the marginalised contour plots of the posterior distributions. The optimal parameter values, estimated as the means of the respective posterior distributions, are $\theta_\mathrm{mean} = \{1.02 \pm 0.05, -27.53 \pm 0.07, 3.59^{+0.40}_{-0.33}, -2.29^{+0.28}_{-0.27} \}$, where the quoted uncertainties represent the $1\sigma$ credible intervals.

Given $\theta_\mathrm{mean}$, we reconstruct the density profile with the definition in Eq.~\eqref{eq:r_prior} and denote it by $\hat{\mathbf{r}}_\mathrm{FRB}$. This is the density-ratio estimator obtained via FRBs. In principle, one could choose a functional form of any arbitrary complexity or even a fully non-parametric form for $\mathbf{r}$, provided the data can constrain such a model robustly. In Section \ref{sub:eff_non_para}, we shortly compare a density estimator with a non-parametric prior on $\mathbf{r}$.

\begin{figure}[!ht]
    \centering
    \includegraphics[width=0.9\linewidth]{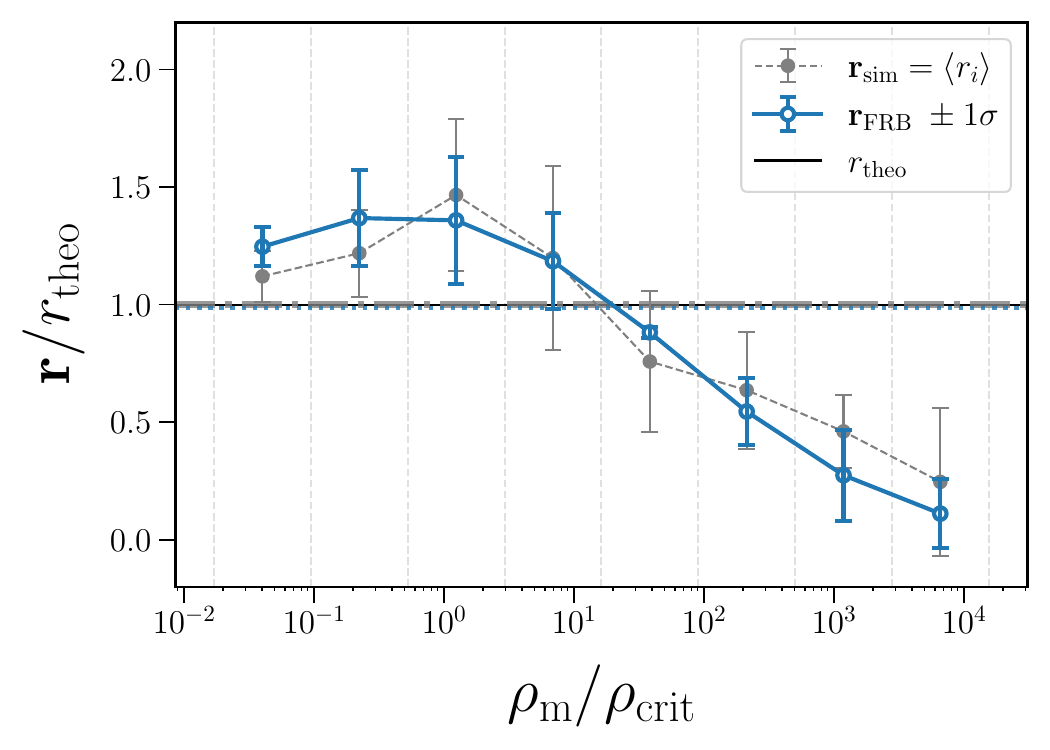}
    \caption{\small The reconstructed $\mathbf{r}$ values from the MCMC analysis calculated using Eq.~\eqref{eq:r_prior} are shown in blue. The vertical dotted lines represent the bin edges, and the grey data points indicate the ground truth from the simulation. The horizontal black line at unity marks $r = r_\mathrm{theo}$. The mass-weighted means of the ground truth and the reconstructed profiles are shown by dotted grey and blue horizontal lines at 1.0007 and 0.9860, respectively.}
    \label{fig:r_reconstruct}
    \vspace{.1cm}
\end{figure}

The profile of $\hat{\mathbf{r}}_\mathrm{FRB}$ is shown in blue points in Fig.~\ref{fig:r_reconstruct}, alongside the ground truth from the simulations in grey (dashed line). While the ground truth is inaccessible in real observational data, including it in the figure serves as a validation of our result. Crucially, our inversion for density estimation functions independently of the true values and requires only the configuration matrix and the DM values. We show the values in units of $r_\mathrm{theo}$. The horizontal solid black line at unity represents $r_\mathrm{theo}$, whereas the grey and blue lines denote the means of $\mathbf{r}_\mathrm{sim}$ and $\mathbf{r}_\mathrm{FRB}$. The mass-weighted mean of the simulated data (grey line) is $\langle r_\mathrm{sim}\rangle_M = 1.0007\,r_\mathrm{theo}$, which confirms that the binned ground truth recovers the electron budget of the snapshot. The analogous value from the reconstruction yields $\langle r_\mathrm{FRB}\rangle_M = (0.986 \pm 0.005)\,r_\mathrm{theo}$, which we indicate with the dotted blue line.

Recalling that $\mathbf{r}$ summarises the ratio of the electron and total matter density per density bin, we observe that the values are greater (smaller) than unity for low (high) density bins. Such behaviour is caused by the interplay of gravitational interactions within the LSS, combined with non-gravitational baryonic feedback via AGNs and supernovae, which drives electrons from the central part of the halo or high-density bins to the outskirts with lower density. Additionally, gas inflows induce shock heating that persists since reionisation due to very long cooling and recombination times in the ionised IGM. Note that this rise at low densities is largely compensated by a fall at higher densities, so that the mass-weighted mean recovers the theoretical value to 1.4 per cent ($\langle r_\mathrm{FRB}\rangle_M = 0.986\,r_\mathrm{theo}$). The theoretical mean is $r_\mathrm{theo} \approx 7.84 \times 10^{22} \;\mathrm{gram}^{-1}$.

Fig.~\ref{fig:r_reconstruct} shows that electrons are mostly localised in low- to mid-density regions, consistent with the global theoretical estimate. The location of the highest value hints at the effects of the gravitational interaction, suggesting most baryons are localised in low-density filamentary regions. The relative decrement of electrons in high-density environments is expected to be largely due to feedback, and, in principle, more bins can describe the baryon redistribution more robustly. Increasing the number of bins requires finer sensitivity to the contribution of individual bins to the total DM, which in turn requires more FRB sightlines. However, the number of LoS within the simulation box is not arbitrary and is limited by the correlation threshold (see Section \ref{sub:eff_corr}). Therefore, for this study, we restrict ourselves to eight density bins. Conversely, real observational data of localised FRBs can avoid this limitation due to larger angular separation.

In the next section, we anchor this density estimator to a more data-driven case using galaxy surveys and provide a quantitative estimate of the electron density in the eight bins.

\subsection{Electron density estimation}
The density estimator using FRBs is the ratio of electron number density and the total matter density, i.e., $\mathbf{r}_\mathrm{FRB} \equiv n_\mathrm{e}/\rho_\mathrm{m}$. Similarly, the density estimator from the galaxy survey, $\mathbf{r}_\mathrm{gal}$, is defined in Eq.~\eqref{eq:r_gal_definition}. Then, the estimated electron number density can be expressed in terms of the two density estimators via
\begin{align}
\label{eq:n_e_recon}
    \hat{n}_\mathrm{e} = \mathbf{r}_\mathrm{FRB} \cdot \mathbf{r}_\mathrm{gal} =  \mathbf{r}_\mathrm{FRB} \cdot \Bar{\rho}_\mathrm{m} \left[1 + \frac{\delta_\mathrm{gal}}{b} \right] \;
\end{align}
where $\hat{n}_\mathrm{e}$ is the reconstructed electron number density, $\mathbf{r}_\mathrm{FRB}$ and $\mathbf{r}_\mathrm{gal}$ are the estimators from FRBs and galaxies, respectively, and finally, $b$ is the bias.

\begin{figure}[!ht]
    \centering
    \includegraphics[width=0.9\linewidth]{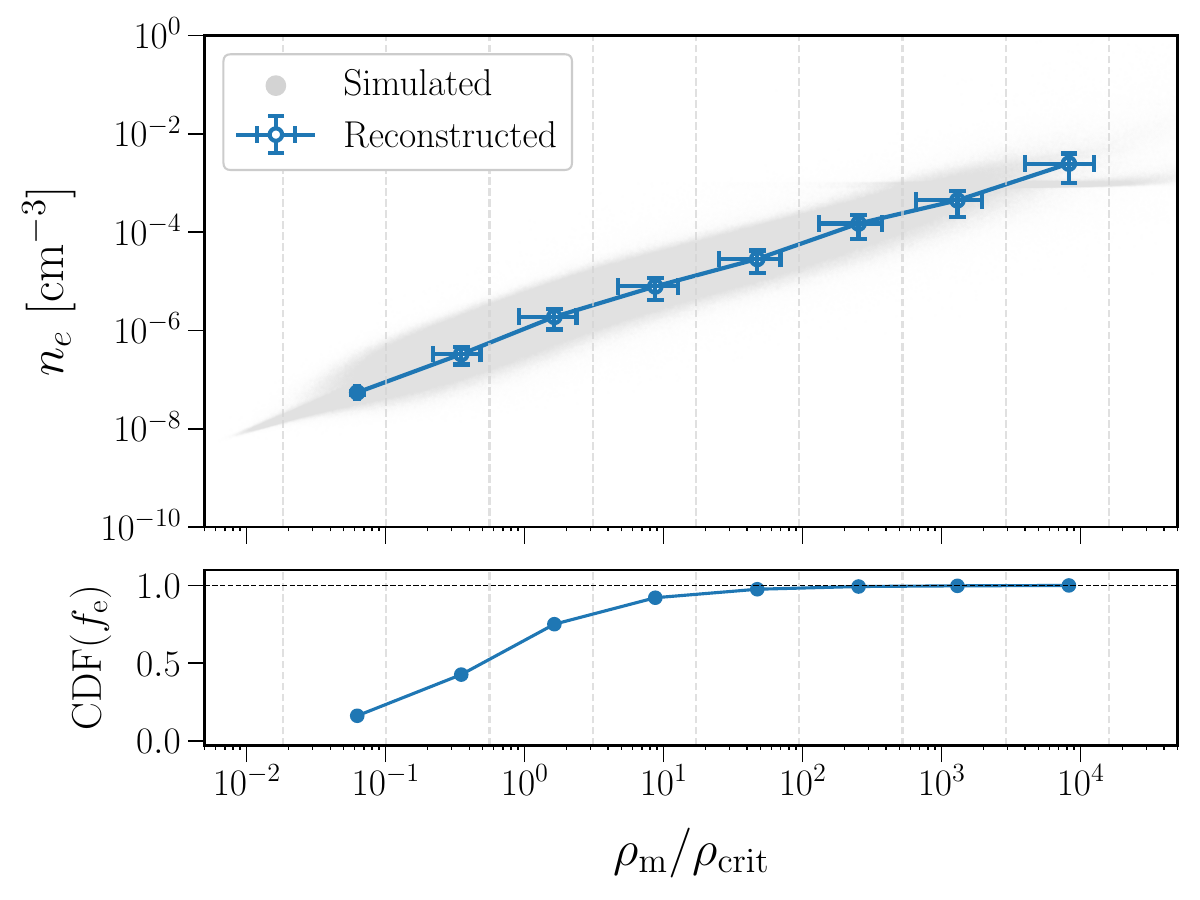}
    \caption{\small Top: electron density inferred by multiplying $\mathbf{r}_\mathrm{FRB}$ from FRBs and $\mathbf{r}_\mathrm{gal}$ from galaxies. In grey, we show the electron density obtained from the snapshot data, with blue dots indicating the reconstructed values. The associated error bars are shown with blue lines. The vertical dotted lines are the bin edges used in FRBs, which are the same as those in Fig.~\ref{fig:r_reconstruct}. Bottom: the cumulative distribution function for Eq.~\eqref{eq:electron_fraction} is shown. This describes the fraction of total electrons in each density bin.}
    \label{fig:n_e_in_bin}
    \vspace{.1cm}
\end{figure}

In the top panel of Fig.~\ref{fig:n_e_in_bin}, we show the result. The grey scattered points represent the electron density calculated from the full snapshot data, which serves as the ground truth in this case. The vertical lines are the bin edges used to classify the LSS. In blue, we show the electron densities, where the connecting line aligns well with the snapshot data across all the density regions.

To quantify the contribution of each density bin to the global electron budget, we can calculate the electron fraction for the $i$-th bin via
\begin{align}
\label{eq:electron_fraction}
    f_{\mathrm{e},\,i} = \frac{\hat{n}_{\mathrm{e},\,i} \, V_i}{\sum (\hat{n}_{\mathrm{e},\,j} V_j)} \;,
\end{align}
where $\hat{n}_{\mathrm{e},\,i}$ and $V_i$ are the electron density and the volume of the $i$-th density bin, respectively and $\sum (\hat{n}_{\mathrm{e},\,j} V_j)$ is the total number of electrons across all bins. Physically, this shows the fraction of the total electron content in each bin.

We show the equivalent cumulative distribution function (CDF) in the bottom panel of Fig.~\ref{fig:n_e_in_bin}. The distribution indicates that most electrons in the LSS are concentrated in low-density regions. Specifically, the first three to four bins contain approximately 75 per cent and 92 per cent of electrons, respectively, consistent with the findings in \citet{martizzi_baryons_2019}.

Further, we calculate the ionised gas fraction or $f_\mathrm{gas}$, for the eight bins using
\begin{align}
\label{eq:f_gas_definition}
    f_{\mathrm{gas},\,i} = \frac{\mu_\mathrm{e} \, m_\mathrm{p}\, n_{\mathrm{e},\,i}}{\rho_{\mathrm{m},\,i}} = \mu_\mathrm{e}\, m_\mathrm{p}\, r_{\mathrm{FRB},\,i} = \frac{f_\mathrm{ion}\,\Omega_\mathrm{b}}{\Omega_\mathrm{m}}\,\frac{r_{\mathrm{FRB},\,i}}{r_\mathrm{theo}} \;,
\end{align}
where the last equality uses $\mu_\mathrm{e} m_\mathrm{p} r_\mathrm{theo} = f_\mathrm{ion}\Omega_\mathrm{b}/\Omega_\mathrm{m} = 0.1495$. Therefore, the gas fractions depend only on the FRB-based estimator and not on the galaxy-based matter density, which enters only the electron densities and $\Omega_\mathrm{b} f_\mathrm{ion}$. The ground truth follows in the same way from $\mathbf{r}_\mathrm{sim}$ (grey points in Fig.~\ref{fig:r_reconstruct}).

Now, following \citet{artale_illustris_classification_2022}, we categorise the eight bins into three regions \citep[see Eq.~(12) in][]{walker_dispersion_2024}. Specifically, void, filament and halo regions are roughly covered by bin 1, bins 2-5 and bins 6-8, respectively. Note that the density boundaries separating these environments do not, in general, coincide with the nine bin edges of our eight logarithmic density bins. Therefore, this mapping is an approximate assignment of the discrete density bins to the three environments. To minimise the effect of their different volume occupancies and focus on their characteristic physical states, we assign equal weight to the density bins and use the arithmetic mean. Therefore, these values should be interpreted as comparative, equal-bin-weighted estimates rather than volume- or mass-weighted integrated gas fractions. The ground truth is grouped and weighted similarly, and the uncertainties are propagated from the per-bin errors as $\sigma_E = (f_\mathrm{ion}\Omega_\mathrm{b}/\Omega_\mathrm{m})\,[\sum_{i\in E}\sigma^2_{r,i}]^{1/2}/(N_E\,r_\mathrm{theo})$, where $N_E$ is the number of bins in environment $E$. Note that this neglects the correlation between bins induced by the parametric profile in Eq.~\eqref{eq:r_prior}, so these uncertainties are approximate.

We obtain $f_\mathrm{gas}^\mathrm{void} = 0.186 \pm 0.012$, $f_\mathrm{gas}^\mathrm{filament} = 0.179 \pm 0.015$ and $f_\mathrm{gas}^\mathrm{halo} = 0.046 \pm 0.014$, compared with the simulation ground truths of 0.167, 0.173 and 0.067, i.e., within $1.6\sigma$, $0.4\sigma$ and $1.5\sigma$, respectively. As we use equal bin weights, the appropriate global check is the mass-weighted mean over all bins, $\sum_i \mu_\mathrm{e} m_\mathrm{p} \hat{n}_{\mathrm{e},i} V_i / \sum_i \rho_{\mathrm{m},i} V_i = 0.147$, compared with $f_\mathrm{ion}\Omega_\mathrm{b}/\Omega_\mathrm{m} = 0.1495$ in the simulation, consistent with $\langle r_\mathrm{FRB}\rangle_M = 0.986\,r_\mathrm{theo}$.

\begin{figure}[!ht]
    \centering
    \includegraphics[width=0.9\linewidth]{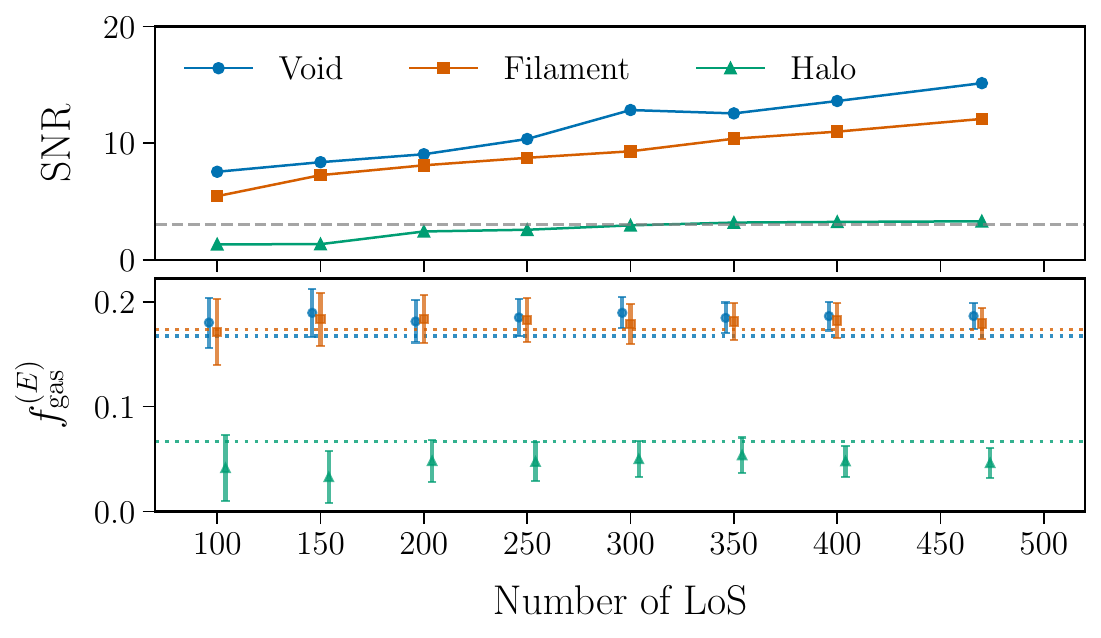}
    \caption{\small Top: SNR values of $f_\mathrm{gas}$ for three regions are shown as a function of the number of LoS. The horizontal line represents the $3\sigma$ threshold. Bottom: The $f_\mathrm{gas}$ parameter for the three regions is shown in the respective colours, where the dotted lines represent the corresponding ground-truth values from the simulation, computed identically from $\mathbf{r}_\mathrm{sim}$. }
    \label{fig:snr_los}
    \vspace{.1cm}
\end{figure}

In Fig.~\ref{fig:snr_los}, we show how the SNR of $f_\mathrm{gas}$ changes based on the number of FRB LoS. The SNR for each of the environments is calculated as the ratio of the mean to the $1\sigma$ uncertainty, i.e., $f^{(E)}_\mathrm{gas}/\sigma_{f^{(E)}_\mathrm{gas}}$, where $E \in \{\text{void, filament, halo} \}$. Since $f_\mathrm{gas}$ depends only on $\mathbf{r}_\mathrm{FRB}$ (Eq.~\ref{eq:f_gas_definition}), the variation is coming from changes in the $\mathbf{r}_\mathrm{FRB}$ estimates at varying numbers of LoS. We note that with reduced LoS, the fidelity of the density estimator, particularly in the high-density bins, worsens (see Section~\ref{sub:eff_corr}). However, averaging $f_\mathrm{gas}$ estimates from multiple bins into three results in a more stable value, with reduced uncertainty, as seen in the bottom panel of Fig.~\ref{fig:snr_los}. The halo estimate lies systematically below the ground truth. As the high-density bins occupy a small volume fraction, their columns of $\mathbf{A}$ carry little weight in the DMs, and after the correlation cut, only a few sightlines through haloes remain. The likelihood is therefore dominated by the low- and intermediate-density bins, and $\mathbf{r}$ at high densities is set largely by the tail of the parametric profile in Eq.~\eqref{eq:r_prior}. Moreover, the sample variance is very high, which implies that a precise measurement of the halo contribution would require a larger number of FRBs and hence a larger simulation volume.

Finally, we compute the density of ionised baryons as the discrete volume integral of the electron densities over all bins via $\Omega_\mathrm{b} f_\mathrm{ion} = \mu_\mathrm{e} m_\mathrm{p} \sum_i \hat{n}_{\mathrm{e},i} V_i / (\rho_\mathrm{crit} V_\mathrm{box})$, where $i$ is the bin index. We obtain $\Omega_\mathrm{b} f_\mathrm{ion} = 0.0455^{+0.0028}_{-0.0026}$, which agrees with the simulation value of $0.0462$ $(= 0.95\times0.0486)$ at the $0.2\sigma$ level. This is consistent with the mass-weighted mean $\langle r_\mathrm{FRB}\rangle_M = 0.986\,r_\mathrm{theo}$, since $0.986 \times 0.0462 = 0.0455$. To calculate $\Omega_\mathrm{b}$, we break the degeneracy using $f_\mathrm{ion}=0.95$, and find $\Omega_\mathrm{b} = 0.0479^{+0.0029}_{-0.0027}$, compared with the true value of 0.0486 ($0.2\sigma$).

\section{Discussion}
\label{sec:discussion}
In this section, we review the effect of various correlation thresholds for the LoS, the choice of a non-parametric prior model for density estimation, and the impact of RSD, grid resolution, and redshift uncertainties on the galactic density estimator. In summary, we assess the sensitivity of the fiducial results for the considered modelling choices.

\subsection{Effect of correlation}
\label{sub:eff_corr}
To quantify the relative independence of the LoS, we compute correlations between all pairs of LoS vectors using Eq.~\eqref{eq:los_corr}, with values of 0 and 1 representing independent and perfectly correlated LoS vectors, respectively. Moreover, in our fiducial calculation, we adopt a correlation requirement of $<0.3$. Here, we quantify its impact on the final reconstruction using nine correlation-threshold cuts ranging from 0.1 to 0.9. Physically, a lower threshold indicates that the LoS are more sparsely distributed, whereas a higher threshold implies greater clustering. For the 0.1 threshold, the sampling preferably selects contributions from the low-to-mid-density regions (bins 1-4), as they occupy a larger volume fraction of the simulation box. At the same time, we undersample the high-density regions (bins 7 and 8). Combined, we are left with a limited number of LoS, resulting in a very low SNR ($\sim 1$) across all bins.

Alternatively, for a very high correlation threshold of 0.9, the sampling and thus the SNR increase in low-density regions. On the other hand, highly correlated LoS vectors may reflect contributions from the same haloes, where the redundancy does not necessarily translate into higher SNR. The corresponding DMs become highly correlated while the diagonal covariance treats them as independent, which leads to multiple counting of the same structure. The estimates in the high-density regions become dominated by a few repeatedly sampled structures. As a result, although the average SNR across all bins is higher, the diagonal-covariance inversion overestimates the amount of independent information and can assign excessive weight to repeatedly sampled structures.

Our test indicates that any threshold below 0.5 avoids over-weighting repeatedly sampled structures in the high-density bins without appreciably reducing the average SNR. However, we select 0.3 as a conservative threshold. Increasing the threshold to 0.5 reduces the uncertainties in the high-density bins by only 3 per cent, which does not justify the additional correlation between sightlines.

The threshold also affects the final reconstructed profile of the density estimator. As pointed out by \citet{walker_dispersion_2024} using the same IllustrisTNG simulation boxes, the void contribution remains almost constant at $\sim 1$ per cent across redshifts. Our results suggest the same, as the model provides a smooth profile, partly due to the parametric modelling choice, rather than picking out the curvature as seen in the albeit low-signal-to-noise grey dashed line in Fig.~\ref{fig:r_reconstruct}. The immediate solution is to increase the number of LoS of localised FRBs. However, because the currently used simulation box size is finite, this inevitably increases correlations among different LoS. Alternatively, one can adopt two distinct LoS schemes to sample the low and high densities independently. This approach ensures that the DM contribution remains sensitive to subtle changes in the low density, while also preventing redundant sampling of the same halo. For a real observation, the angular separations between FRBs are larger, which allows for a larger number of LoS that are mutually independent. Since we aim to apply this method to observational data, we do not further explore the two-way sampling scheme in the present work.

\subsection{Effect of non-parametric prior}
\label{sub:eff_non_para}
In this section, we contrast our results obtained using a parametric prior with those derived from a non-parametric specification. Specifically, we employ hierarchical Bayesian inference with a Gaussian prior centred on the theoretical mean value $r_\mathrm{theo}$, whose covariance is hyperparameterised by the prior strength, smoothness, and correlation length scale. By analytically marginalising over $\mathbf{r}$, the posterior distribution can be expressed as
\begin{align}
    p(\theta\,|\,\mathbf{d})_\mathrm{GP}
    = \mathcal{N}\!\bigl(\mathbf{d}\,\big|\,\mathbf{A} r_\mathrm{theo},\, \mathbf{A}\,\mathbf{K}(\theta)\,\mathbf{A}^\mathrm{T} + \mathbf{C}\bigr) \;,
\end{align}
where $\mathbf{K}(\theta)$ denotes the hyperparameterised covariance matrix. For each posterior sample of $\theta$, we then compute the conditional Gaussian posterior for $\mathbf{r}$ and generate corresponding conditional realisations. The final posterior distribution for $\mathbf{r}$ is obtained by aggregating these conditional samples and marginalising over the hyperparameters.

The results indicate an average correlation length of 2.63 bins (i.e., dex), which enforces relatively smooth variations across the eight bins. Nonetheless, within the non-parametric framework, the overall profile of $\mathbf{r}$ exhibits a markedly stronger contribution from low-density regions than from high-density regions. As a consequence, the mean estimator is systematically biased towards values larger than $r_\mathrm{theo}$.

Such behaviour in a non-parametric model with wide priors is due to model degeneracies. The total DM of an FRB is degenerate among the contributions from the Milky Way, the LSS, and the host galaxy, as seen in Eq.~\eqref{eq:dm_with_mw_lss_host_comps}. In our approach, we simulate only the LSS contribution and then decompose it into contributions from eight LSS regions. This signal is then distributed among the bins. Recalling that the number of LoS is comparatively small for high-density regions, most of the information in the $\mathbf{A}$ operator is concentrated in the low-density bins. Moreover, the contribution to the total DM from individual haloes is much larger than in void regions, leading to a log-normal distribution of observed DM values \citep{walker_dispersion_2024}. This translates to a sharp increase in the respective column for a few LoS. The linear model and its inversion using $\mathbf{d}$ and $\mathbf{A}$ with a non-parametric prior, therefore, dampens the overall high-density contribution in favour of a higher low-density contribution. The parametric prior model in Eq.~\eqref{eq:r_prior}, on the other hand, imposes a mean density estimator or $\log \rho_\mathrm{tran}$ along with $w_\rho$ for the onset and steepness of the turnover and avoids the collapse to smaller values at high-density bins.

Furthermore, the conditional Gaussian posterior can produce unphysical negative values of $\mathbf{r}$. One solution is to use a truncated Gaussian or a log-normal distribution, which makes the marginalisation in the first step non-trivial. Therefore, a more complex non-parametric approach is not warranted over our simulation-driven parametric modelling approach with the current data.

\subsection{Effects on the galaxy-based density estimation}
\subsubsection{RSD and grid resolution}
\label{sub:rsd_and_grid}
To test the effect of RSD, we included the peculiar velocities of the galaxies along the LoS and repeated the density reconstruction at a fixed $512^3$ grid resolution. Compared to the fiducial real-space case, the galaxy density estimate increases by $\sim 8-10$ per cent in low-density bins as RSD partly fills in the voids, and decreases by up to 20 per cent in the high-density bins because of structure smearing along the LoS. The mean absolute change over the eight bins is 14 per cent. As for the result, the density estimator based on FRB ($\mathbf{r}_\mathrm{FRB}$) and the gas fractions in Eq.~\eqref{eq:f_gas_definition} are not affected by RSD. It changes the reconstructed electron density in Eq.~\eqref{eq:n_e_recon} by the same factors, and raises the recovered $\Omega_\mathrm{b} f_\mathrm{ion}$ by $\lesssim 10$ per cent. Since real observations are in redshift space, our real-space estimator is optimistic in this case. RSD is the dominant systematic effect, and dedicated redshift-space calibration is beyond the scope of the work presented here.

The grid resolution of $512^3$ is another fiducial choice, and we repeat the real-space density-field reconstruction at $256^3$ and $1024^3$. This also sets the scales over which the density is averaged, which are $0.80, 0.40$ and $0.20\,h^{-1}$ Mpc. Therefore, the matter-density contrast also increases with increasing resolution, with $\sigma(\delta_m)=16.75$, 25.14 and 35.73, respectively. At the same time, the galaxy field becomes sparser, and the voxel-level correspondence between galaxy and matter overdensities weakens at higher resolution, indicating increasing departures from the simple linear-bias approximation on smaller grid scales. For each resolution, the bias is recalibrated on the same grid. At $1024^3$, applying the bias calibration of $512^3$ systematically underestimates the mean matter density in bins 2-8 by 3-18 per cent. Moreover, in the lowest-density bin, most cells at this high resolution contain no particles, which makes the binned matter density dominated by discreteness. At $256^3$, the estimator underestimates the matter density by 16 per cent on average. With the bin-wise changes, the recovered $\Omega_\mathrm{b} f_\mathrm{ion}$ decreases by 0.4 and 0.7 per cent at $256^3$ and $1024^3$, respectively. This is primarily because the changes in the different bins compensate, leaving the overall shift in $\Omega_\mathrm{b} f_\mathrm{ion}$ small compared with its statistical uncertainty. Finally, we adopt $512^3$ as a compromise between resolving the density field, maintaining adequate sampling of the galaxy distribution, and preserving the linear-bias approximation. The fiducial bias calibration uses the $512^3$ grid.

\subsubsection{Redshift uncertainty}
\label{sub:eff_redshift}
For the fiducial results, we assume a spectroscopic redshift for all galaxies with an upper limit of uncertainty of $\sigma_{\mathrm{spec}} = 0.001 (1+z)$. In realistic surveys, it is unlikely that spectroscopic redshifts can be obtained for all galaxies. Hence, we include photometric redshift information to quantify its effect on density estimation. \textit{Euclid} expects a photometric redshift uncertainty of $\sigma_{\mathrm{phot}} = 0.05 (1+z)$ \citep{laureijs_Euclid_definition_2011}. However, such an error margin has a particularly significant impact on distance measurements at higher redshifts. For instance, as the simulation box has a side length of 205 $h^{-1}$ Mpc, $\Delta z = 0.05$ or the associated error in comoving distance, $\Delta \ell(z=1) \approx 250$ Mpc, is comparable to the simulation box, which implies the density estimates are wrong. For \textit{Euclid}, the spectroscopic redshifts for the wide and deep fields are in the range $0.9 < z < 1.8$ and $0.7 < z < 2.9$ \citep{mellier_Euclid_2025}, where the effect of photometric uncertainty can be minimised with more accurate spectroscopic redshift values.

To test the impact of less complete galaxy surveys, due to the requirement for spectroscopic follow-up, we reduce the number of galaxies by reducing the final subhalo sample by $(i)$ 10 per cent, $(ii)$ 50 per cent, and $(iii)$ 80 per cent. This reduces the number of galaxies along the LoS and introduces errors in the estimation of galactic density. The noticeable effects are in the two extreme bins. A pairwise comparison shows that the low-density estimate for the third sample is 31 per cent and 27 per cent lower than those for the first and second samples, respectively. The same applies to the high-density yields of 53 per cent and 46 per cent. This reveals that galactic density estimation relies on both the number density and precise redshift measurements. Moreover, as there is no significant difference between the 10 per cent and 50 per cent selection, we expect a robust galactic density estimator for real observations with spectroscopic redshift on a subset of the full catalogue.

The electron density calculation relies on the mean matter density, which yields accurate mean values even with small galaxy samples. However, because structure growth is non-Gaussian, density estimates in regions of low galaxy density can yield erroneous estimates of higher-order moments due to interpolation in mass-assignment schemes. This constitutes a loss of cosmological information and can lead to potentially biased estimates.

\section{Summary and conclusion}
\label{sec:conclusion}
In this work, we decomposed the cosmic web into eight density regimes and estimated their (ionised) baryon content relative to the underlying total density field to quantitatively address the missing baryon problem. Below, we summarise our findings:
\begin{enumerate}[label=\roman*.]
    \item We employed fast radio bursts (FRBs) and their dispersion measures (DMs) to quantify the ionised baryon density along the line of sight. To simulate the DM, we used snapshot data from the IllustrisTNG project, a magnetohydrodynamical simulation that incorporates subgrid baryonic feedback, including AGN feedback. Decomposing the cosmic web into eight density bins and calculating the individual contributions to the total DM yielded a linear equation that maps the DM to the density in Eq.~\eqref{eq:frb_decopose_1}. We defined a density estimator in such a system as the ratio between the electron number density and the total matter density, so that it measures the relative abundance of electrons in each bin.
    \vspace{0.8em}
    \item For the simulated DM values and the linear operator, we infer the density estimator using an empirical prior model in Eq.~\eqref{eq:r_prior} within a Bayesian framework. In Fig.~\ref{fig:r_reconstruct}, we show a recovery of 98.60 per cent of the global ionised baryon content, while also accounting for the bin-wise distribution.
    \vspace{0.8em}
    \item Furthermore, to bridge the gap between simulation and observation, we integrated this FRB-based estimator with a mock galaxy catalogue. To this end, we constructed a density estimator along FRB sightlines using galaxies as luminous tracers of the underlying dark matter density field, based on the subhalo catalogue from the IllustrisTNG simulations. The resulting catalogue is a \textit{Euclid}-like mock subset that includes \textit{Euclid}-specific detection criteria and instrumental effects, as described in Section \ref{sub:mock_photometry}. This estimator provides a biased proxy for the total matter density. Therefore, we applied bias correction using the snapshot data.
    \vspace{0.8em}
    \item We obtained the electron density in each bin by combining these two density estimators from FRBs and galaxies. A comparison with the ground truth in Fig.~\ref{fig:n_e_in_bin} reveals an accurate reconstruction. Furthermore, we evaluated the electron fraction and concluded that approximately 75 per cent of the free electrons are located in the first three bins with $\rho_\mathrm{m}/\rho_\mathrm{crit} \leq 2.94$. We find the gas fractions in three substructures of the cosmic web defined in Section \ref{sub:decomposition} as $f_\mathrm{gas}^\mathrm{void} = 0.186 \pm 0.012$, $f_\mathrm{gas}^\mathrm{filament} = 0.179 \pm 0.015$, and $f_\mathrm{gas}^\mathrm{halo} = 0.046 \pm 0.014$, against ground truths of 0.167, 0.173 and 0.067. Finally, we calculate the ionised baryon density as $\Omega_\mathrm{b} f_\mathrm{ion} = 0.0455^{+0.0028}_{-0.0026}$, which agrees with the ground truth of the simulation at the $0.2\sigma$ level, corresponding to $\Omega_\mathrm{b} = 0.0479^{+0.0029}_{-0.0027}$ for $f_\mathrm{ion}=0.95$.
\end{enumerate}

In conclusion, we demonstrate that the linear framework based on FRB dispersion measures, coupled with density estimators, can effectively probe diffuse baryons in the large-scale structure (LSS) of the Universe. By integrating mock observational data, we provide a real-world anchor for our simulations, establishing the groundwork for incorporating future datasets as they become available. The fiducial results not only agree with global estimates of the baryon distribution but also quantify the baryon content in different subregions of the LSS. Furthermore, our modelling approach allows us to trace the effects of baryonic feedback across density regions. This provides an alternative to the more traditional two-point correlation function or power spectrum.

While our framework is robust, it currently relies on simulation-calibrated bias corrections. For real datasets, this could represent the most significant systematic uncertainty. Furthermore, although IllustrisTNG has high resolution, its subgrid feedback modelling is fixed. Applying this methodology to suites that provide variable feedback prescriptions, such as FLAMINGO, would be a critical next step. Such an analysis could determine whether the density estimator itself can be used to directly constrain feedback strength by mapping the flow from high-density regions.
The density estimator, however, could also be directly calibrated against weak-lensing observations, where available.

Looking ahead, we aim to apply the methods developed here to real localised FRBs from CHORD and DSA-2000, combined with galaxy surveys with \textit{Euclid} and LSST, where we expect orders-of-magnitude improvement in number density and redshift information. At larger angular separations, this would also allow for more independent line-of-sight traversal, leading to better constraints with a larger number of density bins.

\begin{acknowledgements}
The authors thank Steven Murray for his helpful comments during the preparation of the manuscript and the funding support from Stellenbosch University. We thank the Center for Information Technology of the University of Groningen for their support and for providing access to the Hábrók high performance computing cluster. Y.-Z. Ma acknowledges South Africa Research Chair Initiative (SARChI; Grant No. RCCA250325306198) from South Africa's Department of Science, Technology and Innovation and National Research Foundation (NRF), and other NRF Grants with No.~150580, No.~CHN22111069370, No.~ERC250324306141, No. AUPP250310302114.

\end{acknowledgements}

\bibliographystyle{bibtex/aa_doi}
\bibliography{bibtex/aa_references}

@ARTICLE{abareshi_overview_2022,
  author        = {Abareshi, B. and Aguilar, J. and Ahlen, S. and Alam, Shadab and Alexander, David M. and Alfarsy, R. and Allen, L. and Prieto, C. Allende and Alves, O. and Ameel, J. and Armengaud, E. and Asorey, J. and Aviles, Alejandro and Bailey, S. and Balaguera-Antolínez, A. and Ballester, O. and Baltay, C. and Bault, A. and Beltran, S. F. and Benavides, B. and BenZvi, S. and Berti, A. and Besuner, R. and Beutler, Florian and Bianchi, D. and Blake, C. and Blanc, P. and Blum, R. and Bolton, A. and Bose, S. and Bramall, D. and Brieden, S. and Brodzeller, A. and Brooks, D. and Brownewell, C. and Buckley-Geer, E. and Cahn, R. N. and Cai, Z. and Canning, R. and Rosell, A. Carnero and Carton, P. and Casas, R. and Castander, F. J. and Cervantes-Cota, J. L. and Chabanier, S. and Chaussidon, E. and Chuang, C. and Circosta, C. and Cole, S. and Cooper, A. P. and Costa, L. da and Cousinou, M.-C. and Cuceu, A. and Davis, T. M. and Dawson, K. and Cruz-Noriega, R. de la and Macorra, A. de la and Mattia, A. de and Costa, J. Della and Demmer, P. and Derwent, M. and Dey, A. and Dey, B. and Dhungana, G. and Ding, Z. and Dobson, C. and Doel, P. and Donald-McCann, J. and Donaldson, J. and Douglass, K. and Duan, Y. and Dunlop, P. and Edelstein, J. and Eftekharzadeh, S. and Eisenstein, D. J. and Enriquez-Vargas, M. and Escoffier, S. and Evatt, M. and Fagrelius, P. and Fan, X. and Fanning, K. and Fawcett, V. A. and Ferraro, S. and Ereza, J. and Flaugher, B. and Font-Ribera, A. and Forero-Romero, J. E. and Frenk, C. S. and Fromenteau, S. and Gänsicke, B. T. and Garcia-Quintero, C. and Garrison, L. and Gaztañaga, E. and Gerardi, F. and Gil-Marín, H. and Gontcho, S. Gontcho A. and Gonzalez-Morales, Alma X. and Gonzalez-de-Rivera, G. and Gonzalez-Perez, V. and Gordon, C. and Graur, O. and Green, D. and Grove, C. and Gruen, D. and Gutierrez, G. and Guy, J. and Hahn, C. and Harris, S. and Herrera, D. and Herrera-Alcantar, Hiram K. and Honscheid, K. and Howlett, C. and Huterer, D. and Iršič, V. and Ishak, M. and Jelinsky, P. and Jiang, L. and Jimenez, J. and Jing, Y. P. and Joyce, R. and Jullo, E. and Juneau, S. and Karaçaylı, N. G. and Karamanis, M. and Karcher, A. and Karim, T. and Kehoe, R. and Kent, S. and Kirkby, D. and Kisner, T. and Kitaura, F. and Koposov, S. E. and Kovács, A. and Kremin, A. and Krolewski, Alex and L'Huillier, B. and Lahav, O. and Lambert, A. and Lamman, C. and Lan, Ting-Wen and Landriau, M. and Lane, S. and Lang, D. and Lange, J. U. and Lasker, J. and Guillou, L. Le and Leauthaud, A. and Suu, A. Le Van and Levi, Michael E. and Li, T. S. and Magneville, C. and Manera, M. and Manser, Christopher J. and Marshall, B. and McCollam, W. and McDonald, P. and Meisner, Aaron M. and Mezcua, J. Mena-Fernández M. and Miller, T. and Miquel, R. and Montero-Camacho, P. and Moon, J. and Martini, J. Paul and Meneses-Rizo, J. and Moustakas, J. and Mueller, E. and Muñoz-Gutiérrez, Andrea and Myers, Adam D. and Nadathur, S. and Najita, J. and Napolitano, L. and Neilsen, E. and Newman, Jeffrey A. and Nie, J. D. and Ning, Y. and Niz, G. and Norberg, P. and Noriega, Hernán E. and O'Brien, T. and Obuljen, A. and Palanque-Delabrouille, N. and Palmese, A. and Zhiwei, P. and Pappalardo, D. and Peng, X. and Percival, W. J. and Perruchot, S. and Pogge, R. and Poppett, C. and Porredon, A. and Prada, F. and Prochaska, J. and Pucha, R. and Pérez-Fernández, A. and Pérez-Ráfols, I. and Rabinowitz, D. and Raichoor, A. and Ramirez-Solano, S. and Ramírez-Pérez, César and Ravoux, C. and Reil, K. and Rezaie, M. and Rocher, A. and Rockosi, C. and Roe, N. A. and Roodman, A. and Ross, A. J. and Rossi, G. and Ruggeri, R. and Ruhlmann-Kleider, V. and Sabiu, C. G. and Safonova, S. and Said, K. and Saintonge, A. and Catonga, Javier Salas and Samushia, L. and Sanchez, E. and Saulder, C. and Schaan, E. and Schlafly, E. and Schlegel, D. and Schmoll, J. and Scholte, D. and Schubnell, M. and Secroun, A. and Seo, H. and Serrano, S. and Sharples, Ray M. and Sholl, Michael J. and Silber, Joseph Harry and Silva, D. R. and Sirk, M. and Siudek, M. and Smith, A. and Sprayberry, D. and Staten, R. and Stupak, B. and Tan, T. and Tarlé, Gregory and Tie, Suk Sien and Tojeiro, R. and Ureña-López, L. A. and Valdes, F. and Valenzuela, O. and Valluri, M. and Vargas-Magaña, M. and Verde, L. and Walther, M. and Wang, B. and Wang, M. S. and Weaver, B. A. and Weaverdyck, C. and Wechsler, R. and Wilson, Michael J. and Yang, J. and Yu, Y. and Yuan, S. and Yèche, Christophe and Zhang, H. and Zhang, K. and Zhao, Cheng and Zhou, Rongpu and Zhou, Zhimin and Zou, H. and Zou, J. and Zou, S. and Zu, Y.},
  title         = {Overview of the {Instrumentation} for the {Dark} {Energy} {Spectroscopic} {Instrument}},
  journal       = {AJ},
  year          = {2022},
  volume        = {164},
  number        = {5},
  pages         = {207},
  doi           = {10.3847/1538-3881/ac882b}
}

@ARTICLE{abbott_ChimeCat2_2026,
  author        = {{FRB Collaboration} and {Abbott}, Thomas and {Andersen}, Bridget C. and {Andrew}, Shion and {Bandura}, Kevin and {Bhardwaj}, Mohit and {Bhusare}, Yash and {Brar}, Charanjot and {Cassanelli}, Tomas and {Chatterjee}, Shami and {Cliche}, Jean-Francois and {Cook}, Amanda M. and {Curtin}, Alice and {Dobbs}, Matt and {Dong}, Fengqiu Adam and {Eadie}, Gwendolyn and {Eftekhari}, Tarraneh and {Fonseca}, Emmanuel and {Gaensler}, B.~M. and {Good}, Deborah and {Halpern}, Mark and {Hessels}, Jason W.~T. and {Ibik}, Adaeze and {Jain}, Naman and {Joseph}, Ronniy C. and {Kader}, Zarif and {Kaspi}, Victoria M. and {Khan}, Afrokk and {Kharel}, Bikash and {Kumar}, Ajay and {Landecker}, T.~L. and {Lang}, Dustin and {Lanman}, Adam E. and {L'Argent}, Magnus and {Lazda}, Mattias and {Leung}, Calvin and {Li}, Dong Zi and {Lintott}, Chris J. and {Main}, Robert and {Masui}, Kiyoshi W. and {Mate}, Sujay and {McGregor}, Kyle and {Mckinven}, Ryan and {Mena-Parra}, Juan and {Meyers}, Bradley W. and {Michilli}, Daniele and {Ng}, Cherry and {Ng}, Mason and {Nimmo}, Kenzie and {Noble}, Gavin and {Pandhi}, Ayush and {Patil}, Swarali S. and {Pearlman}, Aaron B. and {Pen}, Ue-Li and {Pleunis}, Ziggy and {Prochaska}, J. Xavier and {Rafiei-Ravandi}, Masoud and {Ransom}, Scott and {Renard}, Andre and {Sammons}, Mawson W. and {Sand}, Ketan R. and {Scholz}, Paul and {Shah}, Vishwangi and {Shin}, Kaitlyn and {Siegel}, Seth R. and {Sirota}, Sloane and {Smith}, Kendrick and {Stairs}, Ingrid and {Stenning}, David C. and {Tendulkar}, Shriharsh P. and {Vanderlinde}, Keith and {Walmsley}, Mike and {Wang}, Haochen and {Wulf}, Dallas},
  title         = {{The Second CHIME/FRB Catalog of Fast Radio Bursts}},
  journal       = {ApJS},
  year          = {2026},
  volume        = {283},
  number        = {1},
  pages         = {34},
  doi           = {10.3847/1538-4365/ae3828},
  archiveprefix = {arXiv},
  eprint        = {2601.09399},
  primaryclass  = {astro-ph.HE}
}

@ARTICLE{ade_planck_cosmology_2016,
  author        = {{Planck Collaboration} and {Ade}, P.~A.~R. and {Aghanim}, N. and {Arnaud}, M. and {Ashdown}, M. and {Aumont}, J. and {Baccigalupi}, C. and {Banday}, A.~J. and {Barreiro}, R.~B. and {Bartlett}, J.~G. and {Bartolo}, N. and {Battaner}, E. and {Battye}, R. and {Benabed}, K. and {Beno{\^\i}t}, A. and {Benoit-L{\'e}vy}, A. and {Bernard}, J. -P. and {Bersanelli}, M. and {Bielewicz}, P. and {Bock}, J.~J. and {Bonaldi}, A. and {Bonavera}, L. and {Bond}, J.~R. and {Borrill}, J. and {Bouchet}, F.~R. and {Boulanger}, F. and {Bucher}, M. and {Burigana}, C. and {Butler}, R.~C. and {Calabrese}, E. and {Cardoso}, J. -F. and {Catalano}, A. and {Challinor}, A. and {Chamballu}, A. and {Chary}, R. -R. and {Chiang}, H.~C. and {Chluba}, J. and {Christensen}, P.~R. and {Church}, S. and {Clements}, D.~L. and {Colombi}, S. and {Colombo}, L.~P.~L. and {Combet}, C. and {Coulais}, A. and {Crill}, B.~P. and {Curto}, A. and {Cuttaia}, F. and {Danese}, L. and {Davies}, R.~D. and {Davis}, R.~J. and {de Bernardis}, P. and {de Rosa}, A. and {de Zotti}, G. and {Delabrouille}, J. and {D{\'e}sert}, F. -X. and {Di Valentino}, E. and {Dickinson}, C. and {Diego}, J.~M. and {Dolag}, K. and {Dole}, H. and {Donzelli}, S. and {Dor{\'e}}, O. and {Douspis}, M. and {Ducout}, A. and {Dunkley}, J. and {Dupac}, X. and {Efstathiou}, G. and {Elsner}, F. and {En{\ss}lin}, T.~A. and {Eriksen}, H.~K. and {Farhang}, M. and {Fergusson}, J. and {Finelli}, F. and {Forni}, O. and {Frailis}, M. and {Fraisse}, A.~A. and {Franceschi}, E. and {Frejsel}, A. and {Galeotta}, S. and {Galli}, S. and {Ganga}, K. and {Gauthier}, C. and {Gerbino}, M. and {Ghosh}, T. and {Giard}, M. and {Giraud-H{\'e}raud}, Y. and {Giusarma}, E. and {Gjerl{\o}w}, E. and {Gonz{\'a}lez-Nuevo}, J. and {G{\'o}rski}, K.~M. and {Gratton}, S. and {Gregorio}, A. and {Gruppuso}, A. and {Gudmundsson}, J.~E. and {Hamann}, J. and {Hansen}, F.~K. and {Hanson}, D. and {Harrison}, D.~L. and {Helou}, G. and {Henrot-Versill{\'e}}, S. and {Hern{\'a}ndez-Monteagudo}, C. and {Herranz}, D. and {Hildebrandt}, S.~R. and {Hivon}, E. and {Hobson}, M. and {Holmes}, W.~A. and {Hornstrup}, A. and {Hovest}, W. and {Huang}, Z. and {Huffenberger}, K.~M. and {Hurier}, G. and {Jaffe}, A.~H. and {Jaffe}, T.~R. and {Jones}, W.~C. and {Juvela}, M. and {Keih{\"a}nen}, E. and {Keskitalo}, R. and {Kisner}, T.~S. and {Kneissl}, R. and {Knoche}, J. and {Knox}, L. and {Kunz}, M. and {Kurki-Suonio}, H. and {Lagache}, G. and {L{\"a}hteenm{\"a}ki}, A. and {Lamarre}, J. -M. and {Lasenby}, A. and {Lattanzi}, M. and {Lawrence}, C.~R. and {Leahy}, J.~P. and {Leonardi}, R. and {Lesgourgues}, J. and {Levrier}, F. and {Lewis}, A. and {Liguori}, M. and {Lilje}, P.~B. and {Linden-V{\o}rnle}, M. and {L{\'o}pez-Caniego}, M. and {Lubin}, P.~M. and {Mac{\'\i}as-P{\'e}rez}, J.~F. and {Maggio}, G. and {Maino}, D. and {Mandolesi}, N. and {Mangilli}, A. and {Marchini}, A. and {Maris}, M. and {Martin}, P.~G. and {Martinelli}, M. and {Mart{\'\i}nez-Gonz{\'a}lez}, E. and {Masi}, S. and {Matarrese}, S. and {McGehee}, P. and {Meinhold}, P.~R. and {Melchiorri}, A. and {Melin}, J. -B. and {Mendes}, L. and {Mennella}, A. and {Migliaccio}, M. and {Millea}, M. and {Mitra}, S. and {Miville-Desch{\^e}nes}, M. -A. and {Moneti}, A. and {Montier}, L. and {Morgante}, G. and {Mortlock}, D. and {Moss}, A. and {Munshi}, D. and {Murphy}, J.~A. and {Naselsky}, P. and {Nati}, F. and {Natoli}, P. and {Netterfield}, C.~B. and {N{\o}rgaard-Nielsen}, H.~U. and {Noviello}, F. and {Novikov}, D. and {Novikov}, I. and {Oxborrow}, C.~A. and {Paci}, F. and {Pagano}, L. and {Pajot}, F. and {Paladini}, R. and {Paoletti}, D. and {Partridge}, B. and {Pasian}, F. and {Patanchon}, G. and {Pearson}, T.~J. and {Perdereau}, O. and {Perotto}, L. and {Perrotta}, F. and {Pettorino}, V. and {Piacentini}, F. and {Piat}, M. and {Pierpaoli}, E. and {Pietrobon}, D. and {Plaszczynski}, S. and {Pointecouteau}, E. and {Polenta}, G. and {Popa}, L. and {Pratt}, G.~W. and {Pr{\'e}zeau}, G.},
  title         = {{Planck 2015 results. XIII. Cosmological parameters}},
  journal       = {A\&A},
  year          = {2016},
  volume        = {594},
  pages         = {A13},
  eid           = {A13},
  doi           = {10.1051/0004-6361/201525830},
  archiveprefix = {arXiv},
  eprint        = {1502.01589},
  primaryclass  = {astro-ph.CO}
}

@ARTICLE{aghanim_planck_VI_2020,
  author        = {{Planck Collaboration} and {Aghanim}, N. and {Akrami}, Y. and {Ashdown}, M. and {Aumont}, J. and {Baccigalupi}, C. and {Ballardini}, M. and {Banday}, A.~J. and {Barreiro}, R.~B. and {Bartolo}, N. and {Basak}, S. and {Battye}, R. and {Benabed}, K. and {Bernard}, J. -P. and {Bersanelli}, M. and {Bielewicz}, P. and {Bock}, J.~J. and {Bond}, J.~R. and {Borrill}, J. and {Bouchet}, F.~R. and {Boulanger}, F. and {Bucher}, M. and {Burigana}, C. and {Butler}, R.~C. and {Calabrese}, E. and {Cardoso}, J. -F. and {Carron}, J. and {Challinor}, A. and {Chiang}, H.~C. and {Chluba}, J. and {Colombo}, L.~P.~L. and {Combet}, C. and {Contreras}, D. and {Crill}, B.~P. and {Cuttaia}, F. and {de Bernardis}, P. and {de Zotti}, G. and {Delabrouille}, J. and {Delouis}, J. -M. and {Di Valentino}, E. and {Diego}, J.~M. and {Dor{\'e}}, O. and {Douspis}, M. and {Ducout}, A. and {Dupac}, X. and {Dusini}, S. and {Efstathiou}, G. and {Elsner}, F. and {En{\ss}lin}, T.~A. and {Eriksen}, H.~K. and {Fantaye}, Y. and {Farhang}, M. and {Fergusson}, J. and {Fernandez-Cobos}, R. and {Finelli}, F. and {Forastieri}, F. and {Frailis}, M. and {Fraisse}, A.~A. and {Franceschi}, E. and {Frolov}, A. and {Galeotta}, S. and {Galli}, S. and {Ganga}, K. and {G{\'e}nova-Santos}, R.~T. and {Gerbino}, M. and {Ghosh}, T. and {Gonz{\'a}lez-Nuevo}, J. and {G{\'o}rski}, K.~M. and {Gratton}, S. and {Gruppuso}, A. and {Gudmundsson}, J.~E. and {Hamann}, J. and {Handley}, W. and {Hansen}, F.~K. and {Herranz}, D. and {Hildebrandt}, S.~R. and {Hivon}, E. and {Huang}, Z. and {Jaffe}, A.~H. and {Jones}, W.~C. and {Karakci}, A. and {Keih{\"a}nen}, E. and {Keskitalo}, R. and {Kiiveri}, K. and {Kim}, J. and {Kisner}, T.~S. and {Knox}, L. and {Krachmalnicoff}, N. and {Kunz}, M. and {Kurki-Suonio}, H. and {Lagache}, G. and {Lamarre}, J. -M. and {Lasenby}, A. and {Lattanzi}, M. and {Lawrence}, C.~R. and {Le Jeune}, M. and {Lemos}, P. and {Lesgourgues}, J. and {Levrier}, F. and {Lewis}, A. and {Liguori}, M. and {Lilje}, P.~B. and {Lilley}, M. and {Lindholm}, V. and {L{\'o}pez-Caniego}, M. and {Lubin}, P.~M. and {Ma}, Y. -Z. and {Mac{\'\i}as-P{\'e}rez}, J.~F. and {Maggio}, G. and {Maino}, D. and {Mandolesi}, N. and {Mangilli}, A. and {Marcos-Caballero}, A. and {Maris}, M. and {Martin}, P.~G. and {Martinelli}, M. and {Mart{\'\i}nez-Gonz{\'a}lez}, E. and {Matarrese}, S. and {Mauri}, N. and {McEwen}, J.~D. and {Meinhold}, P.~R. and {Melchiorri}, A. and {Mennella}, A. and {Migliaccio}, M. and {Millea}, M. and {Mitra}, S. and {Miville-Desch{\^e}nes}, M. -A. and {Molinari}, D. and {Montier}, L. and {Morgante}, G. and {Moss}, A. and {Natoli}, P. and {N{\o}rgaard-Nielsen}, H.~U. and {Pagano}, L. and {Paoletti}, D. and {Partridge}, B. and {Patanchon}, G. and {Peiris}, H.~V. and {Perrotta}, F. and {Pettorino}, V. and {Piacentini}, F. and {Polastri}, L. and {Polenta}, G. and {Puget}, J. -L. and {Rachen}, J.~P. and {Reinecke}, M. and {Remazeilles}, M. and {Renzi}, A. and {Rocha}, G. and {Rosset}, C. and {Roudier}, G. and {Rubi{\~n}o-Mart{\'\i}n}, J.~A. and {Ruiz-Granados}, B. and {Salvati}, L. and {Sandri}, M. and {Savelainen}, M. and {Scott}, D. and {Shellard}, E.~P.~S. and {Sirignano}, C. and {Sirri}, G. and {Spencer}, L.~D. and {Sunyaev}, R. and {Suur-Uski}, A. -S. and {Tauber}, J.~A. and {Tavagnacco}, D. and {Tenti}, M. and {Toffolatti}, L. and {Tomasi}, M. and {Trombetti}, T. and {Valenziano}, L. and {Valiviita}, J. and {Van Tent}, B. and {Vibert}, L. and {Vielva}, P. and {Villa}, F. and {Vittorio}, N. and {Wandelt}, B.~D. and {Wehus}, I.~K. and {White}, M. and {White}, S.~D.~M. and {Zacchei}, A. and {Zonca}, A.},
  title         = {{Planck 2018 results. VI. Cosmological parameters}},
  journal       = {A\&A},
  year          = {2020},
  volume        = {641},
  pages         = {A6},
  eid           = {A6},
  doi           = {10.1051/0004-6361/201833910},
  archiveprefix = {arXiv},
  eprint        = {1807.06209},
  primaryclass  = {astro-ph.CO}
}

@ARTICLE{artale_illustris_classification_2022,
  author        = {{Artale}, M. Celeste and {Haider}, Markus and {Montero-Dorta}, Antonio D. and {Vogelsberger}, Mark and {Martizzi}, Davide and {Torrey}, Paul and {Bird}, Simeon and {Hernquist}, Lars and {Marinacci}, Federico},
  title         = {{The large-scale distribution of ionized metals in IllustrisTNG}},
  journal       = {MNRAS},
  year          = {2022},
  volume        = {510},
  number        = {1},
  pages         = {399-412},
  doi           = {10.1093/mnras/stab3281},
  archiveprefix = {arXiv},
  eprint        = {2102.01092},
  primaryclass  = {astro-ph.GA}
}

@ARTICLE{ayromlou_feedback_2023,
  author        = {{Ayromlou}, Mohammadreza and {Nelson}, Dylan and {Pillepich}, Annalisa},
  title         = {{Feedback reshapes the baryon distribution within haloes, in halo outskirts, and beyond: the closure radius from dwarfs to massive clusters}},
  journal       = {MNRAS},
  year          = {2023},
  volume        = {524},
  number        = {4},
  pages         = {5391-5410},
  doi           = {10.1093/mnras/stad2046},
  archiveprefix = {arXiv},
  eprint        = {2211.07659},
  primaryclass  = {astro-ph.GA}
}

@ARTICLE{bartelmann_weak_2001,
  author        = {{Bartelmann}, M. and {Schneider}, P.},
  title         = {{Weak gravitational lensing}},
  journal       = {Phys. Rep.},
  year          = {2001},
  volume        = {340},
  number        = {4-5},
  pages         = {291-472},
  doi           = {10.1016/S0370-1573(00)00082-X},
  archiveprefix = {arXiv},
  eprint        = {astro-ph/9912508},
  primaryclass  = {astro-ph}
}

@ARTICLE{batten_cosmic_2021,
  author        = {{Batten}, Adam J. and {Duffy}, Alan R. and {Wijers}, Nastasha A. and {Gupta}, Vivek and {Flynn}, Chris and {Schaye}, Joop and {Ryan-Weber}, Emma},
  title         = {{The cosmic dispersion measure in the EAGLE simulations}},
  journal       = {MNRAS},
  year          = {2021},
  volume        = {505},
  number        = {4},
  pages         = {5356-5369},
  doi           = {10.1093/mnras/stab1528},
  archiveprefix = {arXiv},
  eprint        = {2011.14547},
  primaryclass  = {astro-ph.CO}
}

@ARTICLE{becker_detection_2011,
  author        = {{Becker}, George D. and {Bolton}, James S. and {Haehnelt}, Martin G. and {Sargent}, Wallace L.~W.},
  title         = {{Detection of extended He II reionization in the temperature evolution of the intergalactic medium}},
  journal       = {MNRAS},
  year          = {2011},
  volume        = {410},
  number        = {2},
  pages         = {1096-1112},
  doi           = {10.1111/j.1365-2966.2010.17507.x},
  archiveprefix = {arXiv},
  eprint        = {1008.2622},
  primaryclass  = {astro-ph.CO}
}

@ARTICLE{behroozi_average_2013,
  author        = {{Behroozi}, Peter S. and {Wechsler}, Risa H. and {Conroy}, Charlie},
  title         = {{The Average Star Formation Histories of Galaxies in Dark Matter Halos from z = 0-8}},
  journal       = {ApJ},
  year          = {2013},
  volume        = {770},
  number        = {1},
  pages         = {57},
  eid           = {57},
  doi           = {10.1088/0004-637X/770/1/57},
  archiveprefix = {arXiv},
  eprint        = {1207.6105},
  primaryclass  = {astro-ph.CO}
}

@ARTICLE{behroozi_correlation_2019,
  author        = {{Behroozi}, Peter and {Wechsler}, Risa H. and {Hearin}, Andrew P. and {Conroy}, Charlie},
  title         = {{UNIVERSEMACHINE: The correlation between galaxy growth and dark matter halo assembly from z = 0-10}},
  journal       = {MNRAS},
  year          = {2019},
  volume        = {488},
  number        = {3},
  pages         = {3143-3194},
  doi           = {10.1093/mnras/stz1182},
  archiveprefix = {arXiv},
  eprint        = {1806.07893},
  primaryclass  = {astro-ph.GA}
}

@ARTICLE{behroozi_lack_2013,
  author        = {{Behroozi}, Peter S. and {Wechsler}, Risa H. and {Conroy}, Charlie},
  title         = {{On the Lack of Evolution in Galaxy Star Formation Efficiency}},
  journal       = {ApJL},
  year          = {2013},
  volume        = {762},
  number        = {2},
  pages         = {L31},
  eid           = {L31},
  doi           = {10.1088/2041-8205/762/2/L31},
  archiveprefix = {arXiv},
  eprint        = {1209.3013},
  primaryclass  = {astro-ph.CO}
}

@INBOOK{belsley_1980_regression,
  author        = {David A. Belsley and Edwin Kuh and Roy E. Welsch},
  title         = {Regression diagnostics: Identifying influential data and sources of collinearity},
  year          = {1980},
  pages         = {100--104},
  publisher     = {Wiley}
}

@ARTICLE{bernardi_galaxy_2010,
  author        = {{Bernardi}, M. and {Shankar}, F. and {Hyde}, J.~B. and {Mei}, S. and {Marulli}, F. and {Sheth}, R.~K.},
  title         = {{Galaxy luminosities, stellar masses, sizes, velocity dispersions as a function of morphological type}},
  journal       = {MNRAS},
  year          = {2010},
  volume        = {404},
  number        = {4},
  pages         = {2087-2122},
  doi           = {10.1111/j.1365-2966.2010.16425.x},
  archiveprefix = {arXiv},
  eprint        = {0910.1093},
  primaryclass  = {astro-ph.CO}
}

@ARTICLE{bharadway_nearby_frb_2021,
  author        = {{Bhardwaj}, M. and {Gaensler}, B.~M. and {Kaspi}, V.~M. and {Landecker}, T.~L. and {Mckinven}, R. and {Michilli}, D. and {Pleunis}, Z. and {Tendulkar}, S.~P. and {Andersen}, B.~C. and {Boyle}, P.~J. and {Cassanelli}, T. and {Chawla}, P. and {Cook}, A. and {Dobbs}, M. and {Fonseca}, E. and {Kaczmarek}, J. and {Leung}, C. and {Masui}, K. and {Mnchmeyer}, M. and {Ng}, C. and {Rafiei-Ravandi}, M. and {Scholz}, P. and {Shin}, K. and {Smith}, K.~M. and {Stairs}, I.~H. and {Zwaniga}, A.~V.},
  title         = {{A Nearby Repeating Fast Radio Burst in the Direction of M81}},
  journal       = {ApJL},
  year          = {2021},
  volume        = {910},
  number        = {2},
  pages         = {L18},
  eid           = {L18},
  doi           = {10.3847/2041-8213/abeaa6},
  archiveprefix = {arXiv},
  eprint        = {2103.01295},
  primaryclass  = {astro-ph.HE}
}

@ARTICLE{bisigello_deep_2023,
  author        = {{Euclid Collaboration} and {Bisigello}, L. and {Conselice}, C.~J. and {Baes}, M. and {Bolzonella}, M. and {Brescia}, M. and {Cavuoti}, S. and {Cucciati}, O. and {Humphrey}, A. and {Hunt}, L.~K. and {Maraston}, C. and {Pozzetti}, L. and {Tortora}, C. and {van Mierlo}, S.~E. and {Aghanim}, N. and {Auricchio}, N. and {Baldi}, M. and {Bender}, R. and {Bodendorf}, C. and {Bonino}, D. and {Branchini}, E. and {Brinchmann}, J. and {Camera}, S. and {Capobianco}, V. and {Carbone}, C. and {Carretero}, J. and {Castander}, F.~J. and {Castellano}, M. and {Cimatti}, A. and {Congedo}, G. and {Conversi}, L. and {Copin}, Y. and {Corcione}, L. and {Courbin}, F. and {Cropper}, M. and {Da Silva}, A. and {Degaudenzi}, H. and {Douspis}, M. and {Dubath}, F. and {Duncan}, C.~A.~J. and {Dupac}, X. and {Dusini}, S. and {Farrens}, S. and {Ferriol}, S. and {Frailis}, M. and {Franceschi}, E. and {Franzetti}, P. and {Fumana}, M. and {Garilli}, B. and {Gillard}, W. and {Gillis}, B. and {Giocoli}, C. and {Grazian}, A. and {Grupp}, F. and {Guzzo}, L. and {Haugan}, S.~V.~H. and {Holmes}, W. and {Hormuth}, F. and {Hornstrup}, A. and {Jahnke}, K. and {K{\"u}mmel}, M. and {Kermiche}, S. and {Kiessling}, A. and {Kilbinger}, M. and {Kohley}, R. and {Kunz}, M. and {Kurki-Suonio}, H. and {Ligori}, S. and {Lilje}, P.~B. and {Lloro}, I. and {Maiorano}, E. and {Mansutti}, O. and {Marggraf}, O. and {Markovic}, K. and {Marulli}, F. and {Massey}, R. and {Maurogordato}, S. and {Medinaceli}, E. and {Meneghetti}, M. and {Merlin}, E. and {Meylan}, G. and {Moresco}, M. and {Moscardini}, L. and {Munari}, E. and {Niemi}, S.~M. and {Padilla}, C. and {Paltani}, S. and {Pasian}, F. and {Pedersen}, K. and {Pettorino}, V. and {Polenta}, G. and {Poncet}, M. and {Popa}, L. and {Raison}, F. and {Renzi}, A. and {Rhodes}, J. and {Riccio}, G. and {Rix}, H.-W. and {Romelli}, E. and {Roncarelli}, M. and {Rosset}, C. and {Rossetti}, E. and {Saglia}, R. and {Sapone}, D. and {Sartoris}, B. and {Schneider}, P. and {Scodeggio}, M. and {Secroun}, A. and {Seidel}, G. and {Sirignano}, C. and {Sirri}, G. and {Stanco}, L. and {Tallada-Cresp{\'\i}}, P. and {Tavagnacco}, D. and {Taylor}, A.~N. and {Tereno}, I. and {Toledo-Moreo}, R. and {Torradeflot}, F. and {Tutusaus}, I. and {Valentijn}, E.~A. and {Valenziano}, L. and {Vassallo}, T. and {Wang}, Y. and {Zacchei}, A. and {Zamorani}, G. and {Zoubian}, J. and {Andreon}, S. and {Bardelli}, S. and {Boucaud}, A. and {Colodro-Conde}, C. and {Di Ferdinando}, D. and {Graci{\'a}-Carpio}, J. and {Lindholm}, V. and {Maino}, D. and {Mei}, S. and {Scottez}, V. and {Sureau}, F. and {Tenti}, M. and {Zucca}, E. and {Borlaff}, A.~S. and {Ballardini}, M. and {Biviano}, A. and {Bozzo}, E. and {Burigana}, C. and {Cabanac}, R. and {Cappi}, A. and {Carvalho}, C.~S. and {Casas}, S. and {Castignani}, G. and {Cooray}, A. and {Coupon}, J. and {Courtois}, H.~M. and {Cuby}, J. and {Davini}, S. and {De Lucia}, G. and {Desprez}, G. and {Dole}, H. and {Escartin}, J.~A. and {Escoffier}, S. and {Farina}, M. and {Fotopoulou}, S. and {Ganga}, K. and {Garcia-Bellido}, J. and {George}, K. and {Giacomini}, F. and {Gozaliasl}, G. and {Hildebrandt}, H. and {Hook}, I. and {Huertas-Company}, M. and {Kansal}, V. and {Keihanen}, E. and {Kirkpatrick}, C.~C. and {Loureiro}, A. and {Mac{\'\i}as-P{\'e}rez}, J.~F. and {Magliocchetti}, M. and {Mainetti}, G. and {Marcin}, S. and {Martinelli}, M. and {Martinet}, N. and {Metcalf}, R.~B. and {Monaco}, P. and {Morgante}, G. and {Nadathur}, S. and {Nucita}, A.~A. and {Patrizii}, L. and {Peel}, A. and {Potter}, D. and {Pourtsidou}, A. and {P{\"o}ntinen}, M. and {Reimberg}, P. and {S{\'a}nchez}, A.~G. and {Sakr}, Z. and {Schirmer}, M. and {Sefusatti}, E. and {Sereno}, M. and {Stadel}, J. and {Teyssier}, R. and {Valieri}, C. and {Valiviita}, J. and {Viel}, M.},
  title         = {{Euclid preparation - XXIII. Derivation of galaxy physical properties with deep machine learning using mock fluxes and H-band images}},
  journal       = {MNRAS},
  year          = {2023},
  volume        = {520},
  number        = {3},
  pages         = {3529-3548},
  doi           = {10.1093/mnras/stac3810},
  archiveprefix = {arXiv},
  eprint        = {2206.14944},
  primaryclass  = {astro-ph.GA}
}

@ARTICLE{bissantz_convergence_2007,
  author        = {Bissantz, Nicolai and Hohage, Thorsten and Munk, Axel and Ruymgaart, Frits},
  title         = {Convergence rates of general regularization methods for statistical inverse problems and applications},
  journal       = {SIAM J. Numer. Anal.},
  year          = {2007},
  volume        = {45},
  number        = {6},
  pages         = {2610--2636},
  publisher     = {SIAM},
  doi           = {10.1137/060651884}
}

@ARTICLE{blanton_K_correction_2007,
  author        = {{Blanton}, Michael R. and {Roweis}, Sam},
  title         = {{K-Corrections and Filter Transformations in the Ultraviolet, Optical, and Near-Infrared}},
  journal       = {AJ},
  year          = {2007},
  volume        = {133},
  number        = {2},
  pages         = {734-754},
  doi           = {10.1086/510127},
  archiveprefix = {arXiv},
  eprint        = {astro-ph/0606170},
  primaryclass  = {astro-ph}
}

@ARTICLE{bonjean_gas_2018,
  author        = {{Bonjean}, V. and {Aghanim}, N. and {Salom{'e}}, P. and {Douspis}, M. and {Beelen}, A.},
  title         = {{Gas and galaxies in filaments between clusters of galaxies. The study of A399-A401}},
  journal       = {A\&A},
  year          = {2018},
  volume        = {609},
  pages         = {A49},
  eid           = {A49},
  doi           = {10.1051/0004-6361/201731699},
  archiveprefix = {arXiv},
  eprint        = {1710.08699},
  primaryclass  = {astro-ph.CO}
}

@ARTICLE{booth_interaction_2013,
  author        = {{Booth}, C.~M. and {Schaye}, Joop},
  title         = {{The interaction between feedback from active galactic nuclei and supernovae}},
  journal       = {Sci. Rep.},
  year          = {2013},
  volume        = {3},
  pages         = {1738},
  eid           = {1738},
  doi           = {10.1038/srep01738},
  archiveprefix = {arXiv},
  eprint        = {1203.3802},
  primaryclass  = {astro-ph.CO}
}

@ARTICLE{boquien_CIGALE_2019,
  author        = {{Boquien}, M. and {Burgarella}, D. and {Roehlly}, Y. and {Buat}, V. and {Ciesla}, L. and {Corre}, D. and {Inoue}, A.~K. and {Salas}, H.},
  title         = {{CIGALE: a python Code Investigating GALaxy Emission}},
  journal       = {A\&A},
  year          = {2019},
  volume        = {622},
  pages         = {A103},
  eid           = {A103},
  doi           = {10.1051/0004-6361/201834156},
  archiveprefix = {arXiv},
  eprint        = {1811.03094},
  primaryclass  = {astro-ph.GA}
}

@ARTICLE{caleb_Meerkat_2023,
  author        = {{Caleb}, M. and {Driessen}, L.~N. and {Gordon}, A.~C. and {Tejos}, N. and {Bernales}, L. and {Qiu}, H. and {Chibueze}, J.~O. and {Stappers}, B.~W. and {Rajwade}, K.~M. and {Cavallaro}, F. and {Wang}, Y. and {Kumar}, P. and {Majid}, W.~A. and {Wharton}, R.~S. and {Naudet}, C.~J. and {Bezuidenhout}, M.~C. and {Jankowski}, F. and {Malenta}, M. and {Morello}, V. and {Sanidas}, S. and {Surnis}, M.~P. and {Barr}, E.~D. and {Chen}, W. and {Kramer}, M. and {Fong}, W. and {Kilpatrick}, C.~D. and {Prochaska}, J. Xavier and {Simha}, S. and {Venter}, C. and {Heywood}, I. and {Kundu}, A. and {Schussler}, F.},
  title         = {{A subarcsec localized fast radio burst with a significant host galaxy dispersion measure contribution}},
  journal       = {MNRAS},
  year          = {2023},
  volume        = {524},
  number        = {2},
  pages         = {2064-2077},
  doi           = {10.1093/mnras/stad1839},
  archiveprefix = {arXiv},
  eprint        = {2302.09754},
  primaryclass  = {astro-ph.HE}
}

@ARTICLE{caleb_z_2_FRB_2025,
  author        = {{Caleb}, Manisha and {Nanayakkara}, Themiya and {Stappers}, Benjamin and {Pastor-Marazuela}, In{\'e}s and {Khrykin}, Ilya S. and {Glazebrook}, Karl and {Tejos}, Nicolas and {Prochaska}, J. Xavier and {Rajwade}, Kaustubh and {Mas-Ribas}, Lluis and {Driessen}, Laura N. and {Fong}, Wen-fai and {Gordon}, Alexa C. and {Hoffmann}, Jordan and {James}, Clancy W. and {Jankowski}, Fabian and {Kahinga}, Lordrick and {Kramer}, Michael and {Simha}, Sunil and {Barr}, Ewan D. and {Christiaan Bezuidenhout}, Mechiel and {Deng}, Xihan and {Lin}, Zeren and {Marnoch}, Lachlan and {Martin}, Christopher D. and {Nugent}, Anya and {Shaji}, Kavya and {Tian}, Jun},
  title         = {{A fast radio burst at redshift 2, three billion years after the Big Bang}},
  journal       = {Science},
  year          = {2026},
  pages         = {eadz2675},
  note          = {in press},
  doi           = {10.1126/science.adz2675},
  archiveprefix = {arXiv},
  eprint        = {2508.01648},
  primaryclass  = {astro-ph.HE},
  month         = {oct}
}

@ARTICLE{camps_SKIRT_2020,
  author        = {{Camps}, P. and {Baes}, M.},
  title         = {{SKIRT 9: Redesigning an advanced dust radiative transfer code to allow kinematics, line transfer and polarization by aligned dust grains}},
  journal       = {Astron. Comput.},
  year          = {2020},
  volume        = {31},
  pages         = {100381},
  eid           = {100381},
  doi           = {10.1016/j.ascom.2020.100381},
  archiveprefix = {arXiv},
  eprint        = {2003.00721},
  primaryclass  = {astro-ph.GA}
}

@ARTICLE{carnall_Bagpipes_2018,
  author        = {{Carnall}, A.~C. and {McLure}, R.~J. and {Dunlop}, J.~S. and {Dav{\'e}}, R.},
  title         = {{Inferring the star formation histories of massive quiescent galaxies with BAGPIPES: evidence for multiple quenching mechanisms}},
  journal       = {MNRAS},
  year          = {2018},
  volume        = {480},
  number        = {4},
  pages         = {4379-4401},
  doi           = {10.1093/mnras/sty2169},
  archiveprefix = {arXiv},
  eprint        = {1712.04452},
  primaryclass  = {astro-ph.GA}
}

@ARTICLE{chen_formation_2024,
  author        = {{Chen}, J.~H. and {Jia}, X.~D. and {Dong}, X.~F. and {Wang}, F.~Y.},
  title         = {{The Formation Rate and Luminosity Function of Fast Radio Bursts}},
  journal       = {ApJL},
  year          = {2024},
  volume        = {973},
  number        = {2},
  pages         = {L54},
  eid           = {L54},
  doi           = {10.3847/2041-8213/ad7b39},
  archiveprefix = {arXiv},
  eprint        = {2406.03672},
  primaryclass  = {astro-ph.HE}
}

@ARTICLE{chisari_impact_2018,
  author        = {Chisari, Nora Elisa and Richardson, Mark L. A. and Devriendt, Julien and Dubois, Yohan and Schneider, Aurel and Brun, Amandine M. C. Le and Beckmann, Ricarda S. and Peirani, Sebastien and Slyz, Adrianne and Pichon, Christophe},
  title         = {The impact of baryons on the matter power spectrum from the {Horizon}-{AGN} cosmological hydrodynamical simulation},
  journal       = {MNRAS},
  year          = {2018},
  volume        = {480},
  number        = {3},
  pages         = {3962--3977},
  doi           = {10.1093/mnras/sty2093}
}

@ARTICLE{chisari_modelling_2019,
  author        = {{Chisari}, Nora Elisa and {Mead}, Alexander J. and {Joudaki}, Shahab and {Ferreira}, Pedro G. and {Schneider}, Aurel and {Mohr}, Joseph and {Tr{\"o}ster}, Tilman and {Alonso}, David and {McCarthy}, Ian G. and {Martin-Alvarez}, Sergio and {Devriendt}, Julien and {Slyz}, Adrianne and {van Daalen}, Marcel P.},
  title         = {{Modelling baryonic feedback for survey cosmology}},
  journal       = {OJAp},
  year          = {2019},
  volume        = {2},
  number        = {1},
  pages         = {4},
  eid           = {4},
  doi           = {10.21105/astro.1905.06082},
  archiveprefix = {arXiv},
  eprint        = {1905.06082},
  primaryclass  = {astro-ph.CO}
}

@ARTICLE{chuang_clustering_2016,
  author        = {{Chuang}, Chia-Hsun and {Prada}, Francisco and {Pellejero-Ibanez}, Marcos and {Beutler}, Florian and {Cuesta}, Antonio J. and {Eisenstein}, Daniel J. and {Escoffier}, Stephanie and {Ho}, Shirley and {Kitaura}, Francisco-Shu and {Kneib}, Jean-Paul and {Manera}, Marc and {Nuza}, Sebasti{\'a}n E. and {Rodr{\'\i}guez-Torres}, Sergio and {Ross}, Ashley and {Rubi{\~n}o-Mart{\'\i}n}, J.~A. and {Samushia}, Lado and {Schlegel}, David J. and {Schneider}, Donald P. and {Wang}, Yuting and {Weaver}, Benjamin A. and {Zhao}, Gongbo and {Brownstein}, Joel R. and {Dawson}, Kyle S. and {Maraston}, Claudia and {Olmstead}, Matthew D. and {Thomas}, Daniel},
  title         = {{The clustering of galaxies in the SDSS-III Baryon Oscillation Spectroscopic Survey: single-probe measurements from CMASS anisotropic galaxy clustering}},
  journal       = {MNRAS},
  year          = {2016},
  volume        = {461},
  number        = {4},
  pages         = {3781-3793},
  doi           = {10.1093/mnras/stw1535},
  archiveprefix = {arXiv},
  eprint        = {1312.4889},
  primaryclass  = {astro-ph.CO}
}

@ARTICLE{cline_condition_number_1979,
  author        = {{Cline}, A.~K. and {Moler}, C.~B. and {Stewart}, G.~W. and {Wilkinson}, J.~H.},
  title         = {{An Estimate for the Condition Number of a Matrix}},
  journal       = {SIAM J. Numer. Anal.},
  year          = {1979},
  volume        = {16},
  number        = {2},
  pages         = {368-375},
  doi           = {10.1137/0716029}
}

@ARTICLE{connor_cosmic_2025,
  author        = {{Connor}, Liam and {Ravi}, Vikram and {Sharma}, Kritti and {Ocker}, Stella Koch and {Faber}, Jakob and {Hallinan}, Gregg and {Harnach}, Charlie and {Hellbourg}, Greg and {Hobbs}, Rick and {Hodge}, David and {Hodges}, Mark and {Kosogorov}, Nikita and {Lamb}, James and {Law}, Casey and {Rasmussen}, Paul and {Sherman}, Myles and {Somalwar}, Jean and {Weinreb}, Sander and {Woody}, David and {Konietzka}, Ralf M.},
  title         = {{A gas-rich cosmic web revealed by the partitioning of the missing baryons}},
  journal       = {Nat. Astron.},
  year          = {2025},
  volume        = {9},
  number        = {8},
  pages         = {1226--1239},
  doi           = {10.1038/s41550-025-02566-y},
  archiveprefix = {arXiv},
  eprint        = {2409.16952},
  primaryclass  = {astro-ph.CO}
}

@ARTICLE{contarini_era_2026,
  author        = {Contarini, Sofia and Verza, Giovanni and Pisani, Alice},
  title         = {The era of precision cosmology with voids},
  journal       = {A\&ARv},
  year          = {2026},
  volume        = {34},
  number        = {1},
  pages         = {1},
  doi           = {10.1007/s00159-026-00166-x},
  archiveprefix = {arXiv},
  eprint        = {2601.14362}
}

@ARTICLE{cooke_determination_2018,
  author        = {{Cooke}, Ryan J. and {Pettini}, Max and {Steidel}, Charles C.},
  title         = {{One Percent Determination of the Primordial Deuterium Abundance}},
  journal       = {ApJ},
  year          = {2018},
  volume        = {855},
  number        = {2},
  pages         = {102},
  eid           = {102},
  doi           = {10.3847/1538-4357/aaab53},
  archiveprefix = {arXiv},
  eprint        = {1710.11129},
  primaryclass  = {astro-ph.CO}
}

@ARTICLE{cordes_NE2001_2001,
  author        = {{Cordes}, J.~M. and {Lazio}, T.~J.~W.},
  title         = {{NE2001.I. A New Model for the Galactic Distribution of Free Electrons and its Fluctuations}},
  journal       = {arXiv e-prints},
  year          = {2002},
  pages         = {astro-ph/0207156},
  eid           = {astro-ph/0207156},
  doi           = {10.48550/arXiv.astro-ph/0207156},
  archiveprefix = {arXiv},
  eprint        = {astro-ph/0207156},
  primaryclass  = {astro-ph}
}

@ARTICLE{coupon_galaxy-halo_2015,
  author        = {{Coupon}, J. and {Arnouts}, S. and {van Waerbeke}, L. and {Moutard}, T. and {Ilbert}, O. and {van Uitert}, E. and {Erben}, T. and {Garilli}, B. and {Guzzo}, L. and {Heymans}, C. and {Hildebrandt}, H. and {Hoekstra}, H. and {Kilbinger}, M. and {Kitching}, T. and {Mellier}, Y. and {Miller}, L. and {Scodeggio}, M. and {Bonnett}, C. and {Branchini}, E. and {Davidzon}, I. and {De Lucia}, G. and {Fritz}, A. and {Fu}, L. and {Hudelot}, P. and {Hudson}, M.~J. and {Kuijken}, K. and {Leauthaud}, A. and {Le F{\`e}vre}, O. and {McCracken}, H.~J. and {Moscardini}, L. and {Rowe}, B.~T.~P. and {Schrabback}, T. and {Semboloni}, E. and {Velander}, M.},
  title         = {{The galaxy-halo connection from a joint lensing, clustering and abundance analysis in the CFHTLenS/VIPERS field}},
  journal       = {MNRAS},
  year          = {2015},
  volume        = {449},
  number        = {2},
  pages         = {1352-1379},
  doi           = {10.1093/mnras/stv276},
  archiveprefix = {arXiv},
  eprint        = {1502.02867},
  primaryclass  = {astro-ph.CO}
}

@ARTICLE{cox_analysis_1993,
  author        = {Cox, Dennis D},
  title         = {An analysis of Bayesian inference for nonparametric regression},
  journal       = {Ann. Stat.},
  year          = {1993},
  volume        = {21},
  number        = {2},
  pages         = {903--923},
  doi           = {10.1214/aos/1176349157}
}

@INPROCEEDINGS{crill_SphereX_2020,
  author        = {{Crill}, Brendan P. and {Werner}, Michael and {Akeson}, Rachel and {Ashby}, Matthew and {Bleem}, Lindsey and {Bock}, James J. and {Bryan}, Sean and {Burnham}, Jill and {Byunh}, Joyce and {Chang}, Tzu-Ching and {Chiang}, Yi-Kuan and {Cook}, Walter and {Cooray}, Asantha and {Davis}, Andrew and {Dor{\'e}}, OIivier and {Dowell}, C. Darren and {Dubois-Felsmann}, Gregory and {Eifler}, Tim and {Faisst}, Andreas and {Habib}, Salman and {Heinrich}, Chen and {Heitmann}, Katrin and {Heaton}, Grigory and {Hirata}, Christopher and {Hristov}, Viktor and {Hui}, Howard and {Jeong}, Woong-Seob and {Kang}, Jae Hwan and {Kecman}, Branislav and {Kirkpatrick}, J. Davy and {Korngut}, Phillip M. and {Krause}, Elisabeth and {Lee}, Bomee and {Lisse}, Carey and {Masters}, Daniel and {Mauskopf}, Philip and {Melnick}, Gary and {Miyasaka}, Hiromasa and {Nayyeri}, Hooshang and {Nguyen}, Hien and {{\"O}berg}, Karin and {Padin}, Steve and {Paladini}, Roberta and {Pourrahmani}, Milad and {Pyo}, Jeonghyun and {Smith}, Roger and {Song}, Yong-Seong and {Symons}, Teresa and {Teplitz}, Harry and {Tolls}, Volker and {Unwin}, Steve and {Windhorst}, Rogier and {Yang}, Yujin and {Zemcov}, Michael},
  title         = {{SPHEREx: NASA's near-infrared spectrophotometric all-sky survey}},
  booktitle     = {Space Telescopes and Instrumentation 2020: Optical, Infrared, and Millimeter Wave},
  editor        = {{Lystrup}, Makenzie and {Perrin}, Marshall D.},
  series        = {Society of Photo-Optical Instrumentation Engineers (SPIE) Conference Series},
  year          = {2020},
  volume        = {11443},
  pages         = {114430I},
  eid           = {114430I},
  doi           = {10.1117/12.2567224},
  archiveprefix = {arXiv},
  eprint        = {2404.11017},
  primaryclass  = {astro-ph.IM}
}

@ARTICLE{cropper_Eucild_VIS_2025,
  author        = {{Euclid Collaboration} and {Cropper}, M.~S. and {Al-Bahlawan}, A. and {Amiaux}, J. and {Awan}, S. and {Azzollini}, R. and {Benson}, K. and {Berthe}, M. and {Boucher}, J. and {Bozzo}, E. and {Brockley-Blatt}, C. and {Candini}, G.~P. and {Cara}, C. and {Chaudery}, R.~A. and {Cole}, R.~E. and {Danto}, P. and {Denniston}, J. and {Di Giorgio}, A.~M. and {Dryer}, B. and {Dubois}, J.-P. and {Endicott}, J. and {Farina}, M. and {Galli}, E. and {Genolet}, L. and {Gow}, J.~P.~D. and {Guttridge}, P. and {Hailey}, M. and {Hall}, D. and {Harper}, C. and {Hoekstra}, H. and {Holland}, A.~D. and {Horeau}, B. and {Hu}, D. and {James}, R.~E. and {Khalil}, A. and {King}, R. and {Kitching}, T. and {Kohley}, R. and {Larcheveque}, C. and {Lawrenson}, A. and {Liebing}, P. and {Liu}, S.~J. and {Martignac}, J. and {Massey}, R. and {McCracken}, H.~J. and {Miller}, L. and {Murray}, N. and {Nakajima}, R. and {Niemi}, S.-M. and {Nightingale}, J.~W. and {Paltani}, S. and {Pendem}, A. and {Philippon}, A. and {Plana}, C. and {Pool}, P. and {Pottinger}, S. and {Racca}, G.~D. and {Rhodes}, J. and {Rousseau}, A. and {Ruane}, K. and {Salatti}, M. and {Salvignol}, J.-C. and {Sciortino}, A. and {Short}, A. and {Skottfelt}, J. and {Smit}, S.~J.~A. and {Swindells}, I. and {Szafraniec}, M. and {Thomas}, P.~D. and {Thomas}, W. and {Tommasi}, E. and {Tosti}, S. and {Visticot}, F. and {Walton}, D.~M. and {Willis}, G. and {Winter}, B. and {Aghanim}, N. and {Altieri}, B. and {Amara}, A. and {Andreon}, S. and {Auricchio}, N. and {Aussel}, H. and {Baccigalupi}, C. and {Baldi}, M. and {Balestra}, A. and {Bardelli}, S. and {Basset}, A. and {Bender}, R. and {Bernardeau}, F. and {Bodendorf}, C. and {Boenke}, T. and {Bonino}, D. and {Branchini}, E. and {Brescia}, M. and {Brinchmann}, J. and {Camera}, S. and {Capobianco}, V. and {Carbone}, C. and {Cardone}, V.~F. and {Carretero}, J. and {Casas}, R. and {Casas}, S. and {Castander}, F.~J. and {Castellano}, M. and {Castignani}, G. and {Cavuoti}, S. and {Cimatti}, A. and {Colodro-Conde}, C. and {Congedo}, G. and {Conselice}, C.~J. and {Conversi}, L. and {Copin}, Y. and {Courbin}, F. and {Courtois}, H.~M. and {Crocce}, M. and {Cuby}, J.-G. and {Cuillandre}, J.-C. and {Da Silva}, A. and {Degaudenzi}, H. and {De Lucia}, G. and {Dinis}, J. and {Dolding}, C. and {Douspis}, M. and {Duncan}, C.~A.~J. and {Dupac}, X. and {Dusini}, S. and {Ealet}, A. and {Fabricius}, M. and {Farrens}, S. and {Ferriol}, S. and {Fosalba}, P. and {Fotopoulou}, S. and {Frailis}, M. and {Franceschi}, E. and {Franzetti}, P. and {Frugier}, P.-A. and {Fumana}, M. and {Galeotta}, S. and {Garilli}, B. and {George}, K. and {Gillard}, W. and {Gillis}, B. and {Giocoli}, C. and {G{\'o}mez-Alvarez}, P. and {Granett}, B.~R. and {Grazian}, A. and {Grupp}, F. and {Guzzo}, L. and {Haugan}, S.~V.~H. and {Herent}, O. and {Hoar}, J. and {Holliman}, M.~S. and {Holmes}, W. and {Hook}, I. and {Hormuth}, F. and {Hornstrup}, A. and {Hudelot}, P. and {Ili{\'c}}, S. and {Jahnke}, K. and {Jhabvala}, M. and {Joachimi}, B. and {Keih{\"a}nen}, E. and {Kermiche}, S. and {Kiessling}, A. and {Kilbinger}, M. and {Kubik}, B. and {Kuijken}, K. and {K{\"u}mmel}, M. and {Kunz}, M. and {Kurki-Suonio}, H. and {Lahav}, O. and {Laureijs}, R. and {Ligori}, S. and {Lilje}, P.~B. and {Lindholm}, V. and {Lloro}, I. and {Lorenzo Alvarez}, J. and {Mainetti}, G. and {Maino}, D. and {Maiorano}, E. and {Mansutti}, O. and {Marcin}, S. and {Marggraf}, O. and {Markovic}, K. and {Martinelli}, M. and {Martinet}, N. and {Marulli}, F. and {Masters}, D.~C. and {Maurogordato}, S. and {Medinaceli}, E. and {Mei}, S. and {Melchior}, M. and {Mellier}, Y. and {Meneghetti}, M. and {Merlin}, E. and {Meylan}, G. and {Mohr}, J.~J. and {Moresco}, M. and {Moscardini}, L. and {Neissner}, C.},
  title         = {{Euclid: II. The VIS instrument}},
  journal       = {A\&A},
  year          = {2025},
  volume        = {697},
  pages         = {A2},
  eid           = {A2},
  doi           = {10.1051/0004-6361/202450996},
  archiveprefix = {arXiv},
  eprint        = {2405.13492},
  primaryclass  = {astro-ph.IM}
}

@ARTICLE{cui_ideal_2008,
  author        = {Cui, Weiguang and Liu, Lei and Yang, Xiaohu and Wang, Yu and Feng, Longlong and Springel, Volker},
  title         = {An ideal mass assignment scheme for measuring the {Power} {Spectrum} with {FFTs}},
  journal       = {ApJ},
  year          = {2008},
  volume        = {687},
  number        = {2},
  pages         = {738--744},
  doi           = {10.1086/592079}
}

@ARTICLE{cyburt_BBN_2016,
  author        = {{Cyburt}, Richard H. and {Fields}, Brian D. and {Olive}, Keith A. and {Yeh}, Tsung-Han},
  title         = {{Big bang nucleosynthesis: Present status}},
  journal       = {Rev. Mod. Phys.},
  year          = {2016},
  volume        = {88},
  number        = {1},
  pages         = {015004},
  eid           = {015004},
  doi           = {10.1103/RevModPhys.88.015004},
  archiveprefix = {arXiv},
  eprint        = {1505.01076},
  primaryclass  = {astro-ph.CO}
}

@ARTICLE{darvish_comparative_2015,
  author        = {{Darvish}, Behnam and {Mobasher}, Bahram and {Sobral}, David and {Scoville}, Nicholas and {Aragon-Calvo}, Miguel},
  title         = {{A Comparative Study of Density Field Estimation for Galaxies: New Insights into the Evolution of Galaxies with Environment in COSMOS out to z{\ensuremath{\sim}}3}},
  journal       = {ApJ},
  year          = {2015},
  volume        = {805},
  number        = {2},
  pages         = {121},
  eid           = {121},
  doi           = {10.1088/0004-637X/805/2/121},
  archiveprefix = {arXiv},
  eprint        = {1503.07879},
  primaryclass  = {astro-ph.GA}
}

@ARTICLE{deng_wei_2014_dm_IGM_derivation,
  author        = {{Deng}, Wei and {Zhang}, Bing},
  title         = {{Cosmological Implications of Fast Radio Burst/Gamma-Ray Burst Associations}},
  journal       = {ApJL},
  year          = {2014},
  volume        = {783},
  number        = {2},
  pages         = {L35},
  eid           = {L35},
  doi           = {10.1088/2041-8205/783/2/L35},
  archiveprefix = {arXiv},
  eprint        = {1401.0059},
  primaryclass  = {astro-ph.HE}
}

@ARTICLE{desjacques_large_2018,
  author        = {{Desjacques}, Vincent and {Jeong}, Donghui and {Schmidt}, Fabian},
  title         = {{Large-scale galaxy bias}},
  journal       = {Phys. Rep.},
  year          = {2018},
  volume        = {733},
  pages         = {1-193},
  doi           = {10.1016/j.physrep.2017.12.002},
  archiveprefix = {arXiv},
  eprint        = {1611.09787},
  primaryclass  = {astro-ph.CO}
}

@ARTICLE{dolag_constraints_2015,
  author        = {{Dolag}, K. and {Gaensler}, B.~M. and {Beck}, A.~M. and {Beck}, M.~C.},
  title         = {{Constraints on the distribution and energetics of fast radio bursts using cosmological hydrodynamic simulations}},
  journal       = {MNRAS},
  year          = {2015},
  volume        = {451},
  number        = {4},
  pages         = {4277-4289},
  doi           = {10.1093/mnras/stv1190},
  archiveprefix = {arXiv},
  eprint        = {1412.4829},
  primaryclass  = {astro-ph.CO}
}

@ARTICLE{donoho_statistical_1994,
  author        = {Donoho, David L},
  title         = {Statistical estimation and optimal recovery},
  journal       = {Ann. Stat.},
  year          = {1994},
  volume        = {22},
  number        = {1},
  pages         = {238--270},
  publisher     = {Institute of Mathematical Statistics},
  doi           = {10.1214/aos/1176325367}
}

@ARTICLE{douglass_VAST_2022,
  author        = {{Douglass}, Kelly and {Veyrat}, Dahlia and {O'Neill}, Stephen and {BenZvi}, Segev and {Zaidouni}, Fatima and {Guzzetti}, Michaela},
  title         = {{VAST: the Void Analysis Software Toolkit}},
  journal       = {J. Open Source Softw.},
  year          = {2022},
  volume        = {7},
  number        = {77},
  pages         = {4033},
  eid           = {4033},
  doi           = {10.21105/joss.04033}
}

@ARTICLE{dries_hierarchical_2016,
  author        = {{Dries}, M. and {Trager}, S.~C. and {Koopmans}, L.~V.~E.},
  title         = {{A hierarchical Bayesian approach for reconstructing the initial mass function of single stellar populations}},
  journal       = {MNRAS},
  year          = {2016},
  volume        = {463},
  number        = {1},
  pages         = {886-912},
  doi           = {10.1093/mnras/stw2049},
  archiveprefix = {arXiv},
  eprint        = {1608.05194},
  primaryclass  = {astro-ph.GA}
}

@ARTICLE{eckert_warm_2015,
  author        = {{Eckert}, Dominique and {Jauzac}, Mathilde and {Shan}, Huanyuan and {Kneib}, Jean-Paul and {Erben}, Thomas and {Israel}, Holger and {Jullo}, Eric and {Klein}, Matthias and {Massey}, Richard and {Richard}, Johan and {Tchernin}, C{'e}line},
  title         = {{Warm-hot baryons comprise 5-10 per cent of filaments in the cosmic web}},
  journal       = {Nature},
  year          = {2015},
  volume        = {528},
  number        = {7580},
  pages         = {105-107},
  doi           = {10.1038/nature16058},
  archiveprefix = {arXiv},
  eprint        = {1512.00454},
  primaryclass  = {astro-ph.CO}
}

@ARTICLE{edmond_measurement_2026,
  author        = {{Chaussidon}, Edmond and {Hotinli}, Selim C. and {Ferraro}, Simone and {Smith}, Kendrick and {Chen}, Xinyi and {Aguilar}, J. and {Ahlen}, S. and {Bianchi}, D. and {Brooks}, D. and {Claybaugh}, T. and {Cuceu}, A. and {de la Macorra}, A. and {Dey}, B. and {Doel}, P. and {Font-Ribera}, A. and {Forero-Romero}, J.~E. and {Gazta{\~n}aga}, E. and {Gontcho}, S. Gontcho A and {Gutierrez}, G. and {Guy}, J. and {Herrera-Alcantar}, H.~K. and {Honscheid}, K. and {Howlett}, C. and {Huterer}, D. and {Ishak}, M. and {Joyce}, R. and {Kirkby}, D. and {Kremin}, A. and {Lahav}, O. and {Landriau}, M. and {Le Guillou}, L. and {Manera}, M. and {Meisner}, A. and {Miquel}, R. and {Nadathur}, S. and {Newman}, J.~A. and {Palanque-Delabrouille}, N. and {Percival}, W.~J. and {Prada}, F. and {P{\'e}rez-R{\`a}fols}, I. and {Rossi}, G. and {Samushia}, L. and {Sanchez}, E. and {Schlegel}, D. and {Schubnell}, M. and {Seo}, H. and {Silber}, J. and {Sprayberry}, D. and {Tarl{\'e}}, G. and {Weaver}, B.~A. and {Y{\`e}che}, C. and {Zhou}, R.},
  title         = {{Measurement of the galaxy-velocity power spectrum of DESI tracers with the kinematic Sunyaev-Zeldovich effect using DESI DR2 and ACT DR6}},
  journal       = {Phys. Rev. D},
  year          = {2026},
  note          = {in press},
  doi           = {10.1103/3pn7-k7yl},
  archiveprefix = {arXiv},
  eprint        = {2604.04867},
  primaryclass  = {astro-ph.CO}
}

@ARTICLE{eisert_ERGO_2024,
  author        = {{Eisert}, Lukas and {Bottrell}, Connor and {Pillepich}, Annalisa and {Shimakawa}, Rhythm and {Rodriguez-Gomez}, Vicente and {Nelson}, Dylan and {Angeloudi}, Eirini and {Huertas-Company}, Marc},
  title         = {{ERGO-ML: comparing IllustrisTNG and HSC galaxy images via contrastive learning}},
  journal       = {MNRAS},
  year          = {2024},
  volume        = {528},
  number        = {4},
  pages         = {7411-7439},
  doi           = {10.1093/mnras/stae481},
  archiveprefix = {arXiv},
  eprint        = {2310.19904},
  primaryclass  = {astro-ph.GA}
}

@ARTICLE{erik_new_Yhe_2026,
  author        = {{Aver}, Erik and {Skillman}, Evan D. and {Pogge}, Richard W. and {Rogers}, Noah S.~J. and {Weller}, Miqaela K. and {Olive}, Keith A. and {Berg}, Danielle A. and {Salzer}, John J. and {Miller}, Jr., John H. and {M{\'e}ndez-Delgado}, Jos{\'e} Eduardo},
  title         = {{The LBT Y$_\mathrm{p}$ Project IV: A New Value of the Primordial Helium Abundance}},
  journal       = {ApJ},
  year          = {2026},
  volume        = {1008},
  pages         = {240},
  doi           = {10.3847/1538-4357/ae879b},
  archiveprefix = {arXiv},
  eprint        = {2601.22238},
  primaryclass  = {astro-ph.CO}
}

@ARTICLE{feldbrugge_phase_2025,
  author        = {{Feldbrugge}, Job},
  title         = {{Phase-space Delaunay tessellation field estimator}},
  journal       = {MNRAS},
  year          = {2025},
  volume        = {536},
  number        = {1},
  pages         = {807-815},
  doi           = {10.1093/mnras/stae2627},
  archiveprefix = {arXiv},
  eprint        = {2402.16234},
  primaryclass  = {astro-ph.CO}
}

@ARTICLE{foley_ROMAN_2018,
  author        = {{Foley}, R.~J. and {Koekemoer}, A.~M. and {Spergel}, D.~N. and {Bianco}, F.~B. and {Capak}, P. and {Dai}, L. and {Dore}, O. and {Fazio}, G.~G. and {Ferguson}, H. and {Filippenko}, A.~V. and {Frye}, B. and {Galbany}, L. and {Gawiser}, E. and {Gronwall}, C. and {Hathi}, N.~P. and {Hirata}, C. and {Hounsell}, R. and {Jha}, S.~W. and {Kim}, A.~G. and {Kelly}, P.~L. and {Kruk}, J.~W. and {Malhotra}, S. and {Mandel}, K.~S. and {Margutti}, R. and {Marrone}, D. and {McQuinn}, K.~B.~W. and {Melchior}, P. and {Moustakas}, L. and {Newman}, J.~A. and {Peek}, J.~E.~G. and {Perlmutter}, S. and {Rhodes}, J.~D. and {Robertson}, B. and {Rubin}, D. and {Scolnic}, D. and {Somerville}, R. and {Street}, R. and {Wang}, Y. and {Whalen}, D.~J. and {Windhorst}, R.~A. and {Wollack}, E.~J.},
  title         = {{LSST Observing Strategy White Paper: LSST Observations of WFIRST Deep Fields}},
  journal       = {arXiv e-prints},
  year          = {2018},
  pages         = {arXiv:1812.00514},
  eid           = {arXiv:1812.00514},
  doi           = {10.48550/arXiv.1812.00514},
  archiveprefix = {arXiv},
  eprint        = {1812.00514},
  primaryclass  = {astro-ph.IM}
}

@ARTICLE{fukugita_cosmic_2004,
  author        = {{Fukugita}, Masataka and {Peebles}, P.~J.~E.},
  title         = {{The Cosmic Energy Inventory}},
  journal       = {ApJ},
  year          = {2004},
  volume        = {616},
  number        = {2},
  pages         = {643-668},
  doi           = {10.1086/425155},
  archiveprefix = {arXiv},
  eprint        = {astro-ph/0406095},
  primaryclass  = {astro-ph}
}

@ARTICLE{gaztanaga_bias_1994,
  author        = {{Gaztanaga}, Enrique and {Frieman}, Joshua A.},
  title         = {{Bias and High-Order Galaxy Correlation Functions in the APM Galaxy Survey}},
  journal       = {ApJL},
  year          = {1994},
  volume        = {437},
  pages         = {L13},
  doi           = {10.1086/187671},
  archiveprefix = {arXiv},
  eprint        = {astro-ph/9407079},
  primaryclass  = {astro-ph}
}

@ARTICLE{giannini_DES_gg_2026,
  author        = {{Giannini}, G. and {Camacho-Ciurana}, G. and {Whyley}, A. and {Prat}, J. and {Blazek}, J. and {S{\'a}nchez}, C. and {Zacharegkas}, G. and {Alarcon}, A. and {Legnani}, E. and {Amon}, A. and {Anbajagane}, D. and {Avila}, S. and {Bechtol}, K. and {Becker}, M.~R. and {Bernstein}, G.~M. and {Bocquet}, S. and {Campos}, A. and {Carnero Rosell}, A. and {Cawthon}, R. and {Chang}, C. and {Crocce}, M. and {d'Assignies}, W. and {De Vicente}, J. and {Drlica-Wagner}, A. and {Elvin-Poole}, S. and {Fert{\'e}}, A. and {Gatti}, M. and {Gruen}, D. and {Jarvis}, M. and {Manera}, M. and {Mau}, S. and {McCullough}, J. and {Menanteau}, F. and {Myles}, J. and {Porredon}, A. and {Rodriguez-Monroy}, M. and {Roodman}, A. and {Rykoff}, E.~S. and {Samuroff}, S. and {Sanchez Cid}, D. and {Sevilla-Noarbe}, I. and {Schutt}, T. and {Troxel}, M.~A. and {Weaverdyck}, N. and {Yamamoto}, M. and {Yin}, B. and {Abbott}, T.~M.~C. and {Aguena}, M. and {Allam}, S. and {Alves}, O. and {Andrade-Oliveira}, F. and {Bacon}, D. and {Bertin}, E. and {Brooks}, D. and {Camacho}, H. and {Carretero}, J. and {da Costa}, L.~N. and {da Silva Pereira}, M.~E. and {Davis}, T.~M. and {DePoy}, D.~L. and {Desai}, S. and {Diehl}, H.~T. and {Doel}, P. and {Doux}, C. and {Eifler}, T.~F. and {Everett}, S. and {Evrard}, A.~E. and {Fosalba}, P. and {Frieman}, J. and {Garc{\'\i}a-Bellido}, J. and {Gaztanaga}, E. and {Giles}, P. and {Glazebrook}, K. and {Harrison}, I. and {Hartley}, W.~G. and {Herner}, K. and {Hinton}, S.~R. and {Hollowood}, D.~L. and {Honscheid}, K. and {Huterer}, D. and {Jain}, B. and {James}, D.~J. and {Jeffrey}, N. and {Kacprzak}, T. and {Kent}, S. and {Krause}, E. and {Lahav}, O. and {Lee}, S. and {Marshall}, J.~L. and {Mena-Fern{\'a}ndez}, J. and {Miquel}, R. and {Mohr}, J.~J. and {Muir}, J. and {Nichol}, R.~C. and {Ogando}, R.~L.~C. and {Palmese}, A. and {Paterno}, M. and {Percival}, W.~J. and {Petravick}, D. and {Plazas Malag{\'o}n}, A.~A. and {Raveri}, M. and {Rosenfeld}, R. and {Sanchez}, E. and {Sheldon}, E. and {Shin}, T. and {Allyn. Smith}, J. and {Smith}, M. and {Soares-Santos}, M. and {Suchyta}, E. and {Swanson}, M.~E.~C. and {Tarle}, G. and {Thomas}, D. and {To}, C. and {Tucker}, D.~L. and {Vikram}, V. and {Vincenzi}, M. and {Walker}, A.~R. and {Wiseman}, P. and {Yanny}, B.},
  title         = {{Dark Energy Survey Year 6 Results: Galaxy-galaxy lensing}},
  journal       = {Phys. Rev. D},
  year          = {2026},
  volume        = {113},
  number        = {10},
  pages         = {103503},
  doi           = {10.1103/c87p-j9nt},
  archiveprefix = {arXiv},
  eprint        = {2601.15175},
  primaryclass  = {astro-ph.CO}
}

@ARTICLE{girelli_stellar_2020,
  author        = {{Girelli}, G. and {Pozzetti}, L. and {Bolzonella}, M. and {Giocoli}, C. and {Marulli}, F. and {Baldi}, M.},
  title         = {{The stellar-to-halo mass relation over the past 12 Gyr. I. Standard {\ensuremath{\Lambda}}CDM model}},
  journal       = {A\&A},
  year          = {2020},
  volume        = {634},
  pages         = {A135},
  eid           = {A135},
  doi           = {10.1051/0004-6361/201936329},
  archiveprefix = {arXiv},
  eprint        = {2001.02230},
  primaryclass  = {astro-ph.CO}
}

@ARTICLE{golub_tikhonov_1999,
  author        = {Golub, Gene H and Hansen, Per Christian and O'Leary, Dianne P},
  title         = {Tikhonov regularization and total least squares},
  journal       = {SIAM J. Matrix Anal. Appl.},
  year          = {1999},
  volume        = {21},
  number        = {1},
  pages         = {185--194},
  publisher     = {SIAM},
  doi           = {https://doi.org/10.1137/S0895479897326432}
}

@ARTICLE{goncalves_detectability_2026,
  author        = {{Gon{\c{c}}alves}, Gustavo F. and {Machado}, Rubens E.~G. and {Valen{\c{c}}a}, Raquel R. and {Athanassoula}, E. and {Men{\'e}ndez-Delmestre}, Kar{\'\i}n and {Bueno-Dalpiaz}, Thiago},
  title         = {{The detectability of bars at high redshift: a case study using Euclid-like mock observations of TNG50 simulated galaxies}},
  journal       = {MNRAS},
  year          = {2026},
  volume        = {546},
  number        = {4},
  pages         = {stag263},
  doi           = {10.1093/mnras/stag263},
  archiveprefix = {arXiv},
  eprint        = {2601.16335},
  primaryclass  = {astro-ph.GA}
}

@ARTICLE{gong_morphological_2025,
  author        = {{Gong}, Jun-Yu and {Lin}, Weipeng and {Tang}, Lin and {Lan}, Yanyao},
  title         = {{Mock Observations: Morphological Analysis of Galaxies in TNG100 Simulations}},
  journal       = {ApJS},
  year          = {2025},
  volume        = {279},
  number        = {1},
  pages         = {16},
  eid           = {16},
  doi           = {10.3847/1538-4365/add5e4},
  archiveprefix = {arXiv},
  eprint        = {2504.18042},
  primaryclass  = {astro-ph.GA}
}

@ARTICLE{haider_illustris_lls_2016,
  author        = {{Haider}, M. and {Steinhauser}, D. and {Vogelsberger}, M. and {Genel}, S. and {Springel}, V. and {Torrey}, P. and {Hernquist}, L.},
  title         = {{Large-scale mass distribution in the Illustris simulation}},
  journal       = {MNRAS},
  year          = {2016},
  volume        = {457},
  number        = {3},
  pages         = {3024-3035},
  doi           = {10.1093/mnras/stw077},
  archiveprefix = {arXiv},
  eprint        = {1508.01525},
  primaryclass  = {astro-ph.CO}
}

@INPROCEEDINGS{hallinan_DSA_2019,
  author        = {{Hallinan}, Gregg and {Ravi}, V. and {Weinreb}, S. and {Kocz}, J. and {Huang}, Y. and {Woody}, D.~P. and {Lamb}, J. and {D'Addario}, L. and {Catha}, M. and {Law}, C. and {Kulkarni}, S.~R. and {Phinney}, E.~S. and {Eastwood}, M.~W. and {Bouman}, K. and {McLaughlin}, M. and {Ransom}, S. and {Siemens}, X. and {Cordes}, J. and {Lynch}, R. and {Kaplan}, D. and {Brazier}, A. and {Bhatnagar}, S. and {Myers}, S. and {Walter}, F. and {Gaensler}, B.},
  title         = {{The DSA-2000 {\textemdash} A Radio Survey Camera}},
  booktitle     = {Bulletin of the American Astronomical Society},
  year          = {2019},
  volume        = {51},
  pages         = {255},
  eid           = {255},
  doi           = {10.48550/arXiv.1907.07648},
  archiveprefix = {arXiv},
  eprint        = {1907.07648},
  primaryclass  = {astro-ph.IM}
}

@ARTICLE{hamaus_forecasts_2022,
  author        = {{Hamaus}, N. and {Aubert}, M. and {Pisani}, A. and {Contarini}, S. and {Verza}, G. and {Cousinou}, M.-C. and {Escoffier}, S. and {Hawken}, A. and {Lavaux}, G. and {Pollina}, G. and {Wandelt}, B.~D. and {Weller}, J. and {Bonici}, M. and {Carbone}, C. and {Guzzo}, L. and {Kovacs}, A. and {Marulli}, F. and {Massara}, E. and {Moscardini}, L. and {Ntelis}, P. and {Percival}, W.~J. and {Radinovi{\'c}}, S. and {Sahl{\'e}n}, M. and {Sakr}, Z. and {S{\'a}nchez}, A.~G. and {Winther}, H.~A. and {Auricchio}, N. and {Awan}, S. and {Bender}, R. and {Bodendorf}, C. and {Bonino}, D. and {Branchini}, E. and {Brescia}, M. and {Brinchmann}, J. and {Capobianco}, V. and {Carretero}, J. and {Castander}, F.~J. and {Castellano}, M. and {Cavuoti}, S. and {Cimatti}, A. and {Cledassou}, R. and {Congedo}, G. and {Conversi}, L. and {Copin}, Y. and {Corcione}, L. and {Cropper}, M. and {Da Silva}, A. and {Degaudenzi}, H. and {Douspis}, M. and {Dubath}, F. and {Duncan}, C.~A.~J. and {Dupac}, X. and {Dusini}, S. and {Ealet}, A. and {Ferriol}, S. and {Fosalba}, P. and {Frailis}, M. and {Franceschi}, E. and {Franzetti}, P. and {Fumana}, M. and {Garilli}, B. and {Gillis}, B. and {Giocoli}, C. and {Grazian}, A. and {Grupp}, F. and {Haugan}, S.~V.~H. and {Holmes}, W. and {Hormuth}, F. and {Jahnke}, K. and {Kermiche}, S. and {Kiessling}, A. and {Kilbinger}, M. and {Kitching}, T. and {K{\"u}mmel}, M. and {Kunz}, M. and {Kurki-Suonio}, H. and {Ligori}, S. and {Lilje}, P.~B. and {Lloro}, I. and {Maiorano}, E. and {Marggraf}, O. and {Markovic}, K. and {Massey}, R. and {Maurogordato}, S. and {Melchior}, M. and {Meneghetti}, M. and {Meylan}, G. and {Moresco}, M. and {Munari}, E. and {Niemi}, S.~M. and {Padilla}, C. and {Paltani}, S. and {Pasian}, F. and {Pedersen}, K. and {Pettorino}, V. and {Pires}, S. and {Poncet}, M. and {Popa}, L. and {Pozzetti}, L. and {Rebolo}, R. and {Rhodes}, J. and {Rix}, H. and {Roncarelli}, M. and {Rossetti}, E. and {Saglia}, R. and {Schneider}, P. and {Secroun}, A. and {Seidel}, G. and {Serrano}, S. and {Sirignano}, C. and {Sirri}, G. and {Starck}, J.-L. and {Tallada-Cresp{\'\i}}, P. and {Tavagnacco}, D. and {Taylor}, A.~N. and {Tereno}, I. and {Toledo-Moreo}, R. and {Torradeflot}, F. and {Valentijn}, E.~A. and {Valenziano}, L. and {Wang}, Y. and {Welikala}, N. and {Zamorani}, G. and {Zoubian}, J. and {Andreon}, S. and {Baldi}, M. and {Camera}, S. and {Mei}, S. and {Neissner}, C. and {Romelli}, E.},
  title         = {{Euclid: Forecasts from redshift-space distortions and the Alcock-Paczynski test with cosmic voids}},
  journal       = {A\&A},
  year          = {2022},
  volume        = {658},
  pages         = {A20},
  eid           = {A20},
  doi           = {10.1051/0004-6361/202142073},
  archiveprefix = {arXiv},
  eprint        = {2108.10347},
  primaryclass  = {astro-ph.CO}
}

@ARTICLE{hoekstra_lensing_2008,
  author        = {{Hoekstra}, Henk and {Jain}, Bhuvnesh},
  title         = {{Weak Gravitational Lensing and Its Cosmological Applications}},
  journal       = {Annu. Rev. Nucl. Part. Sci.},
  year          = {2008},
  volume        = {58},
  number        = {1},
  pages         = {99-123},
  doi           = {10.1146/annurev.nucl.58.110707.171151},
  archiveprefix = {arXiv},
  eprint        = {0805.0139},
  primaryclass  = {astro-ph}
}

@ARTICLE{hu_impact_2024,
  author        = {{Hu}, Teng and {Khaire}, Vikram and {Hennawi}, Joseph F. and {O{\~n}orbe}, Jose and {Walther}, Michael and {Lukic}, Zarija and {Davies}, Frederick},
  title         = {{The impact of the WHIM on the IGM thermal state determined from the low-z Lyman {\ensuremath{\alpha}} forest}},
  journal       = {MNRAS},
  year          = {2024},
  volume        = {527},
  number        = {4},
  pages         = {11338-11359},
  doi           = {10.1093/mnras/stad3846},
  archiveprefix = {arXiv},
  eprint        = {2308.14738},
  primaryclass  = {astro-ph.CO}
}

@ARTICLE{hung_establishing_2020,
  author        = {{Hung}, D. and {Lemaux}, B.~C. and {Gal}, R.~R. and {Tomczak}, A.~R. and {Lubin}, L.~M. and {Cucciati}, O. and {Pelliccia}, D. and {Shen}, L. and {Le F{\`e}vre}, O. and {Wu}, P.-F. and {Kocevski}, D.~D. and {Mei}, S. and {Squires}, G.~K.},
  title         = {{Establishing a new technique for discovering large-scale structure using the ORELSE survey}},
  journal       = {MNRAS},
  year          = {2020},
  volume        = {491},
  number        = {4},
  pages         = {5524-5554},
  doi           = {10.1093/mnras/stz3164},
  archiveprefix = {arXiv},
  eprint        = {1905.09298},
  primaryclass  = {astro-ph.GA}
}

@ARTICLE{inoue_2004_dm_derivation,
  author        = {{Inoue}, Susumu},
  title         = {{Probing the cosmic reionization history and local environment of gamma-ray bursts through radio dispersion}},
  journal       = {MNRAS},
  year          = {2004},
  volume        = {348},
  number        = {3},
  pages         = {999-1008},
  doi           = {10.1111/j.1365-2966.2004.07359.x},
  archiveprefix = {arXiv},
  eprint        = {astro-ph/0309364},
  primaryclass  = {astro-ph}
}

@ARTICLE{ioka_2003_dm_derivation,
  author        = {{Ioka}, Kunihito},
  title         = {{The Cosmic Dispersion Measure from Gamma-Ray Burst Afterglows: Probing the Reionization History and the Burst Environment}},
  journal       = {ApJL},
  year          = {2003},
  volume        = {598},
  number        = {2},
  pages         = {L79-L82},
  doi           = {10.1086/380598},
  archiveprefix = {arXiv},
  eprint        = {astro-ph/0309200},
  primaryclass  = {astro-ph}
}

@ARTICLE{ivanov_precision_2022,
  author        = {{Ivanov}, Mikhail M. and {Philcox}, Oliver H.~E. and {Nishimichi}, Takahiro and {Simonovi{\'c}}, Marko and {Takada}, Masahiro and {Zaldarriaga}, Matias},
  title         = {{Precision analysis of the redshift-space galaxy bispectrum}},
  journal       = {Phys. Rev. D},
  year          = {2022},
  volume        = {105},
  number        = {6},
  pages         = {063512},
  eid           = {063512},
  doi           = {10.1103/PhysRevD.105.063512},
  archiveprefix = {arXiv},
  eprint        = {2110.10161},
  primaryclass  = {astro-ph.CO}
}

@ARTICLE{ivezic_LSST_2019,
  author        = {{Ivezi{\'c}}, {\v{Z}}eljko and {Kahn}, Steven M. and {Tyson}, J. Anthony and {Abel}, Bob and {Acosta}, Emily and {Allsman}, Robyn and {Alonso}, David and {AlSayyad}, Yusra and {Anderson}, Scott F. and {Andrew}, John and {Angel}, James Roger P. and {Angeli}, George Z. and {Ansari}, Reza and {Antilogus}, Pierre and {Araujo}, Constanza and {Armstrong}, Robert and {Arndt}, Kirk T. and {Astier}, Pierre and {Aubourg}, {\'E}ric and {Auza}, Nicole and {Axelrod}, Tim S. and {Bard}, Deborah J. and {Barr}, Jeff D. and {Barrau}, Aurelian and {Bartlett}, James G. and {Bauer}, Amanda E. and {Bauman}, Brian J. and {Baumont}, Sylvain and {Bechtol}, Ellen and {Bechtol}, Keith and {Becker}, Andrew C. and {Becla}, Jacek and {Beldica}, Cristina and {Bellavia}, Steve and {Bianco}, Federica B. and {Biswas}, Rahul and {Blanc}, Guillaume and {Blazek}, Jonathan and {Blandford}, Roger D. and {Bloom}, Josh S. and {Bogart}, Joanne and {Bond}, Tim W. and {Booth}, Michael T. and {Borgland}, Anders W. and {Borne}, Kirk and {Bosch}, James F. and {Boutigny}, Dominique and {Brackett}, Craig A. and {Bradshaw}, Andrew and {Brandt}, William Nielsen and {Brown}, Michael E. and {Bullock}, James S. and {Burchat}, Patricia and {Burke}, David L. and {Cagnoli}, Gianpietro and {Calabrese}, Daniel and {Callahan}, Shawn and {Callen}, Alice L. and {Carlin}, Jeffrey L. and {Carlson}, Erin L. and {Chandrasekharan}, Srinivasan and {Charles-Emerson}, Glenaver and {Chesley}, Steve and {Cheu}, Elliott C. and {Chiang}, Hsin-Fang and {Chiang}, James and {Chirino}, Carol and {Chow}, Derek and {Ciardi}, David R. and {Claver}, Charles F. and {Cohen-Tanugi}, Johann and {Cockrum}, Joseph J. and {Coles}, Rebecca and {Connolly}, Andrew J. and {Cook}, Kem H. and {Cooray}, Asantha and {Covey}, Kevin R. and {Cribbs}, Chris and {Cui}, Wei and {Cutri}, Roc and {Daly}, Philip N. and {Daniel}, Scott F. and {Daruich}, Felipe and {Daubard}, Guillaume and {Daues}, Greg and {Dawson}, William and {Delgado}, Francisco and {Dellapenna}, Alfred and {de Peyster}, Robert and {de Val-Borro}, Miguel and {Digel}, Seth W. and {Doherty}, Peter and {Dubois}, Richard and {Dubois-Felsmann}, Gregory P. and {Durech}, Josef and {Economou}, Frossie and {Eifler}, Tim and {Eracleous}, Michael and {Emmons}, Benjamin L. and {Fausti Neto}, Angelo and {Ferguson}, Henry and {Figueroa}, Enrique and {Fisher-Levine}, Merlin and {Focke}, Warren and {Foss}, Michael D. and {Frank}, James and {Freemon}, Michael D. and {Gangler}, Emmanuel and {Gawiser}, Eric and {Geary}, John C. and {Gee}, Perry and {Geha}, Marla and {Gessner}, Charles J.~B. and {Gibson}, Robert R. and {Gilmore}, D. Kirk and {Glanzman}, Thomas and {Glick}, William and {Goldina}, Tatiana and {Goldstein}, Daniel A. and {Goodenow}, Iain and {Graham}, Melissa L. and {Gressler}, William J. and {Gris}, Philippe and {Guy}, Leanne P. and {Guyonnet}, Augustin and {Haller}, Gunther and {Harris}, Ron and {Hascall}, Patrick A. and {Haupt}, Justine and {Hernandez}, Fabio and {Herrmann}, Sven and {Hileman}, Edward and {Hoblitt}, Joshua and {Hodgson}, John A. and {Hogan}, Craig and {Howard}, James D. and {Huang}, Dajun and {Huffer}, Michael E. and {Ingraham}, Patrick and {Innes}, Walter R. and {Jacoby}, Suzanne H. and {Jain}, Bhuvnesh and {Jammes}, Fabrice and {Jee}, M. James and {Jenness}, Tim and {Jernigan}, Garrett and {Jevremovi{\'c}}, Darko and {Johns}, Kenneth and {Johnson}, Anthony S. and {Johnson}, Margaret W.~G. and {Jones}, R. Lynne and {Juramy-Gilles}, Claire and {Juri{\'c}}, Mario and {Kalirai}, Jason S. and {Kallivayalil}, Nitya J. and {Kalmbach}, Bryce and {Kantor}, Jeffrey P. and {Karst}, Pierre and {Kasliwal}, Mansi M. and {Kelly}, Heather and {Kessler}, Richard and {Kinnison}, Veronica and {Kirkby}, David and {Knox}, Lloyd and {Kotov}, Ivan V. and {Krabbendam}, Victor L. and {Krughoff}, K. Simon and {Kub{\'a}nek}, Petr and {Kuczewski}, John and {Kulkarni}, Shri and {Ku}, John and {Kurita}, Nadine R. and {Lage}, Craig S. and {Lambert}, Ron and {Lange}, Travis and {Langton}, J. Brian and {Le Guillou}, Laurent and {Levine}, Deborah and {Liang}, Ming and {Lim}, Kian-Tat and {Lintott}, Chris J. and {Long}, Kevin E. and {Lopez}, Margaux and {Lotz}, Paul J. and {Lupton}, Robert H. and {Lust}, Nate B. and {MacArthur}, Lauren A. and {Mahabal}, Ashish and {Mandelbaum}, Rachel and {Markiewicz}, Thomas W. and {Marsh}, Darren S. and {Marshall}, Philip J. and {Marshall}, Stuart and {May}, Morgan and {McKercher}, Robert and {McQueen}, Michelle and {Meyers}, Joshua and {Migliore}, Myriam and {Miller}, Michelle and {Mills}, David J.},
  title         = {{LSST: From Science Drivers to Reference Design and Anticipated Data Products}},
  journal       = {ApJ},
  year          = {2019},
  volume        = {873},
  number        = {2},
  pages         = {111},
  eid           = {111},
  doi           = {10.3847/1538-4357/ab042c},
  archiveprefix = {arXiv},
  eprint        = {0805.2366},
  primaryclass  = {astro-ph}
}

@ARTICLE{jaroszynski_fast_2019,
  author        = {Jaroszynski, M},
  title         = {Fast radio bursts and cosmological tests},
  journal       = {MNRAS},
  year          = {2019},
  volume        = {484},
  number        = {2},
  pages         = {1637--1644},
  doi           = {10.1093/mnras/sty3529}
}

@ARTICLE{jasche_bayesian_2013,
  author        = {Jasche, Jens and Wandelt, Benjamin D.},
  title         = {Bayesian physical reconstruction of initial conditions from large scale structure surveys},
  journal       = {MNRAS},
  year          = {2013},
  volume        = {432},
  number        = {2},
  pages         = {894--913},
  doi           = {10.1093/mnras/stt449}
}

@ARTICLE{johnson_Prospector_2021,
  author        = {{Johnson}, Benjamin D. and {Leja}, Joel and {Conroy}, Charlie and {Speagle}, Joshua S.},
  title         = {{Stellar Population Inference with Prospector}},
  journal       = {ApJS},
  year          = {2021},
  volume        = {254},
  number        = {2},
  pages         = {22},
  eid           = {22},
  doi           = {10.3847/1538-4365/abef67},
  archiveprefix = {arXiv},
  eprint        = {2012.01426},
  primaryclass  = {astro-ph.GA}
}

@ARTICLE{jones_THEMIS_2017,
  author        = {{Jones}, A.~P. and {K{\"o}hler}, M. and {Ysard}, N. and {Bocchio}, M. and {Verstraete}, L.},
  title         = {{The global dust modelling framework THEMIS}},
  journal       = {A\&A},
  year          = {2017},
  volume        = {602},
  pages         = {A46},
  eid           = {A46},
  doi           = {10.1051/0004-6361/201630225},
  archiveprefix = {arXiv},
  eprint        = {1703.00775},
  primaryclass  = {astro-ph.GA}
}

@ARTICLE{kaiser_clustering_1987,
  author        = {{Kaiser}, Nick},
  title         = {{Clustering in real space and in redshift space}},
  journal       = {MNRAS},
  year          = {1987},
  volume        = {227},
  pages         = {1-21},
  doi           = {10.1093/mnras/227.1.1}
}

@ARTICLE{khrykin_cosmic_2024,
  author        = {{Khrykin}, Ilya S. and {Sorini}, Daniele and {Lee}, Khee-Gan and {Dav{\'e}}, Romeel},
  title         = {{The cosmic baryon partition between the IGM and CGM in the SIMBA simulations}},
  journal       = {MNRAS},
  year          = {2024},
  volume        = {529},
  number        = {1},
  pages         = {537-549},
  doi           = {10.1093/mnras/stae525},
  archiveprefix = {arXiv},
  eprint        = {2310.01496},
  primaryclass  = {astro-ph.GA}
}

@ARTICLE{khrykin_FLIMFLAM_2024,
  author        = {Khrykin, Ilya S. and Ata, Metin and Lee, Khee-Gan and Simha, Sunil and Huang, Yuxin and Prochaska, J. Xavier and Tejos, Nicolas and Bannister, Keith W. and Cooke, Jeff and Day, Cherie K. and Deller, Adam and Glowacki, Marcin and Gordon, Alexa C. and James, Clancy W. and Marnoch, Lachlan and Shannon, Ryan M. and Zhang, Jielai and Bernales-Cortes, Lucas},
  title         = {{FLIMFLAM} {DR1}: {The} {First} {Constraints} on the {Cosmic} {Baryon} {Distribution} from 8 {FRB} sightlines},
  journal       = {ApJ},
  year          = {2024},
  volume        = {973},
  number        = {2},
  pages         = {151},
  doi           = {10.3847/1538-4357/ad6567},
  archiveprefix = {arXiv},
  eprint        = {2402.00505}
}

@ARTICLE{konar_constraining_2025,
  author        = {Konar, Koustav and Reischke, Robert and Hagstotz, Steffen and Nicola, Andrina and Hildebrandt, Hendrik},
  title         = {Constraining the dispersion measure redshift relation with simulation-based inference},
  journal       = {OJAp},
  year          = {2025},
  volume        = {8},
  doi           = {10.33232/001c.142524}
}

@ARTICLE{konietzka_ray_tracing_2025,
  author        = {{Konietzka}, Ralf M. and {Connor}, Liam and {Semenov}, Vadim A. and {Beane}, Angus and {Springel}, Volker and {Hernquist}, Lars},
  title         = {{Ray-tracing Fast Radio Bursts Through IllustrisTNG: Cosmological Dispersion Measures from Redshift 0 to 5.5}},
  journal       = {ApJ},
  year          = {2026},
  volume        = {1010},
  number        = {2},
  pages         = {122},
  doi           = {10.3847/1538-4357/ae99e8},
  archiveprefix = {arXiv},
  eprint        = {2507.07090},
  primaryclass  = {astro-ph.CO}
}

@ARTICLE{kouroumpatzakis_SFR_2023,
  author        = {{Kouroumpatzakis}, K. and {Zezas}, A. and {Kyritsis}, E. and {Salim}, S. and {Svoboda}, J.},
  title         = {{Star formation rate and stellar mass calibrations based on infrared photometry and their dependence on stellar population age and extinction}},
  journal       = {A\&A},
  year          = {2023},
  volume        = {673},
  pages         = {A16},
  eid           = {A16},
  doi           = {10.1051/0004-6361/202346054},
  archiveprefix = {arXiv},
  eprint        = {2303.10013},
  primaryclass  = {astro-ph.GA}
}

@ARTICLE{kraljic_Euclid_3D_2026,
  author        = {{Euclid Collaboration} and {Kraljic}, K. and {Laigle}, C. and {Balogh}, M. and {Jablonka}, P. and {Kuchner}, U. and {Malavasi}, N. and {Sarron}, F. and {Pichon}, C. and {De Lucia}, G. and {Bethermin}, M. and {Durret}, F. and {Fumagalli}, M. and {Gouin}, C. and {Magliocchetti}, M. and {Sorce}, J.~G. and {Cucciati}, O. and {Fontanot}, F. and {Hirschmann}, M. and {Kang}, Y. and {Spinelli}, M. and {Aghanim}, N. and {Amara}, A. and {Andreon}, S. and {Auricchio}, N. and {Baccigalupi}, C. and {Baldi}, M. and {Bardelli}, S. and {Biviano}, A. and {Branchini}, E. and {Brescia}, M. and {Brinchmann}, J. and {Camera}, S. and {Ca{\~n}as-Herrera}, G. and {Capobianco}, V. and {Carbone}, C. and {Carretero}, J. and {Casas}, R. and {Casas}, S. and {Castander}, F.~J. and {Castellano}, M. and {Castignani}, G. and {Cavuoti}, S. and {Chambers}, K.~C. and {Cimatti}, A. and {Colodro-Conde}, C. and {Congedo}, G. and {Conselice}, C.~J. and {Conversi}, L. and {Copin}, Y. and {Courbin}, F. and {Courtois}, H.~M. and {Da Silva}, A. and {Degaudenzi}, H. and {de la Torre}, S. and {Dole}, H. and {Douspis}, M. and {Dubath}, F. and {Duncan}, C.~A.~J. and {Dupac}, X. and {Dusini}, S. and {Escoffier}, S. and {Farina}, M. and {Farinelli}, R. and {Ferriol}, S. and {Finelli}, F. and {Fosalba}, P. and {Fourmanoit}, N. and {Frailis}, M. and {Franceschi}, E. and {Fumana}, M. and {Galeotta}, S. and {George}, K. and {Gillard}, W. and {Gillis}, B. and {Giocoli}, C. and {Gracia-Carpio}, J. and {Grazian}, A. and {Grupp}, F. and {Haugan}, S.~V.~H. and {Holmes}, W. and {Hormuth}, F. and {Hornstrup}, A. and {Jahnke}, K. and {Jhabvala}, M. and {Joachimi}, B. and {Keih{\"a}nen}, E. and {Kermiche}, S. and {Kiessling}, A. and {Kilbinger}, M. and {Kubik}, B. and {K{\"u}mmel}, M. and {Kunz}, M. and {Kurki-Suonio}, H. and {Le Brun}, A.~M.~C. and {Ligori}, S. and {Lilje}, P.~B. and {Lindholm}, V. and {Lloro}, I. and {Mainetti}, G. and {Maino}, D. and {Maiorano}, E. and {Mansutti}, O. and {Marcin}, S. and {Marggraf}, O. and {Martinelli}, M. and {Martinet}, N. and {Marulli}, F. and {Massey}, R. and {Maurogordato}, S. and {Medinaceli}, E. and {Mei}, S. and {Mellier}, Y. and {Meneghetti}, M. and {Merlin}, E. and {Meylan}, G. and {Mora}, A. and {Moresco}, M. and {Moscardini}, L. and {Nakajima}, R. and {Neissner}, C. and {Niemi}, S.-M. and {Padilla}, C. and {Paltani}, S. and {Pasian}, F. and {Pedersen}, K. and {Percival}, W.~J. and {Pettorino}, V. and {Pires}, S. and {Polenta}, G. and {Poncet}, M. and {Popa}, L.~A. and {Pozzetti}, L. and {Raison}, F. and {Rebolo}, R. and {Renzi}, A. and {Rhodes}, J. and {Riccio}, G. and {Romelli}, E. and {Roncarelli}, M. and {Rosset}, C. and {Rossetti}, E. and {Saglia}, R. and {Sakr}, Z. and {S{\'a}nchez}, A.~G. and {Sapone}, D. and {Sartoris}, B. and {Schneider}, P. and {Schrabback}, T. and {Scodeggio}, M. and {Secroun}, A. and {Sefusatti}, E. and {Seidel}, G. and {Seiffert}, M. and {Serrano}, S. and {Simon}, P. and {Sirignano}, C. and {Sirri}, G. and {Stanco}, L. and {Steinwagner}, J. and {Tallada-Cresp{\'\i}}, P. and {Taylor}, A.~N. and {Teplitz}, H.~I. and {Tereno}, I. and {Tessore}, N. and {Toft}, S. and {Toledo-Moreo}, R. and {Torradeflot}, F. and {Tutusaus}, I. and {Valenziano}, L. and {Valiviita}, J. and {Vassallo}, T. and {Verdoes Kleijn}, G. and {Veropalumbo}, A. and {Vibert}, D. and {Wang}, Y. and {Weller}, J. and {Zacchei}, A. and {Zamorani}, G. and {Zucca}, E. and {Allevato}, V. and {Ballardini}, M. and {Bolzonella}, M. and {Bozzo}, E. and {Burigana}, C. and {Cabanac}, R. and {Calabrese}, M. and {Cappi}, A. and {Di Ferdinando}, D. and {Escartin Vigo}, J.~A. and {Gabarra}, L. and {Hartley}, W.~G. and {Mart{\'\i}n-Fleitas}, J. and {Matthew}, S. and {Mauri}, N. and {Metcalf}, R.~B. and {Nucita}, A.~A. and {Pezzotta}, A. and {P{\"o}ntinen}, M. and {Porciani}, C.},
  title         = {{Euclid preparation. 3D reconstruction of the cosmic web with simulated Euclid Deep spectroscopic samples}},
  journal       = {A\&A},
  year          = {2026},
  volume        = {708},
  pages         = {A164},
  doi           = {10.1051/0004-6361/202557286},
  archiveprefix = {arXiv},
  eprint        = {2601.10709},
  primaryclass  = {astro-ph.GA}
}

@ARTICLE{laureijs_Euclid_definition_2011,
  author        = {{Laureijs}, R. and {Amiaux}, J. and {Arduini}, S. and {Augu{\`e}res}, J. -L. and {Brinchmann}, J. and {Cole}, R. and {Cropper}, M. and {Dabin}, C. and {Duvet}, L. and {Ealet}, A. and {Garilli}, B. and {Gondoin}, P. and {Guzzo}, L. and {Hoar}, J. and {Hoekstra}, H. and {Holmes}, R. and {Kitching}, T. and {Maciaszek}, T. and {Mellier}, Y. and {Pasian}, F. and {Percival}, W. and {Rhodes}, J. and {Saavedra Criado}, G. and {Sauvage}, M. and {Scaramella}, R. and {Valenziano}, L. and {Warren}, S. and {Bender}, R. and {Castander}, F. and {Cimatti}, A. and {Le F{\`e}vre}, O. and {Kurki-Suonio}, H. and {Levi}, M. and {Lilje}, P. and {Meylan}, G. and {Nichol}, R. and {Pedersen}, K. and {Popa}, V. and {Rebolo Lopez}, R. and {Rix}, H. -W. and {Rottgering}, H. and {Zeilinger}, W. and {Grupp}, F. and {Hudelot}, P. and {Massey}, R. and {Meneghetti}, M. and {Miller}, L. and {Paltani}, S. and {Paulin-Henriksson}, S. and {Pires}, S. and {Saxton}, C. and {Schrabback}, T. and {Seidel}, G. and {Walsh}, J. and {Aghanim}, N. and {Amendola}, L. and {Bartlett}, J. and {Baccigalupi}, C. and {Beaulieu}, J. -P. and {Benabed}, K. and {Cuby}, J. -G. and {Elbaz}, D. and {Fosalba}, P. and {Gavazzi}, G. and {Helmi}, A. and {Hook}, I. and {Irwin}, M. and {Kneib}, J. -P. and {Kunz}, M. and {Mannucci}, F. and {Moscardini}, L. and {Tao}, C. and {Teyssier}, R. and {Weller}, J. and {Zamorani}, G. and {Zapatero Osorio}, M.~R. and {Boulade}, O. and {Foumond}, J.~J. and {Di Giorgio}, A. and {Guttridge}, P. and {James}, A. and {Kemp}, M. and {Martignac}, J. and {Spencer}, A. and {Walton}, D. and {Bl{\"u}mchen}, T. and {Bonoli}, C. and {Bortoletto}, F. and {Cerna}, C. and {Corcione}, L. and {Fabron}, C. and {Jahnke}, K. and {Ligori}, S. and {Madrid}, F. and {Martin}, L. and {Morgante}, G. and {Pamplona}, T. and {Prieto}, E. and {Riva}, M. and {Toledo}, R. and {Trifoglio}, M. and {Zerbi}, F. and {Abdalla}, F. and {Douspis}, M. and {Grenet}, C. and {Borgani}, S. and {Bouwens}, R. and {Courbin}, F. and {Delouis}, J. -M. and {Dubath}, P. and {Fontana}, A. and {Frailis}, M. and {Grazian}, A. and {Koppenh{\"o}fer}, J. and {Mansutti}, O. and {Melchior}, M. and {Mignoli}, M. and {Mohr}, J. and {Neissner}, C. and {Noddle}, K. and {Poncet}, M. and {Scodeggio}, M. and {Serrano}, S. and {Shane}, N. and {Starck}, J. -L. and {Surace}, C. and {Taylor}, A. and {Verdoes-Kleijn}, G. and {Vuerli}, C. and {Williams}, O.~R. and {Zacchei}, A. and {Altieri}, B. and {Escudero Sanz}, I. and {Kohley}, R. and {Oosterbroek}, T. and {Astier}, P. and {Bacon}, D. and {Bardelli}, S. and {Baugh}, C. and {Bellagamba}, F. and {Benoist}, C. and {Bianchi}, D. and {Biviano}, A. and {Branchini}, E. and {Carbone}, C. and {Cardone}, V. and {Clements}, D. and {Colombi}, S. and {Conselice}, C. and {Cresci}, G. and {Deacon}, N. and {Dunlop}, J. and {Fedeli}, C. and {Fontanot}, F. and {Franzetti}, P. and {Giocoli}, C. and {Garcia-Bellido}, J. and {Gow}, J. and {Heavens}, A. and {Hewett}, P. and {Heymans}, C. and {Holland}, A. and {Huang}, Z. and {Ilbert}, O. and {Joachimi}, B. and {Jennins}, E. and {Kerins}, E. and {Kiessling}, A. and {Kirk}, D. and {Kotak}, R. and {Krause}, O. and {Lahav}, O. and {van Leeuwen}, F. and {Lesgourgues}, J. and {Lombardi}, M. and {Magliocchetti}, M. and {Maguire}, K. and {Majerotto}, E. and {Maoli}, R. and {Marulli}, F. and {Maurogordato}, S. and {McCracken}, H. and {McLure}, R. and {Melchiorri}, A. and {Merson}, A. and {Moresco}, M. and {Nonino}, M. and {Norberg}, P. and {Peacock}, J. and {Pello}, R. and {Penny}, M. and {Pettorino}, V. and {Di Porto}, C. and {Pozzetti}, L. and {Quercellini}, C. and {Radovich}, M. and {Rassat}, A. and {Roche}, N. and {Ronayette}, S. and {Rossetti}, E.},
  title         = {{Euclid Definition Study Report}},
  journal       = {arXiv e-prints},
  year          = {2011},
  pages         = {arXiv:1110.3193},
  eid           = {arXiv:1110.3193},
  doi           = {10.48550/arXiv.1110.3193},
  archiveprefix = {arXiv},
  eprint        = {1110.3193},
  primaryclass  = {astro-ph.CO}
}

@ARTICLE{law_DSA110_2024,
  author        = {{Law}, Casey J. and {Sharma}, Kritti and {Ravi}, Vikram and {Chen}, Ge and {Catha}, Morgan and {Connor}, Liam and {Faber}, Jakob T. and {Hallinan}, Gregg and {Harnach}, Charlie and {Hellbourg}, Greg and {Hobbs}, Rick and {Hodge}, David and {Hodges}, Mark and {Lamb}, James W. and {Rasmussen}, Paul and {Sherman}, Myles B. and {Shi}, Jun and {Simard}, Dana and {Squillace}, Reynier and {Weinreb}, Sander and {Woody}, David P. and {Yurk}, Nitika Yadlapalli},
  title         = {{Deep Synoptic Array Science: First FRB and Host Galaxy Catalog}},
  journal       = {ApJ},
  year          = {2024},
  volume        = {967},
  number        = {1},
  pages         = {29},
  eid           = {29},
  doi           = {10.3847/1538-4357/ad3736},
  archiveprefix = {arXiv},
  eprint        = {2307.03344},
  primaryclass  = {astro-ph.HE}
}

@ARTICLE{leauthaud_theoretical_2011,
  author        = {{Leauthaud}, Alexie and {Tinker}, Jeremy and {Behroozi}, Peter S. and {Busha}, Michael T. and {Wechsler}, Risa H.},
  title         = {{A Theoretical Framework for Combining Techniques that Probe the Link Between Galaxies and Dark Matter}},
  journal       = {ApJ},
  year          = {2011},
  volume        = {738},
  number        = {1},
  pages         = {45},
  eid           = {45},
  doi           = {10.1088/0004-637X/738/1/45},
  archiveprefix = {arXiv},
  eprint        = {1103.2077},
  primaryclass  = {astro-ph.CO}
}

@ARTICLE{legrand_cosmos_2019,
  author        = {{Legrand}, L. and {McCracken}, H.~J. and {Davidzon}, I. and {Ilbert}, O. and {Coupon}, J. and {Aghanim}, N. and {Douspis}, M. and {Capak}, P.~L. and {Le F{\`e}vre}, O. and {Milvang-Jensen}, B.},
  title         = {{The COSMOS-UltraVISTA stellar-to-halo mass relationship: new insights on galaxy formation efficiency out to z {\ensuremath{\sim}} 5}},
  journal       = {MNRAS},
  year          = {2019},
  volume        = {486},
  number        = {4},
  pages         = {5468-5481},
  doi           = {10.1093/mnras/stz1198},
  archiveprefix = {arXiv},
  eprint        = {1810.10557},
  primaryclass  = {astro-ph.GA}
}

@ARTICLE{leja_new_2020,
  author        = {{Leja}, Joel and {Speagle}, Joshua S. and {Johnson}, Benjamin D. and {Conroy}, Charlie and {van Dokkum}, Pieter and {Franx}, Marijn},
  title         = {{A New Census of the 0.2 < z < 3.0 Universe. I. The Stellar Mass Function}},
  journal       = {ApJ},
  year          = {2020},
  volume        = {893},
  number        = {2},
  pages         = {111},
  eid           = {111},
  doi           = {10.3847/1538-4357/ab7e27},
  archiveprefix = {arXiv},
  eprint        = {1910.04168},
  primaryclass  = {astro-ph.GA}
}

@ARTICLE{li_WL_2025,
  author        = {{Li}, Shun-Sheng and {Hoekstra}, Henk and {Kuijken}, Konrad and {Schaller}, Matthieu and {Schaye}, Joop},
  title         = {{Weak lensing constraints on the stellar-to-halo mass relation of galaxy groups with simulation-informed scatter}},
  journal       = {A\&A},
  year          = {2025},
  volume        = {700},
  pages         = {A202},
  eid           = {A202},
  doi           = {10.1051/0004-6361/202452892},
  archiveprefix = {arXiv},
  eprint        = {2411.03152},
  primaryclass  = {astro-ph.GA}
}

@ARTICLE{libeskind_tracking_2018,
  author        = {{Libeskind}, Noam I. and {van de Weygaert}, Rien and {Cautun}, Marius and {Falck}, Bridget and {Tempel}, Elmo and {Abel}, Tom and {Alpaslan}, Mehmet and {Arag{\'o}n-Calvo}, Miguel A. and {Forero-Romero}, Jaime E. and {Gonzalez}, Roberto and {Gottl{\"o}ber}, Stefan and {Hahn}, Oliver and {Hellwing}, Wojciech A. and {Hoffman}, Yehuda and {Jones}, Bernard J.~T. and {Kitaura}, Francisco and {Knebe}, Alexander and {Manti}, Serena and {Neyrinck}, Mark and {Nuza}, Sebasti{\'a}n E. and {Padilla}, Nelson and {Platen}, Erwin and {Ramachandra}, Nesar and {Robotham}, Aaron and {Saar}, Enn and {Shandarin}, Sergei and {Steinmetz}, Matthias and {Stoica}, Radu S. and {Sousbie}, Thierry and {Yepes}, Gustavo},
  title         = {{Tracing the cosmic web}},
  journal       = {MNRAS},
  year          = {2018},
  volume        = {473},
  number        = {1},
  pages         = {1195-1217},
  doi           = {10.1093/mnras/stx1976},
  archiveprefix = {arXiv},
  eprint        = {1705.03021},
  primaryclass  = {astro-ph.CO}
}

@ARTICLE{lorimer_2007_first_frb_detection,
  author        = {{Lorimer}, D.~R. and {Bailes}, M. and {McLaughlin}, M.~A. and {Narkevic}, D.~J. and {Crawford}, F.},
  title         = {{A Bright Millisecond Radio Burst of Extragalactic Origin}},
  journal       = {Science},
  year          = {2007},
  volume        = {318},
  number        = {5851},
  pages         = {777},
  doi           = {10.1126/science.1147532},
  archiveprefix = {arXiv},
  eprint        = {0709.4301},
  primaryclass  = {astro-ph}
}

@ARTICLE{lower_how_2020,
  author        = {{Lower}, Sidney and {Narayanan}, Desika and {Leja}, Joel and {Johnson}, Benjamin D. and {Conroy}, Charlie and {Dav{\'e}}, Romeel},
  title         = {{How Well Can We Measure the Stellar Mass of a Galaxy: The Impact of the Assumed Star Formation History Model in SED Fitting}},
  journal       = {ApJ},
  year          = {2020},
  volume        = {904},
  number        = {1},
  pages         = {33},
  eid           = {33},
  doi           = {10.3847/1538-4357/abbfa7},
  archiveprefix = {arXiv},
  eprint        = {2006.03599},
  primaryclass  = {astro-ph.GA}
}

@ARTICLE{mackey_emcee_2013,
  author        = {{Foreman-Mackey}, Daniel and {Hogg}, David W. and {Lang}, Dustin and {Goodman}, Jonathan},
  title         = {{emcee: The MCMC Hammer}},
  journal       = {PASP},
  year          = {2013},
  volume        = {125},
  number        = {925},
  pages         = {306},
  doi           = {10.1086/670067},
  archiveprefix = {arXiv},
  eprint        = {1202.3665},
  primaryclass  = {astro-ph.IM}
}

@ARTICLE{macquart_census_2020,
  author        = {Macquart, J.-P. and Prochaska, J. X. and McQuinn, M. and Bannister, K. W. and Bhandari, S. and Day, C. K. and Deller, A. T. and Ekers, R. D. and James, C. W. and Marnoch, L. and Osłowski, S. and Phillips, C. and Ryder, S. D. and Scott, D. R. and Shannon, R. M. and Tejos, N.},
  title         = {A census of baryons in the {Universe} from localized fast radio bursts},
  journal       = {Nature},
  year          = {2020},
  volume        = {581},
  number        = {7809},
  pages         = {391--395},
  doi           = {10.1038/s41586-020-2300-2}
}

@ARTICLE{madhava_cosmology_ksz_2019,
  author        = {{Madhavacheril}, Mathew S. and {Battaglia}, Nicholas and {Smith}, Kendrick M. and {Sievers}, Jonathan L.},
  title         = {{Cosmology with the kinematic Sunyaev-Zeldovich effect: Breaking the optical depth degeneracy with fast radio bursts}},
  journal       = {Phys. Rev. D},
  year          = {2019},
  volume        = {100},
  number        = {10},
  pages         = {103532},
  eid           = {103532},
  doi           = {10.1103/PhysRevD.100.103532}
}

@ARTICLE{marazuela_Apertif_FRB_2025,
  author        = {{Pastor-Marazuela}, In{\'e}s and {van Leeuwen}, Joeri and {Bilous}, Anna and {Connor}, Liam and {Maan}, Yogesh and {Oostrum}, Leon and {Petroff}, Emily and {Vohl}, Dany and {Hess}, Kelley M. and {Orr{\`u}}, Emanuela and {Sclocco}, Alessio and {Wang}, Yuyang},
  title         = {{Comprehensive analysis of the Apertif fast radio burst sample: Similarities with young energetic neutron stars}},
  journal       = {A\&A},
  year          = {2025},
  volume        = {693},
  pages         = {A279},
  eid           = {A279},
  doi           = {10.1051/0004-6361/202450953},
  archiveprefix = {arXiv},
  eprint        = {2406.00482},
  primaryclass  = {astro-ph.HE}
}

@ARTICLE{marinacci_illustris_haloes_2018,
  author        = {{Marinacci}, Federico and {Vogelsberger}, Mark and {Pakmor}, R{\"u}diger and {Torrey}, Paul and {Springel}, Volker and {Hernquist}, Lars and {Nelson}, Dylan and {Weinberger}, Rainer and {Pillepich}, Annalisa and {Naiman}, Jill and {Genel}, Shy},
  title         = {{First results from the IllustrisTNG simulations: radio haloes and magnetic fields}},
  journal       = {MNRAS},
  year          = {2018},
  volume        = {480},
  number        = {4},
  pages         = {5113-5139},
  doi           = {10.1093/mnras/sty2206},
  archiveprefix = {arXiv},
  eprint        = {1707.03396},
  primaryclass  = {astro-ph.CO}
}

@ARTICLE{martizzi_baryons_2019,
  author        = {Martizzi, Davide and Vogelsberger, Mark and Artale, Maria Celeste and Haider, Markus and Torrey, Paul and Marinacci, Federico and Nelson, Dylan and Pillepich, Annalisa and Weinberger, Rainer and Hernquist, Lars and Naiman, Jill and Springel, Volker},
  title         = {Baryons in the {CosmicWeb} of {IllustrisTNG} - {I}: {Gas} in {Knots}, {Filaments}, {Sheets} and {Voids}},
  journal       = {MNRAS},
  year          = {2019},
  volume        = {486},
  number        = {3},
  pages         = {3766--3787},
  doi           = {10.1093/mnras/stz1106}
}

@ARTICLE{mcConnel_ASKAP_2016,
  author        = {{McConnell}, D. and {Allison}, J.~R. and {Bannister}, K. and {Bell}, M.~E. and {Bignall}, H.~E. and {Chippendale}, A.~P. and {Edwards}, P.~G. and {Harvey-Smith}, L. and {Hegarty}, S. and {Heywood}, I. and {Hotan}, A.~W. and {Indermuehle}, B.~T. and {Lenc}, E. and {Marvil}, J. and {Popping}, A. and {Raja}, W. and {Reynolds}, J.~E. and {Sault}, R.~J. and {Serra}, P. and {Voronkov}, M.~A. and {Whiting}, M. and {Amy}, S.~W. and {Axtens}, P. and {Ball}, L. and {Bateman}, T.~J. and {Bock}, D.~C.-J. and {Bolton}, R. and {Brodrick}, D. and {Brothers}, M. and {Brown}, A.~J. and {Bunton}, J.~D. and {Cheng}, W. and {Cornwell}, T. and {DeBoer}, D. and {Feain}, I. and {Gough}, R. and {Gupta}, N. and {Guzman}, J.~C. and {Hampson}, G.~A. and {Hay}, S. and {Hayman}, D.~B. and {Hoyle}, S. and {Humphreys}, B. and {Jacka}, C. and {Jackson}, C.~A. and {Jackson}, S. and {Jeganathan}, K. and {Joseph}, J. and {Koribalski}, B.~S. and {Leach}, M. and {Lensson}, E.~S. and {MacLeod}, A. and {Mackay}, S. and {Marquarding}, M. and {McClure-Griffiths}, N.~M. and {Mirtschin}, P. and {Mitchell}, D. and {Neuhold}, S. and {Ng}, A. and {Norris}, R. and {Pearce}, S. and {Qiao}, R.~Y. and {Schinckel}, A.~E.~T. and {Shields}, M. and {Shimwell}, T.~W. and {Storey}, M. and {Troup}, E. and {Turner}, B. and {Tuthill}, J. and {Tzioumis}, A. and {Wark}, R.~M. and {Westmeier}, T. and {Wilson}, C. and {Wilson}, T.},
  title         = {{The Australian Square Kilometre Array Pathfinder: Performance of the Boolardy Engineering Test Array}},
  journal       = {PASA},
  year          = {2016},
  volume        = {33},
  pages         = {e042},
  eid           = {e042},
  doi           = {10.1017/pasa.2016.37},
  archiveprefix = {arXiv},
  eprint        = {1608.00750},
  primaryclass  = {astro-ph.IM}
}

@ARTICLE{medlock_probing_2024,
  author        = {{Medlock}, Isabel and {Nagai}, Daisuke and {Singh}, Priyanka and {Oppenheimer}, Benjamin and {Angl{\'e}s-Alc{\'a}zar}, Daniel and {Villaescusa-Navarro}, Francisco},
  title         = {{Probing the Circumgalactic Medium with Fast Radio Bursts: Insights from CAMELS}},
  journal       = {ApJ},
  year          = {2024},
  volume        = {967},
  number        = {1},
  pages         = {32},
  eid           = {32},
  doi           = {10.3847/1538-4357/ad3070},
  archiveprefix = {arXiv},
  eprint        = {2403.02313},
  primaryclass  = {astro-ph.GA}
}

@ARTICLE{mellier_Euclid_2025,
  author        = {{Euclid Collaboration} and {Mellier}, Y. and {Abdurro'uf} and {Acevedo Barroso}, J.~A. and {Ach{\'u}carro}, A. and {Adamek}, J. and {Adam}, R. and {Addison}, G.~E. and {Aghanim}, N. and {Aguena}, M. and {Ajani}, V. and {Akrami}, Y. and {Al-Bahlawan}, A. and {Alavi}, A. and {Albuquerque}, I.~S. and {Alestas}, G. and {Alguero}, G. and {Allaoui}, A. and {Allen}, S.~W. and {Allevato}, V. and {Alonso-Tetilla}, A.~V. and {Altieri}, B. and {Alvarez-Candal}, A. and {Alvi}, S. and {Amara}, A. and {Amendola}, L. and {Amiaux}, J. and {Andika}, I.~T. and {Andreon}, S. and {Andrews}, A. and {Angora}, G. and {Angulo}, R.~E. and {Annibali}, F. and {Anselmi}, A. and {Anselmi}, S. and {Arcari}, S. and {Archidiacono}, M. and {Aric{\`o}}, G. and {Arnaud}, M. and {Arnouts}, S. and {Asgari}, M. and {Asorey}, J. and {Atayde}, L. and {Atek}, H. and {Atrio-Barandela}, F. and {Aubert}, M. and {Aubourg}, E. and {Auphan}, T. and {Auricchio}, N. and {Aussel}, B. and {Aussel}, H. and {Avelino}, P.~P. and {Avgoustidis}, A. and {Avila}, S. and {Awan}, S. and {Azzollini}, R. and {Baccigalupi}, C. and {Bachelet}, E. and {Bacon}, D. and {Baes}, M. and {Bagley}, M.~B. and {Bahr-Kalus}, B. and {Balaguera-Antolinez}, A. and {Balbinot}, E. and {Balcells}, M. and {Baldi}, M. and {Baldry}, I. and {Balestra}, A. and {Ballardini}, M. and {Ballester}, O. and {Balogh}, M. and {Ba{\~n}ados}, E. and {Barbier}, R. and {Bardelli}, S. and {Baron}, M. and {Barreiro}, T. and {Barrena}, R. and {Barriere}, J.-C. and {Barros}, B.~J. and {Barthelemy}, A. and {Bartolo}, N. and {Basset}, A. and {Battaglia}, P. and {Battisti}, A.~J. and {Baugh}, C.~M. and {Baumont}, L. and {Bazzanini}, L. and {Beaulieu}, J.-P. and {Beckmann}, V. and {Belikov}, A.~N. and {Bel}, J. and {Bellagamba}, F. and {Bella}, M. and {Bellini}, E. and {Benabed}, K. and {Bender}, R. and {Benevento}, G. and {Bennett}, C.~L. and {Benson}, K. and {Bergamini}, P. and {Bermejo-Climent}, J.~R. and {Bernardeau}, F. and {Bertacca}, D. and {Berthe}, M. and {Berthier}, J. and {Bethermin}, M. and {Beutler}, F. and {Bevillon}, C. and {Bhargava}, S. and {Bhatawdekar}, R. and {Bianchi}, D. and {Bisigello}, L. and {Biviano}, A. and {Blake}, R.~P. and {Blanchard}, A. and {Blazek}, J. and {Blot}, L. and {Bosco}, A. and {Bodendorf}, C. and {Boenke}, T. and {B{\"o}hringer}, H. and {Boldrini}, P. and {Bolzonella}, M. and {Bonchi}, A. and {Bonici}, M. and {Bonino}, D. and {Bonino}, L. and {Bonvin}, C. and {Bon}, W. and {Booth}, J.~T. and {Borgani}, S. and {Borlaff}, A.~S. and {Borsato}, E. and {Bose}, B. and {Botticella}, M.~T. and {Boucaud}, A. and {Bouche}, F. and {Boucher}, J.~S. and {Boutigny}, D. and {Bouvard}, T. and {Bouwens}, R. and {Bouy}, H. and {Bowler}, R.~A.~A. and {Bozza}, V. and {Bozzo}, E. and {Branchini}, E. and {Brando}, G. and {Brau-Nogue}, S. and {Brekke}, P. and {Bremer}, M.~N. and {Brescia}, M. and {Breton}, M.-A. and {Brinchmann}, J. and {Brinckmann}, T. and {Brockley-Blatt}, C. and {Brodwin}, M. and {Brouard}, L. and {Brown}, M.~L. and {Bruton}, S. and {Bucko}, J. and {Buddelmeijer}, H. and {Buenadicha}, G. and {Buitrago}, F. and {Burger}, P. and {Burigana}, C. and {Busillo}, V. and {Busonero}, D. and {Cabanac}, R. and {Cabayol-Garcia}, L. and {Cagliari}, M.~S. and {Caillat}, A. and {Caillat}, L. and {Calabrese}, M. and {Calabro}, A. and {Calderone}, G. and {Calura}, F. and {Camacho Quevedo}, B. and {Camera}, S. and {Campos}, L. and {Ca{\~n}as-Herrera}, G. and {Candini}, G.~P. and {Cantiello}, M. and {Capobianco}, V. and {Cappellaro}, E. and {Cappelluti}, N. and {Cappi}, A. and {Caputi}, K.~I. and {Cara}, C. and {Carbone}, C. and {Cardone}, V.~F. and {Carella}, E. and {Carlberg}, R.~G. and {Carle}, M. and {Carminati}, L. and {Caro}, F. and {Carrasco}, J.~M. and {Carretero}, J. and {Carrilho}, P. and {Carron Duque}, J. and {Carry}, B.},
  title         = {{Euclid: I. Overview of the Euclid mission}},
  journal       = {A\&A},
  year          = {2025},
  volume        = {697},
  pages         = {A1},
  eid           = {A1},
  doi           = {10.1051/0004-6361/202450810},
  archiveprefix = {arXiv},
  eprint        = {2405.13491},
  primaryclass  = {astro-ph.CO}
}

@ARTICLE{mitchell_how_2013,
  author        = {{Mitchell}, Peter D. and {Lacey}, Cedric G. and {Baugh}, Carlton M. and {Cole}, Shaun},
  title         = {{How well can we really estimate the stellar masses of galaxies from broad-band photometry?}},
  journal       = {MNRAS},
  year          = {2013},
  volume        = {435},
  number        = {1},
  pages         = {87-114},
  doi           = {10.1093/mnras/stt1280},
  archiveprefix = {arXiv},
  eprint        = {1303.7228},
  primaryclass  = {astro-ph.CO}
}

@ARTICLE{mobasher_critical_2015,
  author        = {{Mobasher}, Bahram and {Dahlen}, Tomas and {Ferguson}, Henry C. and {Acquaviva}, Viviana and {Barro}, Guillermo and {Finkelstein}, Steven L. and {Fontana}, Adriano and {Gruetzbauch}, Ruth and {Johnson}, Seth and {Lu}, Yu and {Papovich}, Casey J. and {Pforr}, Janine and {Salvato}, Mara and {Somerville}, Rachel S. and {Wiklind}, Tommy and {Wuyts}, Stijn and {Ashby}, Matthew L.~N. and {Bell}, Eric and {Conselice}, Christopher J. and {Dickinson}, Mark E. and {Faber}, Sandra M. and {Fazio}, Giovanni and {Finlator}, Kristian and {Galametz}, Audrey and {Gawiser}, Eric and {Giavalisco}, Mauro and {Grazian}, Andrea and {Grogin}, Norman A. and {Guo}, Yicheng and {Hathi}, Nimish and {Kocevski}, Dale and {Koekemoer}, Anton M. and {Koo}, David C. and {Newman}, Jeffrey A. and {Reddy}, Naveen and {Santini}, Paola and {Wechsler}, Risa H.},
  title         = {{A Critical Assessment of Stellar Mass Measurement Methods}},
  journal       = {ApJ},
  year          = {2015},
  volume        = {808},
  number        = {1},
  pages         = {101},
  eid           = {101},
  doi           = {10.1088/0004-637X/808/1/101},
  archiveprefix = {arXiv},
  eprint        = {1505.01501},
  primaryclass  = {astro-ph.GA}
}

@ARTICLE{naiman_illustris_chemical_2018,
  author        = {{Naiman}, Jill P. and {Pillepich}, Annalisa and {Springel}, Volker and {Ramirez-Ruiz}, Enrico and {Torrey}, Paul and {Vogelsberger}, Mark and {Pakmor}, R{\"u}diger and {Nelson}, Dylan and {Marinacci}, Federico and {Hernquist}, Lars and {Weinberger}, Rainer and {Genel}, Shy},
  title         = {{First results from the IllustrisTNG simulations: a tale of two elements - chemical evolution of magnesium and europium}},
  journal       = {MNRAS},
  year          = {2018},
  volume        = {477},
  number        = {1},
  pages         = {1206-1224},
  doi           = {10.1093/mnras/sty618},
  archiveprefix = {arXiv},
  eprint        = {1707.03401},
  primaryclass  = {astro-ph.GA}
}

@ARTICLE{nelson_illustris_galaxy_2018,
  author        = {{Nelson}, Dylan and {Pillepich}, Annalisa and {Springel}, Volker and {Weinberger}, Rainer and {Hernquist}, Lars and {Pakmor}, R{\"u}diger and {Genel}, Shy and {Torrey}, Paul and {Vogelsberger}, Mark and {Kauffmann}, Guinevere and {Marinacci}, Federico and {Naiman}, Jill},
  title         = {{First results from the IllustrisTNG simulations: the galaxy colour bimodality}},
  journal       = {MNRAS},
  year          = {2018},
  volume        = {475},
  number        = {1},
  pages         = {624-647},
  doi           = {10.1093/mnras/stx3040},
  archiveprefix = {arXiv},
  eprint        = {1707.03395},
  primaryclass  = {astro-ph.GA}
}

@ARTICLE{nicastro_missing_2018,
  author        = {{Nicastro}, F. and {Kaastra}, J. and {Krongold}, Y. and {Borgani}, S. and {Branchini}, E. and {Cen}, R. and {Dadina}, M. and {Danforth}, C.~W. and {Elvis}, M. and {Fiore}, F. and {Gupta}, A. and {Mathur}, S. and {Mayya}, D. and {Paerels}, F. and {Piro}, L. and {Rosa-Gonzalez}, D. and {Schaye}, J. and {Shull}, J.~M. and {Torres-Zafra}, J. and {Wijers}, N. and {Zappacosta}, L.},
  title         = {{Observations of the missing baryons in the warm-hot intergalactic medium}},
  journal       = {Nature},
  year          = {2018},
  volume        = {558},
  number        = {7710},
  pages         = {406-409},
  doi           = {10.1038/s41586-018-0204-1},
  archiveprefix = {arXiv},
  eprint        = {1806.08395},
  primaryclass  = {astro-ph.GA}
}

@ARTICLE{niu_repeating_2022,
  author        = {{Niu}, C.-H. and {Aggarwal}, K. and {Li}, D. and {Zhang}, X. and {Chatterjee}, S. and {Tsai}, C.-W. and {Yu}, W. and {Law}, C.~J. and {Burke-Spolaor}, S. and {Cordes}, J.~M. and {Zhang}, Y.-K. and {Ocker}, S.~K. and {Yao}, J.-M. and {Wang}, P. and {Feng}, Y. and {Niino}, Y. and {Bochenek}, C. and {Cruces}, M. and {Connor}, L. and {Jiang}, J.-A. and {Dai}, S. and {Luo}, R. and {Li}, G.-D. and {Miao}, C.-C. and {Niu}, J.-R. and {Anna-Thomas}, R. and {Sydnor}, J. and {Stern}, D. and {Wang}, W.-Y. and {Yuan}, M. and {Yue}, Y.-L. and {Zhou}, D.-J. and {Yan}, Z. and {Zhu}, W.-W. and {Zhang}, B.},
  title         = {{A repeating fast radio burst associated with a persistent radio source}},
  journal       = {Nature},
  year          = {2022},
  volume        = {606},
  number        = {7916},
  pages         = {873-877},
  doi           = {10.1038/s41586-022-04755-5},
  archiveprefix = {arXiv},
  eprint        = {2110.07418},
  primaryclass  = {astro-ph.HE}
}

@ARTICLE{ocker_ne2025_2026,
  author        = {Ocker, S. K. and Cordes, J. M.},
  title         = {{NE2025}: {An} {Updated} {Electron} {Density} {Model} for the {Galactic} {Interstellar} {Medium}},
  journal       = {ApJ},
  year          = {2026},
  volume        = {1002},
  number        = {1},
  pages         = {3},
  doi           = {10.3847/1538-4357/ae5825},
  archiveprefix = {arXiv},
  eprint        = {2602.11838}
}

@BOOK{rybicki_1979_radiative,
  author        = {{Rybicki}, George B. and {Lightman}, Alan P.},
  title         = {{Radiative processes in astrophysics}},
  year          = {1979},
  publisher     = {Wiley},
  address       = {New York}
}

\begin{appendix}
\nolinenumbers

\section{Details on mock catalogue}
\subsection{Catalogue using photometric data}
\label{ap:photometry}
Here we present a detailed overview of the mock catalogue created in Section \ref{sub:mock_photometry}.  We focus on the VIS band, and all the instrument specifications are obtained from the \texttt{SurveyCodex}\footnote{\url{https://github.com/LSSTDESC/surveycodex}} package. The selection depends on \textit{Euclid}'s detection limit of 24.5 ($10 \sigma$) in the VIS and 24.0 ($5 \sigma$) in the NISP band \citep{romelli_Euclid_2025, cropper_Eucild_VIS_2025}.

\begin{enumerate}
    \item We begin with the IllustrisTNG subhalo catalogue and the absolute magnitudes ($M$) in eight bands. First, subhaloes with floored absolute magnitudes ($10^{37}$) are removed from the full IllustrisTNG catalogue.

    \item For a given absolute magnitude, the apparent magnitude ($m$) is calculated using the luminosity distance, which is known from the underlying cosmology and the redshift of the subhalo catalogue.

    \item The SED fitting is carried out using \texttt{Kcorrect}, which expects input in linearised magnitudes or maggies, calculated from the apparent magnitudes ($m$) as $m = -2.5 \log_{10}(\mathrm{maggies})$.

    \item The SED fitting step returns the rest-frame SED and the K correction values via coefficient fitting as shown in Fig.~\ref{fig:sed_fitting}.
    \begin{figure}[!ht]
    \centering
    \includegraphics[width=0.9\linewidth]{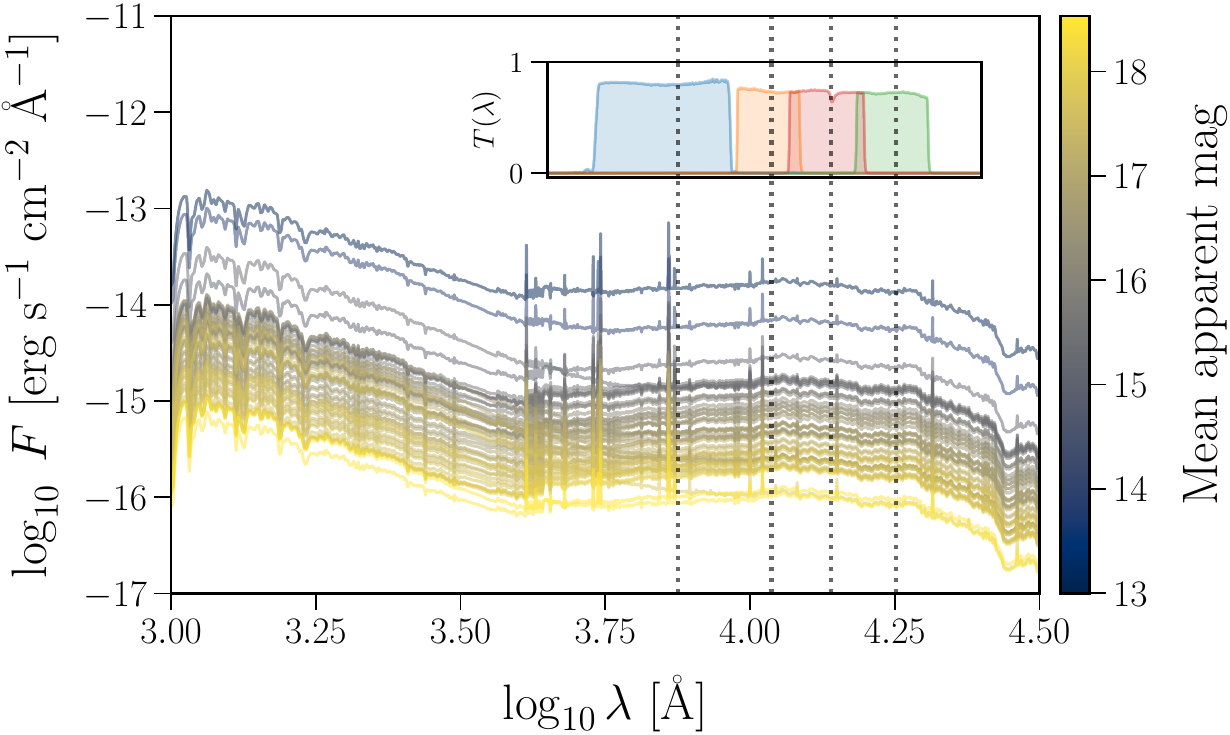}
    \caption{\small The SEDs of the first 50 subhaloes are shown. The colour bar shows the average apparent magnitudes of all the IllustrisTNG photometric bands. The throughputs of the four \textit{Euclid} filters, which act as convolution kernels, are presented in the inset. The vertical dotted lines are the effective wavelengths of bands VIS, Y, J and H, respectively.}
    \label{fig:sed_fitting}
    \vspace{.1cm}
    \end{figure}

    \item We then calculate the rest-frame flux and redshift it along with the wavelength.

    \item We convolve with \textit{Euclid}'s filter response curves from \texttt{Speclite} and calculate the AB magnitudes in the four \textit{Euclid} bands. We show the apparent magnitudes from IllustrisTNG and their equivalent in \textit{Euclid} bands in Fig.~\ref{fig:bands}.
    \begin{figure}[!ht]
    \centering
    \includegraphics[width=0.9\linewidth]{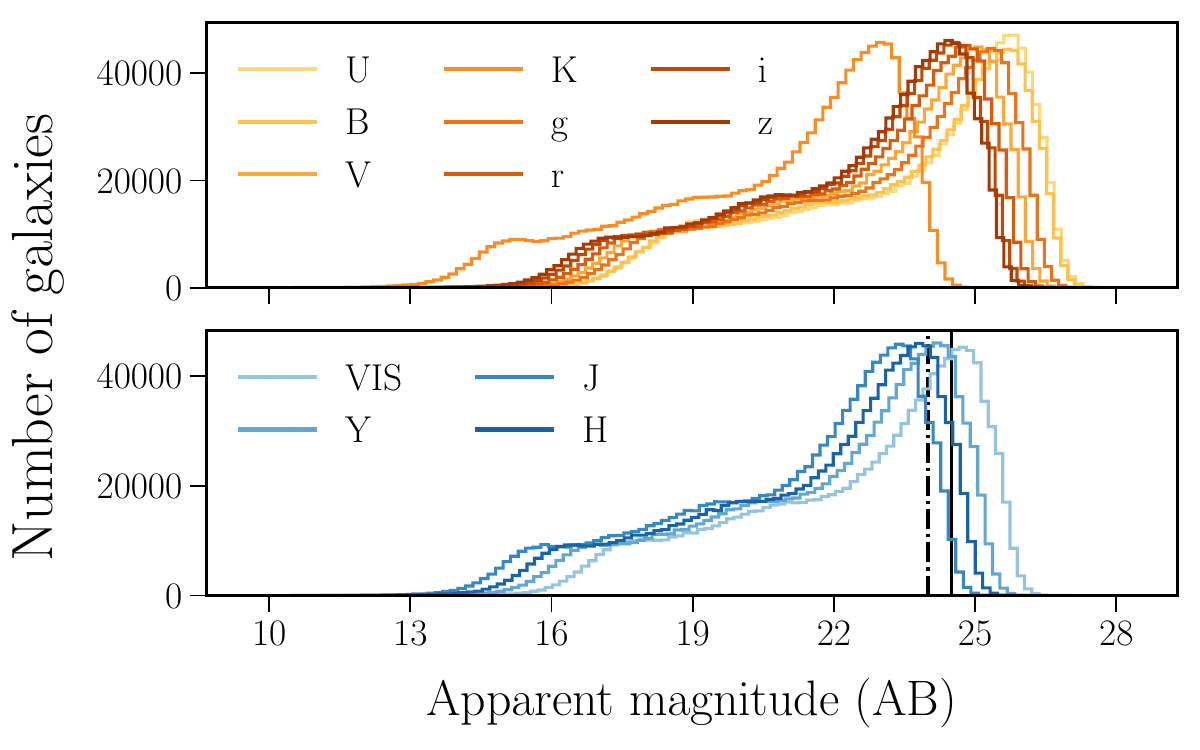}
    \caption{\small Top: Histogram of apparent magnitudes in IllustrisTNG photometric bands. Bottom: Equivalent bands in a \textit{Euclid}-like survey obtained via SED fitting. The two vertical lines represent the magnitude cuts in \textit{Euclid} at 24.0 (dot-dash) for NISP and 24.5 (solid) for the VIS band.}
    \label{fig:bands}
    \vspace{.1cm}
    \end{figure}

    \item Instead of a hard selection cut at \textit{Euclid}'s magnitude limit, we base our selection criteria on the number of electrons ($N_\mathrm{e}$) via
    \begin{align}
        N_\mathrm{e}=10^{-0.4(m-m_\mathrm{zp})}\times t_\mathrm{exp} \;,
    \end{align}
    where $m$ is the apparent magnitude, $m_\mathrm{zp}$ = 25.91 is the VIS zero-point value and $t_\mathrm{exp} = 565$ s is the exposure time. With this step, we can include instrument-specific noise.

    \item The sky brightness is similarly given as
    \begin{align}
        S_\mathrm{pix} = 10^{-0.4(\mu_\mathrm{sky}-m_\mathrm{zp})}\times t_{\mathrm{exp}} \times (\Delta \theta)^2 \;,
    \end{align}
    where $\mu_\mathrm{sky} = 22.3203$ is \textit{Euclid}'s sky brightness. Note that we multiply by the area, $(\Delta \theta)^2$, so that $S_\mathrm{pix}$ is in electrons pixel$^{-1}$.

    \item To simulate the galaxy brightness profile, we utilise \texttt{GALSIM}\footnote{\url{https://github.com/GalSim-developers/GalSim}} \citep{rowe_GALSIM_2015}. Specifically, a S\'{e}rsic profile is used with an index of 1, and we set the half-light radius equal to the half-mass stellar radius from IllustrisTNG, accessed via the ``SubhaloHalfmassRadType'' field. The total flux is anchored to the number of electrons in the exposure to maintain uniformity.

    \item Next, we add two convolution kernels
    \begin{itemize}
        \item PSF: The point spread function (PSF) for \textit{Euclid} is Gaussian with a full width at half maximum (FWHM) of 0.17''.
        \item Optical: The optical effects are included as the Airy disc model in \texttt{GALSIM} using the effective wavelength of 750.637 nm for the VIS filter, a diameter of 1.2 m, and an obscuration of 0.13.
    \end{itemize}

    \item The final true image is the convolution of the S\'{e}rsic profile with the PSF and the optical effects.

    \item Noise: We add a mean-zero Gaussian read noise of 4.4 electrons\footnote{\url{https://www.Euclid-ec.org/public/mission/vis/}} along with Poisson noise.

    \item We choose a fixed aperture radius of 0\farcs6 and the corresponding flux is
    \begin{align}
        \sigma_\mathrm{Flux} = \sqrt{F_\mathrm{aper} + N_\mathrm{pix}S_\mathrm{pix} + N_\mathrm{pix}\sigma^2_\mathrm{read}} \;,
    \end{align}
    where $F_\mathrm{aper}$ is the measured source electrons inside the aperture, $S_\mathrm{pix}$ is the sky electrons/pixel, $\sigma_\mathrm{read}$ is the read noise and $N_\mathrm{pix}$ is the number of pixels in the aperture, i.e.,
   \begin{align}
        N_\mathrm{pix} = \pi \times \left[ \frac{0.6''}{r_\mathrm{PS}} \right]^2 \;,
    \end{align}
    where $r_\mathrm{PS}=$0.1 arcsec/pixel is \textit{Euclid}'s pixel scale.

    \item The signal-to-noise ratio (SNR) is then
    \begin{align}
        \mathrm{SNR} = \frac{F_\mathrm{aper}}{\sigma_\mathrm{Flux}} \;.
    \end{align}

    \item In the penultimate step, we impose the VIS detection criteria of a magnitude cut at $I_\mathrm{E} \leq 24.5$ and SNR $\geq 10$ on the full subhalo catalogue \citep{scaramella_Euclid_2022}. A subhalo that satisfies both criteria simultaneously is considered a true detection and included in the final catalogue. See Fig.~\ref{fig:detection} for a visual representation of the pipeline, where red/green indicates if the sample is rejected/selected.

    \begin{figure}[!ht]
    \centering
    \includegraphics[width=0.9\linewidth]{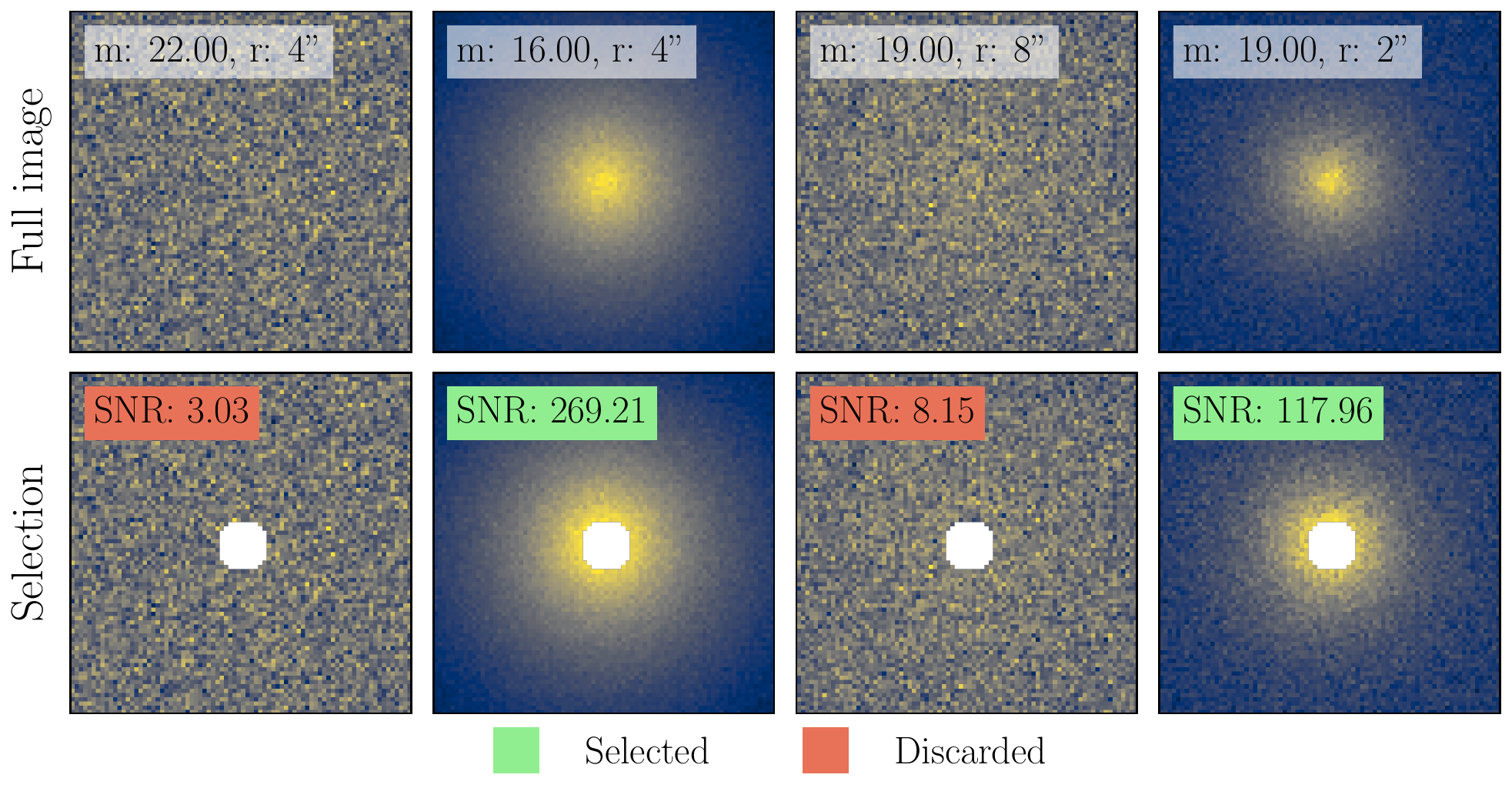}
    \caption{\small A visual representation of the \textit{Euclid} detection pipeline. The top row is the full simulated image for a given apparent magnitude and stellar half-mass radius. The bottom row shows the masked selection of subhaloes, with the corresponding SNR values displayed. The green and red colours in the bottom panel indicate whether the sample is selected. In the first two columns, the magnitude is increased while the stellar half-mass radius is held fixed, thereby raising the SNR above the detection threshold. The third and fourth columns show the effect of this radius for a fixed magnitude. As the flux is concentrated in a smaller area, the SNR exceeds the detection limit.}
    \label{fig:detection}
    \vspace{.1cm}
    \end{figure}

    \item As the sample selection in real \textit{Euclid} observations is not complete, we remove 40 per cent of the selected samples following \citet{hamaus_forecasts_2022, kraljic_Euclid_3D_2026}. The remaining 60 per cent is our final mock catalogue.
\end{enumerate}
For SED fitting, other alternatives are available, such as Bagpipes \citep{carnall_Bagpipes_2018}, CIGALE \citep{boquien_CIGALE_2019}, and Prospector \citep{johnson_Prospector_2021}\footnote{A comprehensive list can be found at \url{http://www.sedfitting.org/Fitting.html}}. Moreover, note that we have not included any complex post-processing of the images before computing the flux. A more realistic and physically accurate approach is to use a radiative transfer code such as SKIRT\footnote{\url{https://skirt.ugent.be/}} \citep{camps_SKIRT_2020} and apply dust models \citep{stevens_dust_2005, jones_THEMIS_2017}. For \textit{Euclid}-specific data, another alternative is to use the dedicated package Phosphoros that calculates the photometric values and the bandpass effects \citep{paltani_Phosphoros_2024}.

However, as we eventually create a 3D density field with the galaxies, the required level of accuracy does not warrant a more complex detection pipeline. Such detailed applications can be found in \citet{rodriguez_optical_2019, eisert_ERGO_2024, gong_morphological_2025, goncalves_detectability_2026}.

\subsection{Stellar and halo mass estimation}
\label{ap:stellar_and_halo_mass}
For our sample catalogue, IllustrisTNG provides the stellar and halo mass as ground truths. For a real galaxy survey, these are not accessible trivially and need to be inferred from observable properties. To showcase a more realistic application, we use the luminosities and the apparent magnitudes in the four \textit{Euclid} bands in Appendix \ref{ap:photometry} to first estimate the stellar mass.

Conversion from the apparent magnitudes to solar luminosities requires the absolute solar magnitudes, which are taken from \citet{wilmer_absolute_2018}\footnote{\url{https://mips.as.arizona.edu/~cnaw/sun.html}}. One of the most conventional ways to obtain the stellar mass is to employ SED fitting \citetext{see e.g. \citealp{mitchell_how_2013, kouroumpatzakis_SFR_2023}; and for a review \citealp{mobasher_critical_2015}}. However, such a dedicated application is not the focus of our study. With access to the true stellar mass from the simulation, we instead employ a random-forest regression to estimate the mass-to-light ratio, $M/L$. Typically, one assumes $M/L = 0.4$-$0.6$ for galaxies with a non-zero star-formation rate (SFR) and $M/L = 0.8$-$1$ for quenched galaxies. In our model, we include the specific star formation rate and colour information via $m_i - m_j$, where $\{i,j\}$ denote different filters, as features that allow a dynamical $M/L$.

\begin{figure}[!ht]
    \centering
    \includegraphics[width=0.9\linewidth]{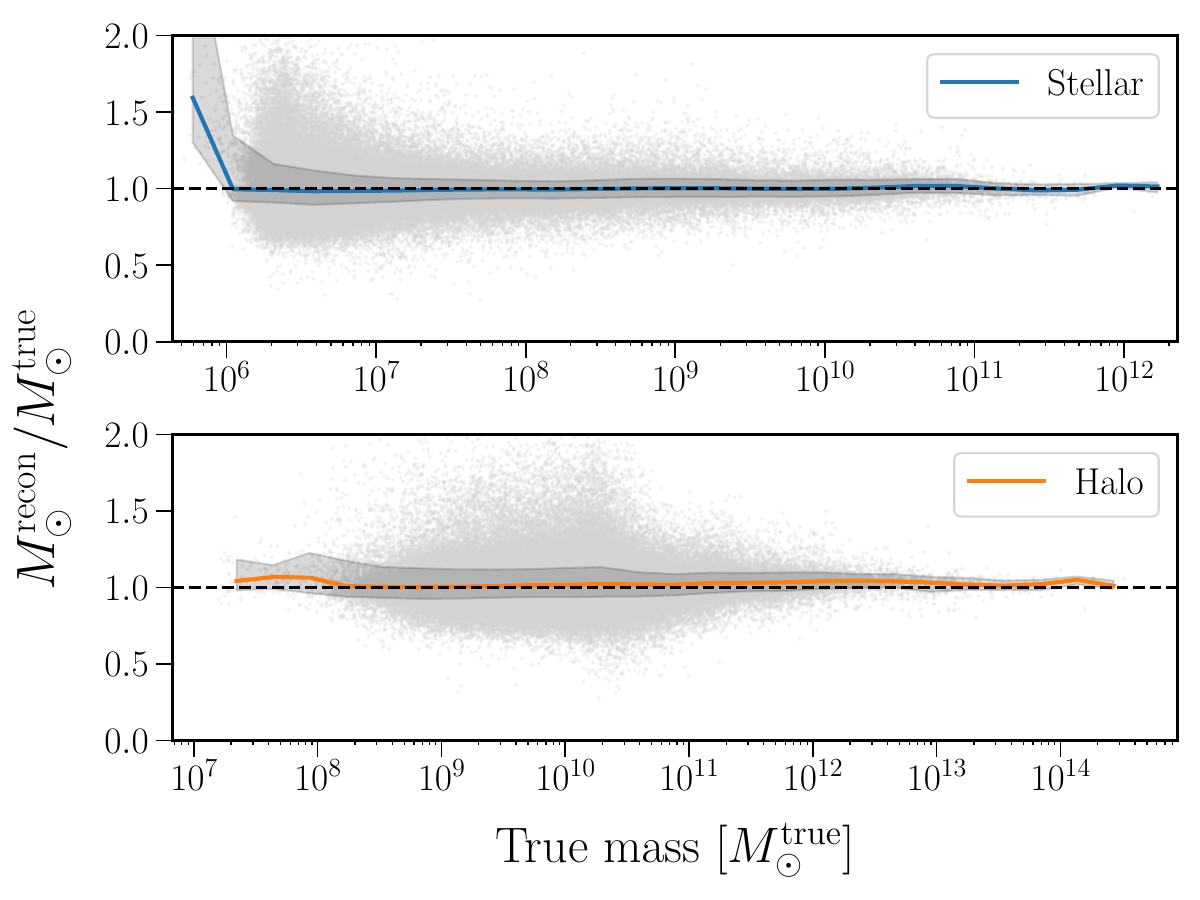}
    \caption{\small Shown are the ratios of the reconstructed and true stellar (top panel) and halo (bottom panel) masses against the true values from IllustrisTNG. The horizontal dotted lines are at unity and represent a perfect reconstruction. The grey scatter points in both panels show the calculated values, where the means are indicated by solid lines. The shaded areas are the 68 per cent error margins around the mean.}
    \label{fig:stellar}
    \vspace{.1cm}
\end{figure}

Further, we split the full dataset into 7:3 for training and testing. As the final target, we use $60$ per cent of the true stellar mass, i.e., we assume that the true stellar mass for $60$ per cent of the galaxies is known. Post-training, we obtain the stellar mass for the full sample, and the result is shown in the top panel of Fig.~\ref{fig:stellar}.

To go from stellar mass to the halo mass, we need the stellar-to-halo mass relation \citep[SHMR,][]{girelli_stellar_2020}, which can be inferred, e.g. via subhalo abundance matching \citep[SHAM,][]{legrand_cosmos_2019}, weak lensing \citep{li_WL_2025, zacharegkas_constraining_2025}, neural networks \citep{villanueva_inferring_2022} or a halo occupation distribution prescription \citep[HOD,][]{leauthaud_theoretical_2011, coupon_galaxy-halo_2015}. Here, we can use the relation from \citet{behroozi_lack_2013, behroozi_average_2013}\footnote{\url{https://www.peterbehroozi.com/data.html}} and interpolate to estimate the halo mass for a given stellar mass. Note that this relation is non-trivially redshift-dependent \citep{legrand_cosmos_2019}. However, with \textit{Euclid}'s stringent spectroscopic and photometric uncertainty limits, the errors only from redshifts are negligible. The largest source of error in this conversion (0.15-0.25 dex) is the intrinsic scatter of the stellar mass at a fixed halo mass. The reconstructed halo masses are shown in the bottom panel of Fig.~\ref{fig:stellar}.

\end{appendix}
\end{document}